\documentclass[11pt]{article}
\usepackage{epsfig}\parskip 5pt plus 1pt
\usepackage{bbm}
\usepackage{amsmath}
\usepackage{amssymb}
\usepackage{amsfonts}
\usepackage{mathrsfs}
\usepackage{graphicx}
\usepackage{cite}
\usepackage{color}
\newcommand{\be}{\begin{equation}}
\newcommand{\ee}{\end{equation}}
\newcommand{\bea}{\begin{eqnarray}}
\newcommand{\eea}{\end{eqnarray}}

\begin{document}
\begin{titlepage}
\vspace*{-2cm}
\flushright{ULB-TH/08-17}\\
\vspace{-4mm}
\flushright{NSF-KITP-08-79}
\vskip 1.5cm
\begin{center}
{\Large \bf Is leptogenesis falsifiable at LHC?}
\end{center}
\vskip 0.5cm
\begin{center}
{\large Jean-Marie Fr\`ere$^{a,b}$}\footnote{frere@ulb.ac.be},
{\large Thomas Hambye$^a$}\footnote{thambye@ulb.ac.be} and
{\large Gilles Vertongen$^a$}\footnote{gilles.vertongen@ulb.ac.be}\\
\vskip .7cm
$^a$Service de Physique Th\'eorique,\\
\vspace{0.5mm}
Universit\'e Libre de Bruxelles, 1050 Brussels, Belgium\\
\vspace{2.5mm}
$^b$KITP, University of California, Santa Barbara CA93106, USA

\end{center}
\vskip 0.5cm

\begin{abstract}
It is well known that the leptogenesis mechanism offers an
attractive possibility to explain the baryon asymmetry of the
universe. Its particular robustness however comes with one major
difficulty: it will be very hard if not impossible to \emph{test}
experimentally in a foreseeable future, as most of the mechanics
\emph{typically} takes place at high energy or \emph{results} from
suppressed interactions, without unavoidable low-energy
implications. An alternate approach is taken by asking: can it be at
least falsified? We show that \emph{possible} discoveries at current
and future colliders, most notably that of right-handed gauge
interactions, would indeed forbid at least the "canonical"
leptogenesis mechanisms, namely those based on right-handed neutrino
decay. General lower bounds for successful leptogenesis on 
the mass of the right-handed gauge boson 
$W_R$ are given. Other possibilities to falsify leptogenesis, including 
from the observation of a $Z'$,
are also considered.

\end{abstract}
\end{titlepage}
\setcounter{footnote}{0}
\vskip2truecm

\newpage
\section{Introduction}

The recent evidence for neutrino masses has brought forward
leptogenesis \cite{fy} as a very attractive mechanism to explain the baryon
asymmetry of the universe. Along this mechanism, the baryon asymmetry of the universe is explained by the same interactions as
the ones which can explain the neutrino masses. In the most
straightforward seesaw model, which assumes right-handed neutrinos
in addition to the standard model particles, both neutrino masses
and leptogenesis originate from the Yukawa interactions and lepton
number violating  Majorana masses of the right-handed neutrinos
\begin{equation}
{\cal L} \owns
- \overline{L} \,{\widetilde H} \, {Y_\nu^\dagger}  \, N
-\frac{1}{2}\,\overline{N} \, {m_N} \,{{N}^c}
+\text{h.c.}
\end{equation}
where $L$ stands for the lepton weak doublets
and $\tilde H$
is related to the standard Brout-Englert-Higgs (hereafter simply Higgs) doublet $H \equiv (H^+, H^0)$
by $\tilde H = i \tau_{2} H^*$.

However, testing this mechanism will be a very difficult task for
several reasons. If the right-handed neutrinos have a hierarchical
mass spectrum, due to neutrino mass constraints, leptogenesis
through $N$ decay can lead
to the observed amount of baryon asymmetry e.g.~only if it involves
right-handed neutrinos with masses above $\sim 10^8$~GeV \cite{predi,di}.
As a result they cannot be produced at colliders.
Moreover there are many more parameters in the Yukawa coupling
matrices which can play an important role for leptogenesis, than
there are (not too suppressed) low energy observables which could
constrain these parameters.\footnote{A possible
exception to that arises in the supersymmetric case from the effects of
Yukawa couplings on the running of the slepton masses \cite{raidal}. This
nevertheless assumes that universality of lepton soft mass terms must be present
(an assumption which requires to be tested) and, for any real test
of leptogenesis, would require to observe a long series of rare
leptonic decays not necessarily expected to be all close to the
present corresponding experimental bounds.}

If the right-handed neutrinos have instead a quasi-degenerate
spectrum (for at least 2 of them), leptogenesis can be efficient at
lower scales \cite{lowscale} but generically in this case the neutrino mass
constraints require suppressed values of Yukawa couplings, which
hampers their production at colliders.

For leptogenesis to be both efficient and tested at low energy, not only
is a  quasi-degeneracy between 2 right-handed neutrinos required,
but also a special flavour structure which allows for larger Yukawa
couplings while preserving  the light neutrino mass
constraints,\footnote{This case can be realized 
if the Yukawa induced dimension 6 operator coefficients are unsuppressed (decoupling from
the suppressed neutrino mass dimension 5 ones). This 
does not necessarily require cancellations of the various entries. It
requires that some of entries are smaller than others, as in the
inverse seesaw, see e.g.~\cite{ValleConcha,Smirnov,ABBGH}. But it e.g.~leads
only to lepton conserving channels with rather large background at
LHC \cite{delAguila}.} and/or a right-handed neutrino production
mechanisms other than through the Yukawas and associated neutrino
mixings.

In this paper we consider the problem of testing leptogenesis
mechanisms  the other way around. While they cannot confirm
leptogenesis, could low energy observations at least  exclude it? We
propose one particularly clear possibility, namely the observation
of  a right-handed charged gauge boson $W_R$. It is known that for
high mass right-handed neutrinos and $W_R$, around $10^{10}$ GeV or
higher, the $W_R$ can have suppression effects on leptogenesis through dilution
and scattering, but, in the specific case of reheating after inflation, they can also
boost the $N$ abundances \cite{Carlier,Cosme,2sarkar} and hence relax the constraints on Yukawa couplings.
Not surprisingly, with a low scale $W_R$ the suppression effects are dramatically enhanced. Actually,
see section 2, they turn out to be so strongly enhanced that, even
with a maximal CP asymmetry of order unity, leptogenesis cannot be a
sufficient cause of the matter excess anymore.

Right-handed gauge interactions lead in particular to much larger
suppression effects at low scale than left-handed interactions do in
other contexts (i.e.~than in leptogenesis from scalar
\cite{typeIIlepto,typeIIleptoeffic} or fermion \cite{typeIIIlepto}
triplet decays, whose efficiency have been calculated in
Refs.~\cite{typeIIleptoeffic,typeIIIlepto}). This is due to the fact
that at the difference of triplets,  a single $N$ can interact
through $W_R$ exchange with fermions which are all in thermal
equilibrium, which induces more efficient, and hence dangerous,
scatterings and decays. In particular, some of the scatterings
involving the $W_R$ turn out to induce a very large suppression due
to the fact that they do not decouple through a Boltzmann
suppression. The production of $N$'s through a light $W_R$, often
presented as the easiest way to produce $N$'s, is therefore
incompatible with successful leptogenesis, and even enhanced $N$
production from reheating cannot compensate for the large
suppression. The lower bounds on the mass of the $W_R$, required for successful leptogenesis,
are given in section 3. 

The possible discovery of a low-energy $W_R$ has recently been the
object of several analysis by LHC collaborations
\cite{Wlhcferrari,Wlhckras,LHCstudies}. It should be feasible up to
$m_{W_R} \sim 3$-5~TeV (see more details, and additional possible
searches, in section 7).

The observation of a $W_R$ is not the only possibility to exclude canonical neutrino decay leptogenesis
from current energy data.
 We give a list of other possibilities in section 5, considering in particular the implications
 of the observation of a $Z'$ at LHC.
The case of other leptogenesis seesaw models with not only or
without right-handed neutrinos is briefly considered in section 6.

\section{Leptogenesis in presence of a low scale $W_R$}

As well known the net rate of baryon asymmetry is given in any
leptogenesis model by 3 ingredients, the CP asymmetry of the
decaying particle, $\varepsilon_N$ for a right-handed neutrino, the
Boltzmann equations which determine the efficiency $\eta$ and the
$L$ to $B$ sphaleron conversion rate, which we denote by $r_{{\cal
L} \rightarrow {\cal B}}$.
Let us first discuss and present our results for the case where the lepton asymmetry is
created from the decay of a single right-handed 
neutrino, $N$.\footnote{We
will not consider finite temperature effects which are not
expected to change our conclusions.}
Later on we will discuss the generalization to more right-handed neutrinos.
In this case, from these 3 ingredients the net baryon asymmetry produced by the
$N$ decays is:
\begin{equation}
Y_{\cal B} = Y_{\cal L} \, r_{{\cal L}\rightarrow {\cal B}} = \varepsilon_N \,\eta \, Y_N^{eq}(T\gg m_N) \, r_{{\cal L}\rightarrow {\cal B}}.
\end{equation}
with $Y_i\equiv n_i/s$, $Y_{\cal B}\equiv Y_{B}-Y_{\bar{B}}$,
$Y_{\cal L}\equiv Y_L-Y_{\bar{L}}$, $n_i$ the comoving number
density of the species "i", "eq" refering to the equilibrium number
density,  and $s$ the comoving entropy
density. For a particle previously in thermal equilibrium,
the efficiency is unity by definition in absence of
any washout effect from inverse decays or scatterings.
If all lepton asymmetry has been produced before the sphaleron decoupling
at the electroweak phase transition and if the sphalerons have had the time
to thermalize  completely the $L$ abundance,  the conversion ratio between lepton and
baryon number is given by \cite{Khlebnikov:1988sr}
\begin{equation}\label{LtoBfactor}
r_{{\cal L}\rightarrow {\cal B}}= -\frac{8\,n_f + 4 \,n_H}{22\,n_f + 13 \,n_H}=-\frac{28}{79},
\end{equation}
where the last equality refers to the SM value, with $n_f$ the number of fermion
families and $n_H$ the number of Higgs doublets.

In the right-handed neutrino decay leptogenesis model without any $W_R$, the CP-asymmetry is defined by
\begin{equation}
\varepsilon_N \equiv \frac{\Gamma(N\rightarrow L H) - \Gamma(N \rightarrow \bar{L} H^*)}{\Gamma(N\rightarrow L H) + \Gamma(N \rightarrow \bar{L} H^*)}. \label{CPasym}
\end{equation}
while the evolution of the comoving abundances is given as a
function of $z\equiv m_N/T$ by the Boltzmann equations:
\begin{eqnarray}
zH(z)s\, Y'_{N} &=& -\left(\frac{Y_{N}}{Y_{N}^{\rm eq}}-1 \right) \left(\gamma_N^{(l)} + 2 \gamma_{Hs} + 4\gamma_{Ht}\right)\\
zH(z)s\, Y'_{\cal L} &=&\gamma_N^{(l)} \left[\varepsilon_{N} \left(\frac{Y_{N}}{Y_{N}^{\rm eq}}-1\right) - \frac{Y_{\cal L}}{2\,Y_{L}^{\rm eq}}\right] -2 \frac{Y_{\cal L}}{Y_{L}^{\rm eq}}\left(\gamma_{Ns}^{\rm sub}+\gamma_{Nt}+\gamma_{Ht} + \gamma_{Hs}\,\frac{Y_{N}}{Y_{N}^{\rm eq}}\right)
\label{Boltzstand2}
\end{eqnarray}
where $'$ denotes the derivative with respect to $z$. The thermally averaged reaction rate
\begin{equation}
\gamma^{(l)}_N = n_N^{eq}(z)~\frac{K_1(z)}{K_2(z)}~\Gamma_N^{(l)},
\end{equation}
parametrizes the effects of Yukawa induced decays and inverse decays
with $\Gamma^{(l)}_{N}=\Gamma(N\rightarrow L H) + \Gamma(N \rightarrow \bar{L} H^*)=\frac{1}{8 \pi}|Y_\nu|^2 m_N$, and $K_{1,2}$ Bessel functions.
The other $\gamma$'s take into account the effects of the various
scatterings through a $H$ or a $N$ in the $s$ or $t$ channels. They are related to the corresponding cross sections in the following way
\begin{eqnarray}
\gamma(a\,b\leftrightarrow 1\,2) &=& \iint d\bar{p}_a d\bar{p}_b f_a^{eq}f_b^{eq} \iint d\bar{p}_1 d\bar{p}_2 (2\pi)^4 \delta^4(p_a+p_b-p_1-p_2) |{\cal M}|^2\\
&=& \frac{T}{64~\pi^4} \int_{s_{min}}^{\infty} ds ~\sqrt{s}~\hat{\sigma}(s)~K_1\left(\frac{\sqrt{s}}{T} \right)
\label{ScatRates}
\end{eqnarray}
with $\hat{\sigma} = 2\,s^{-1}\, \lambda^2[s,m_a^2,m_b^2]\, \sigma(s)$ the reduced cross section, $\lambda[a,b,c]\equiv \sqrt{(a-b-c)^2 -4bc}$ and $s_{min} = \max[(m_a+m_b)^2,(m_1+m_2)^2]$. The analytic expression of the reduced cross sections can be found in Refs.~\cite{gammas1,gammas2}.\footnote{Note that for simplicity we have neglected the subdominant effects of scatterings of the type $N+L \leftrightarrow H +(\gamma,Z,W_L)$ \cite{gammas2}. We also neglect as in ref.~\cite{gammas2} the effects of Yukawa coupling induced $NN \leftrightarrow LL,HH$ processes which have little effects too.}
$\gamma_{Ns}^{\rm sub} = \gamma_{Ns} -\gamma_{N}^{(l)}/4$ in Eq.~(\ref{Boltzstand2}) refers to the substracted scattering through a $N$ in the $s$ channel (i.e.~taking out the contribution of the on-shell propagator in order to avoid double counting with the inverse decay contribution \cite{gammas2}).

The above, now traditional approach assumes that $N$ are introduced
in an isolated way in the model. In many unifying groups (left-right
symmetric \cite{SU2R}, Pati-Salam \cite{PatiSalam}, $SO(10)$
\cite{SO10} or larger) the presence of the $N$ can be nicely
justified as it is precisely the ingredient required to unify all
fermions. These groups however do not introduce the $N$ in such an
isolated way and moreover link the $N$ and $W_R$ masses to the same
$SU(2)_R$ breaking scale $v_R$.\footnote{More complicated breaking mechanisms
could add extra contributions to the gauge boson masses: all mass
contributions to $N$ will also contribute to $W_R$, but the opposite
is not necessarily true.} It is thus a (generally unwarranted)
assumption to neglect the effect of $SU(2)_R$ gauge 
bosons. If $m_{W_R}$ is smaller than $\sim 10^{13}$~GeV, these
effects must be explicitly incorporated for any $N$ whose mass is
not several orders of magnitude below the one of the $W_R$ \cite{Cosme}.

The key interactions of the $W_R$ \cite{SU2R,PatiSalam} are the
\begin{equation}
{\cal L} \owns \frac{g}{\sqrt{2}}
W_R^{\mu} \left( \bar{u}_R \gamma_{\mu} d_R + \bar{N} \gamma_{\mu} \,l_R \right)
\end{equation}
gauge ones. ($N$ and the right-handed charged leptons ($l_R=e_R,\mu_R,\tau_R$), and $u_R$ and $d_R$,  are members of a same $SU(2)_R$ doublet).

Their effects for leptogenesis  can be incorporated by modifying the Boltzmann
equations in the following way:
\begin{eqnarray}
zH(z)s\, Y'_{N} &=& -\left(\frac{Y_{N}}{Y_{N}^{\rm eq}}-1 \right) \left(\gamma_{N}^{(l)} + \gamma_{N}^{(W_R)} + 2 \gamma_{Hs} + 4\gamma_{Ht}+ 2 \gamma_{Nu} + 2 \gamma_{Nd} + 2 \gamma_{Ne} \right)\nonumber\\
& &- \left(\left(\frac{Y_N}{Y_N^{eq}}\right)^2 - 1 \right) \gamma_{NN}\label{NBoGauge}\\
zH(z)s\, Y'_{\cal L} &=&\gamma_{N}^{(l)} \varepsilon_{N} \left(\frac{Y_{N}}{Y_{N}^{\rm eq}}-1\right) - \left(\gamma_{N}^{(l)}+ \gamma_{N}^{(W_R)}\right)\frac{Y_{\cal L}}{2\,Y_{L}^{\rm eq}}\nonumber\\
&&
-\frac{Y_{\cal L}}{Y_{L}^{\rm eq}}\left(2\,\gamma_{Ns}^{\rm sub}+2\,\gamma_{Nt}+2\,\gamma_{Ht} + 2\,\gamma_{Hs}\,\frac{Y_{N}}{Y_{N}^{\rm eq}}
+ \,\gamma_{Nu}
+ \,\gamma_{Nd}
+ \,\gamma_{Ne}\,\frac{Y_{N}}{Y_{N}^{\rm eq}}\right) \,\,\label{LBoGauge}
\end{eqnarray}
with the CP asymmetry unchanged, as given by Eq.~(\ref{CPasym}). In
these Boltzmann equations there are essentially 2 types of effects
induced by the $W_R$, both suppressing the produced lepton
asymmetry: from the presence of alternate decay channels for the
heavy neutrinos, $\gamma_N^{(W_R)}$, and from scatterings, $\gamma_{Nu,d,e}$, see below.

\subsection{Decay effect: dilution and wash-out}

It is useful to distinguish 2 cases depending on the mass hierarchy between $N$ and $W_R$.

a) Case $m_{W_R}> m_{N}$: in this case the decay
of $N$ to leptons or antileptons plus Higgs particles remains
the only possible 2 body decay channels
but a series of three body decay channels with a \textit{virtual} $W_R$
is now possible:
$N \rightarrow l_R  q_R \bar{q}'_R$ or $N \rightarrow \bar{l}_R \bar{q}_R q'_R $
with $l=e,\mu,\tau$, $q=u,c,t$,  $q'=d,s,b$. We obtain:
\begin{equation}
\Gamma(N \rightarrow l_R  q_R \bar{q}'_R)= \frac{3\,g_R^4}{2^9\,\pi^3\,m_N^3} \int_0^{m_N^2} dm_{12}^2 ~\frac{\left(m_N^6-3 m_N^2 m_{12}^4 +2 m_{12}^6\right)}{\left(m_{W_R}^2-m_{12}^2\right)^2+m_{W_R}^2 \Gamma_{W_R}^2\left(m_{12}^2\right)}
\end{equation}
Given the potentially large value of the gauge to Yukawa couplings
ratio, the three body decays can compete with the Yukawa two body
decay. Since the gauge interactions do not provide any CP-violation
and are flavor blind, it can be shown that they do not provide any
new relevant source of CP-asymmetry. But still the gauge
interaction-induced 3 body decays appear in both Boltzmann
equations, Eqs.~(\ref{NBoGauge})-(\ref{LBoGauge}), with
\begin{equation}
\gamma_{N}^{(W_R)}=n_N^{eq}(z)~\frac{K_1(z)}{K_2(z)}~\Gamma_{N}^{(W_R)}.\label{DecayTot}
\end{equation}
where $\Gamma_N^{(W_R)}$ is the total three body decay width.

Unlike in leptogenesis without $W_R$, not all decays
participate in the creation of the asymmetry but only a fraction
$\Gamma^{(l)}_{N}/\Gamma_{N_{Tot}}$ does. This shows up in
the Boltzmann equations through the fact that Eq.~(\ref{NBoGauge}) involves $\Gamma_{N_{Tot}}=\Gamma_N^{(l)}+ \Gamma_N^{(W_R)}$
while the CP-asymmetry in Eq.~(\ref{LBoGauge}) is multiplied only by $\Gamma^{(l)}_{N}$.\footnote{In Eq.~(\ref{LBoGauge}), we made the choice
to keep Eq.~(\ref{CPasym}) as  definition for the CP-asymmetry. In
its denominator, it involves only the Yukawa driven decay rather
than the total decay width, $ \Gamma_{N_{Tot}}$. Therefore this CP asymmetry
doesn't correspond anymore, as in standard leptogenesis, to the
averaged $\Delta L$ which is created each time a $N$ decays. However
this definition is convenient for several reasons. It makes explicit
the fact that the gauge decay does not induce any lepton asymmetry. Moreover in this way, all
(competing) suppression effects, including the dilution one, are put
together in the efficiency, not in the CP-asymmetry. It also allows
to take the simple upper bound $\varepsilon < 1$ for any numerical
calculations.}
This dilution effect leads
automatically to an upper bound on the efficiency. The bound $\eta
<1$, which applies in standard leptogenesis for thermal $N$'s
becomes:
\begin{equation}
\eta <  \Gamma^{(l)}_{N}/\Gamma_{N_{Tot}}
\label{etabound}
\end{equation}
As a numerical example, for $m_{N}\sim 1$~TeV, with Yukawa couplings of order $10^{-6}$, so that
$m_\nu \sim Y_\nu^2 v^2/m_{N} \sim 10^{-1}$~eV, and with $m_{W_R}\sim 3(4)$~TeV we obtain the
 large suppression factor $\Gamma^{(l)}_{N}/\Gamma_{N_{Tot}}= 7\cdot10^{-7}(2\cdot10^{-6})$,
 consistent with leptogenesis only if the CP-asymmetry is of order unity, which requires maximal enhancement of the asymmetry (i.e. right handed neutrino mass splittings of order of their decay widths).

In addition to this dilution effect, the three body decay $\gamma_N^{(W_R)}$ reaction density
also induces a  $L$ asymmetry washout effect from inverse decays (proportional to $Y_{\cal L}$ 
in Eq.~(\ref{LBoGauge})) which can also be large.

b) Case $m_{W_R} < m_{N}$: in this case\footnote{A $N$ much heavier
than $W_R$ is in general not expected in the left-right symmetric
model or extensions given the fact that, as said above, both $W_R$
and $N_R$ have a mass proportional to the $SU(2)_R$ breaking scale
$v_R$, and given the fact that $m_{W_R}\sim g v_R$ with $g$ the
ordinary gauge coupling which is of order unity.} the direct 2 body
decays $N \rightarrow W_R l_R$ are allowed which leads to an even
larger dilution and washout effect for low $m_N$. For example with
$m_N\simeq 1$~TeV, $Y_\nu \simeq 10^{-6}$ and $m_{W_R}\simeq 800$~GeV,
we get $\Gamma^{(l)}_{N}/\Gamma_{N_{Tot}}= 4\cdot10^{-9}$, which
means that the dilution effect makes leptogenesis basically hopeless
at this scale, even with the maximum value $\varepsilon_N=1$. In the
following we will consider only the case where  $m_{W_R} \gtrsim
m_{N}$ (this corresponds to the situation where a discovery of the
$W_R$ and $N$ at LHC would occur through same sign dilepton channel
\cite{Wlhcferrari,Wlhckras,cmsreport}, see section 6).

\subsection{Gauge scattering effect}

\begin{figure}[h!]
\centering
\begin{tabular}{ccccccccccccccc}
\raisebox{1cm}{(a)}&
\includegraphics[width=3.2cm]{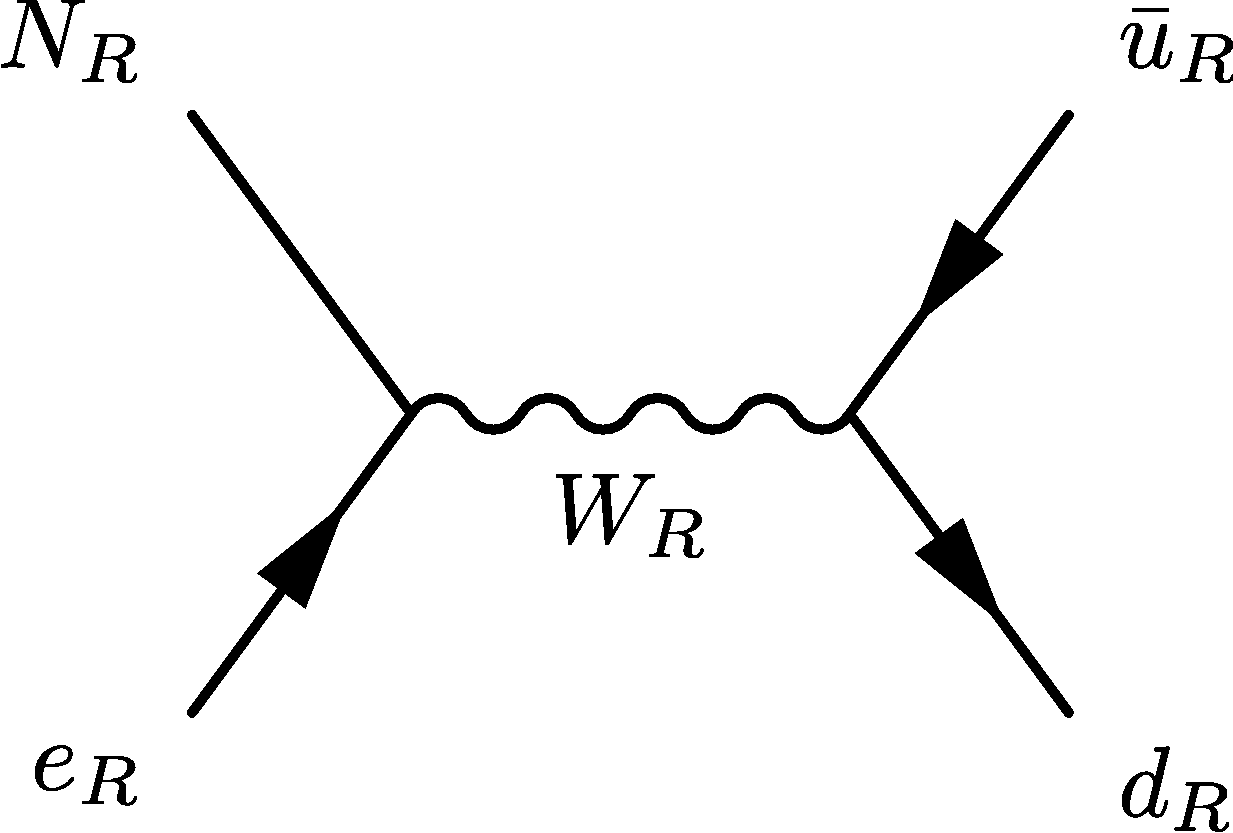}&
\includegraphics[width=3.2cm]{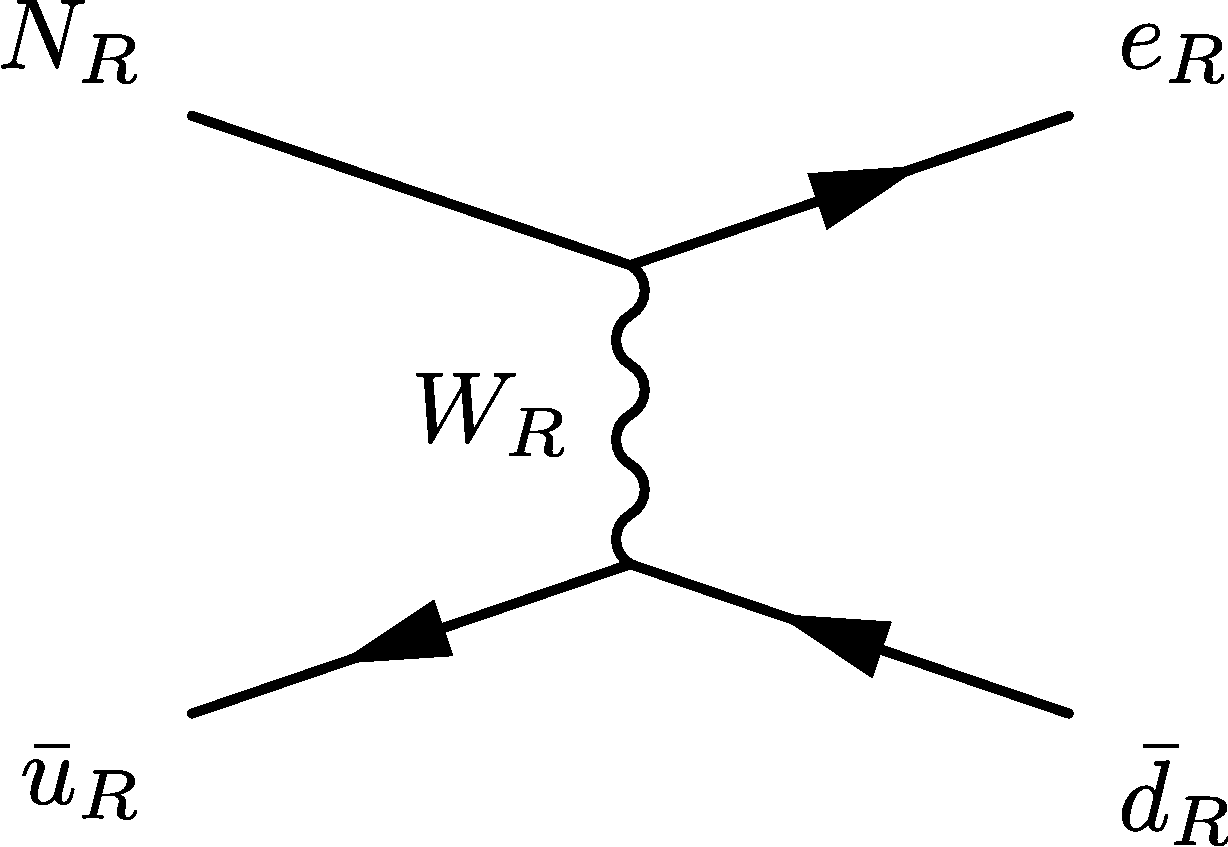}&
\includegraphics[width=3.2cm]{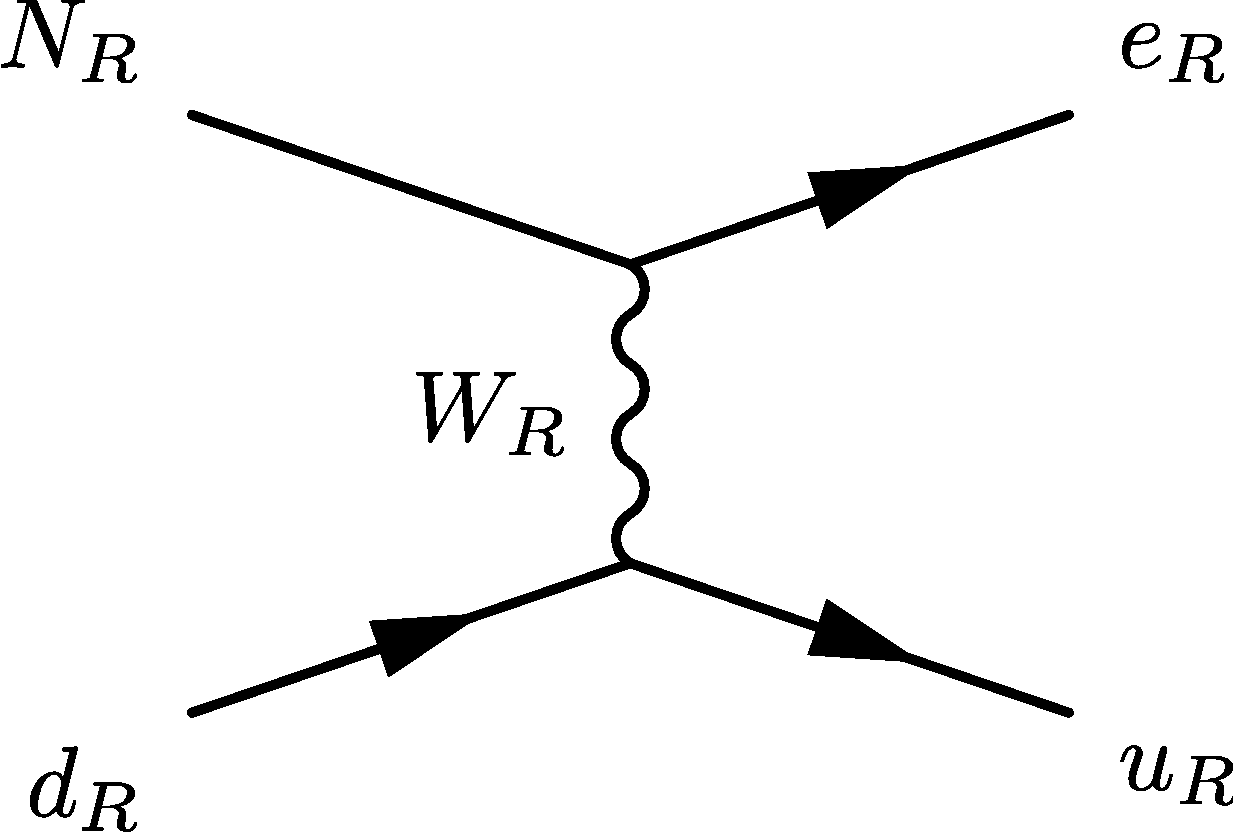}&
\includegraphics[width=3.2cm]{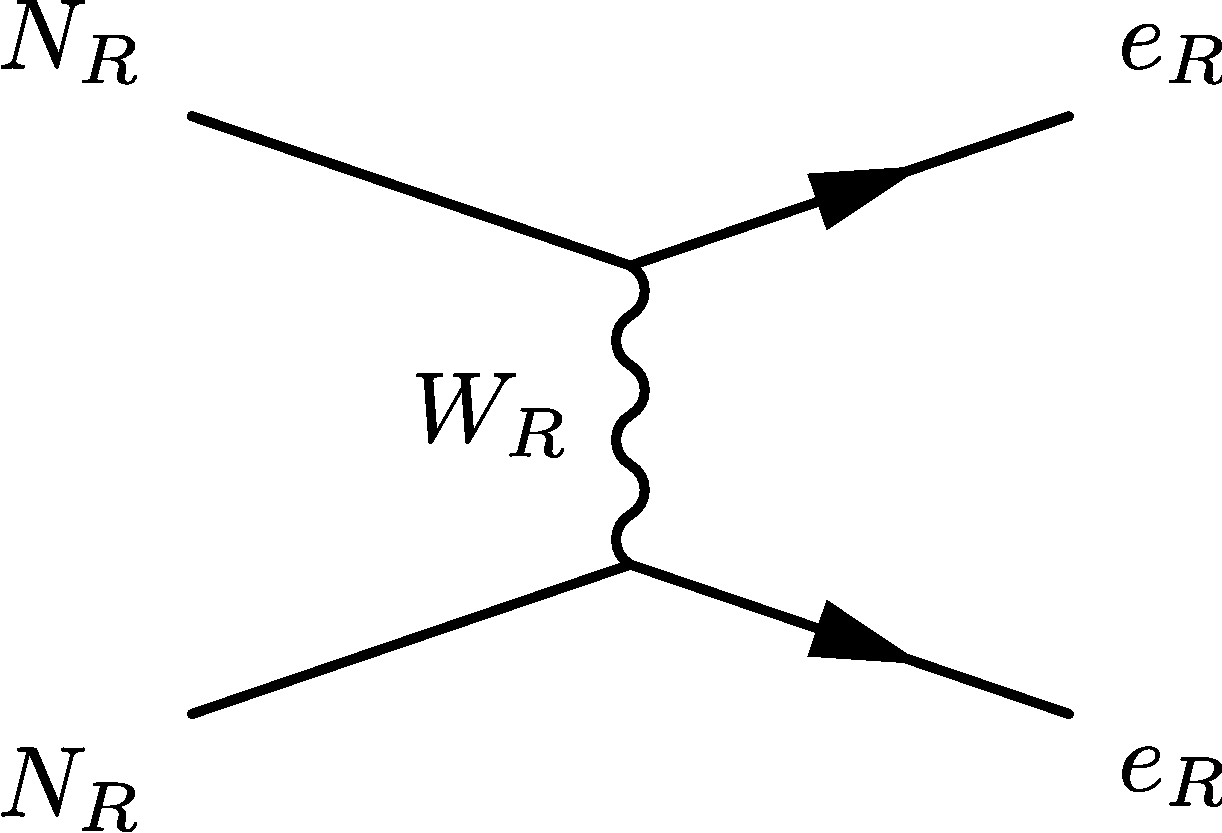}&
\\
\raisebox{1cm}{(b)}&
\includegraphics[width=3.2cm]{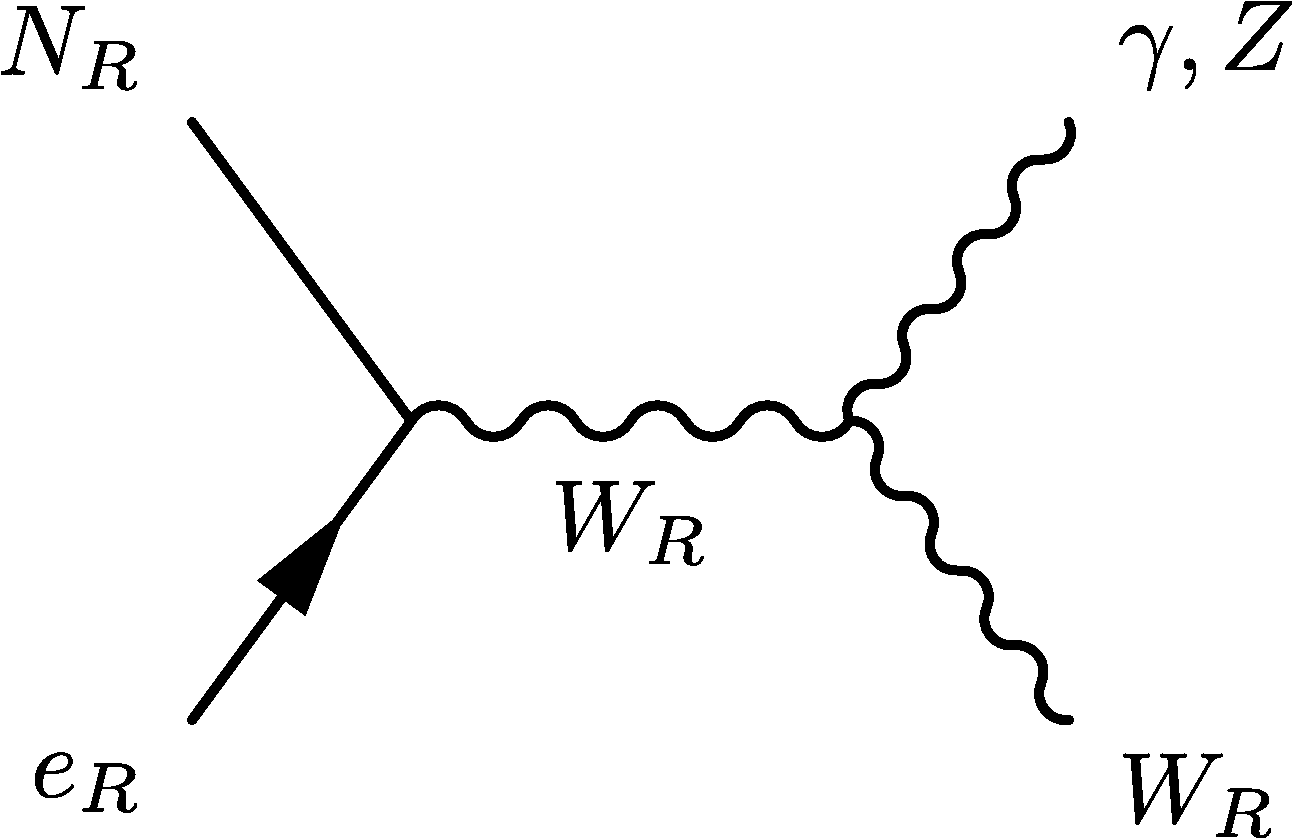}&
\includegraphics[width=3.2cm]{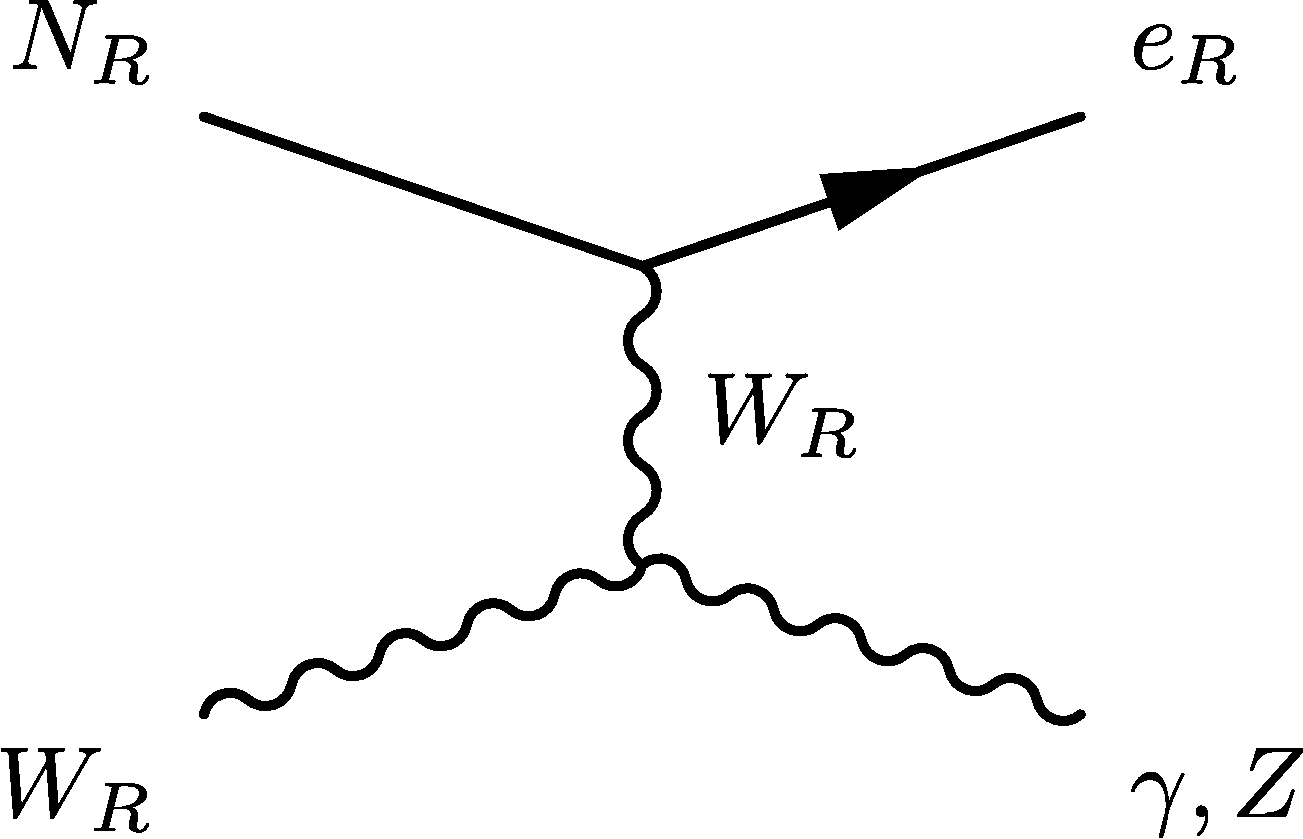}&
\includegraphics[width=3.2cm]{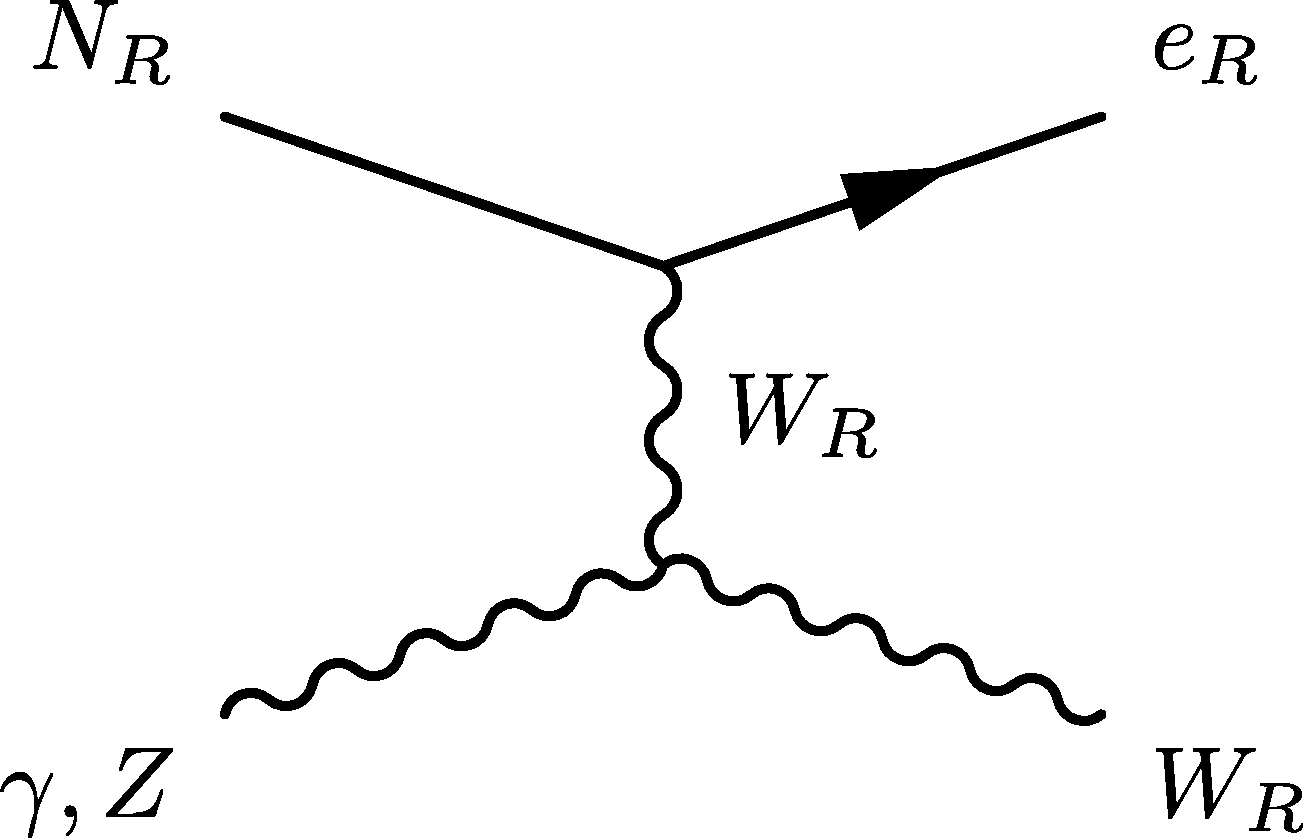}&
\includegraphics[width=3.2cm]{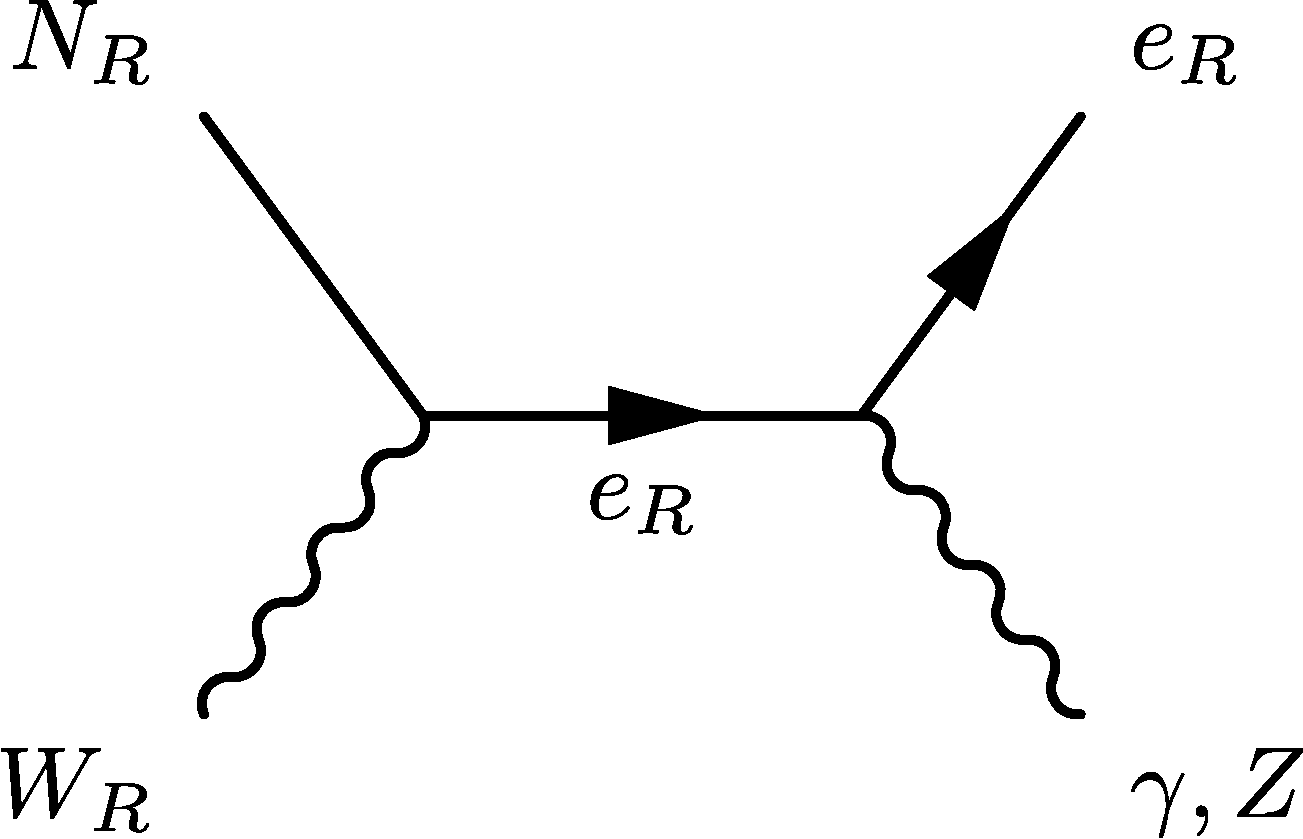}&
\\
&
\includegraphics[width=3.2cm]{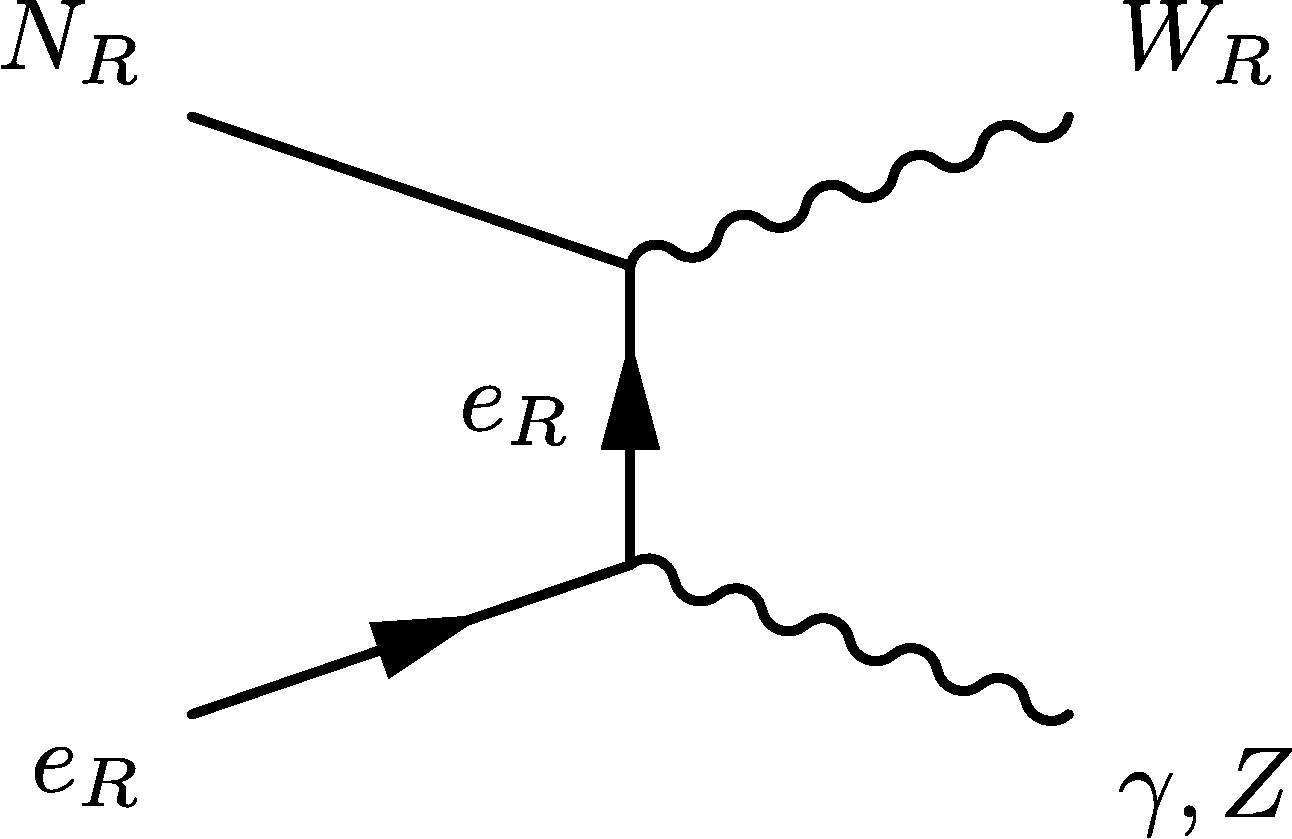}&
\includegraphics[width=3.2cm]{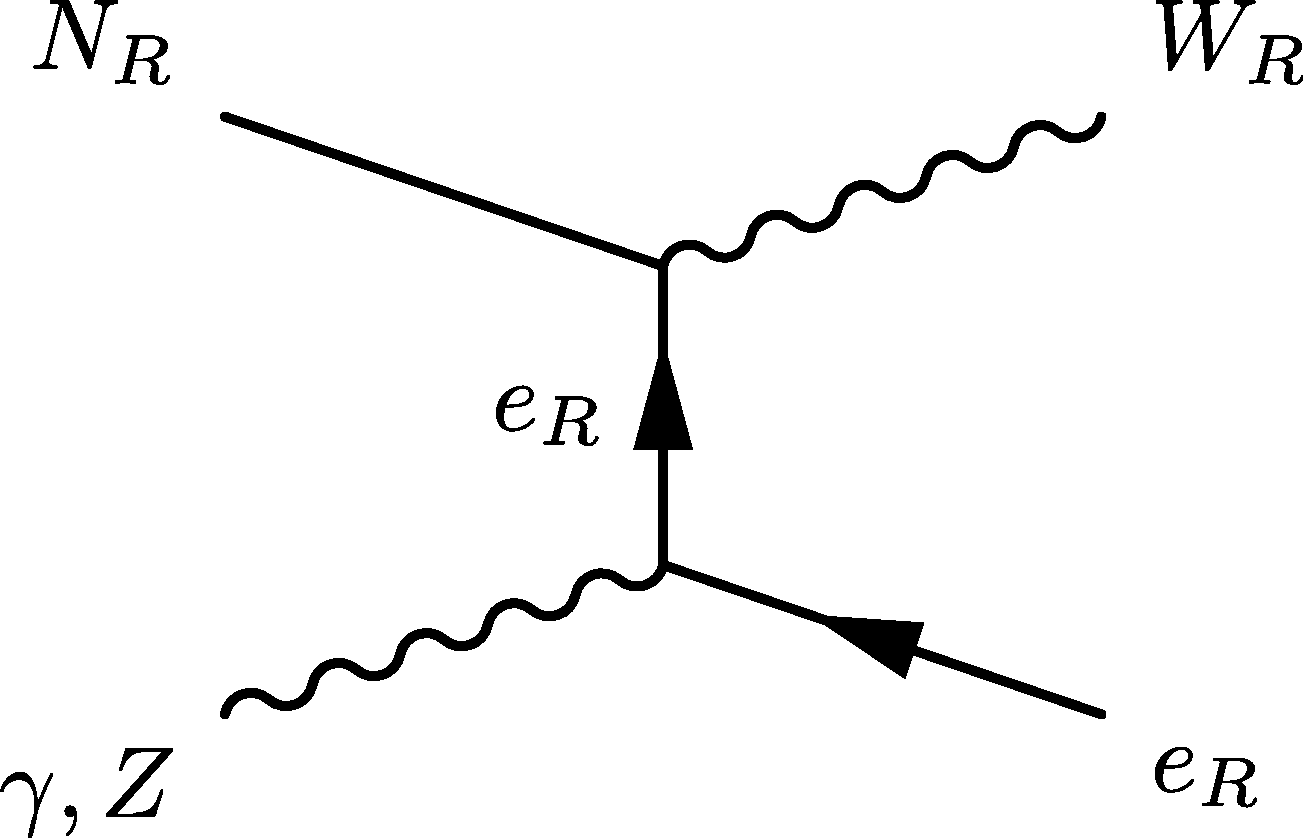}&
\includegraphics[width=3.2cm]{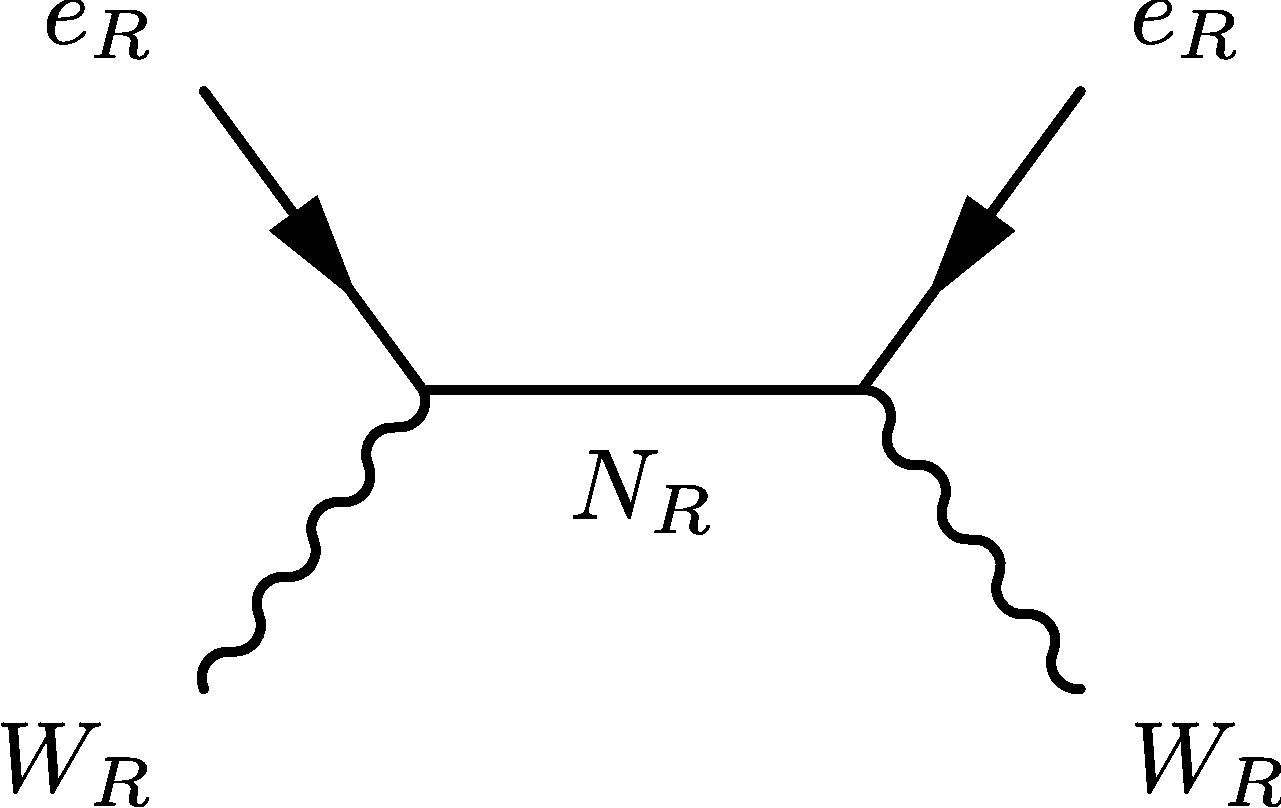}&
\includegraphics[width=3.2cm]{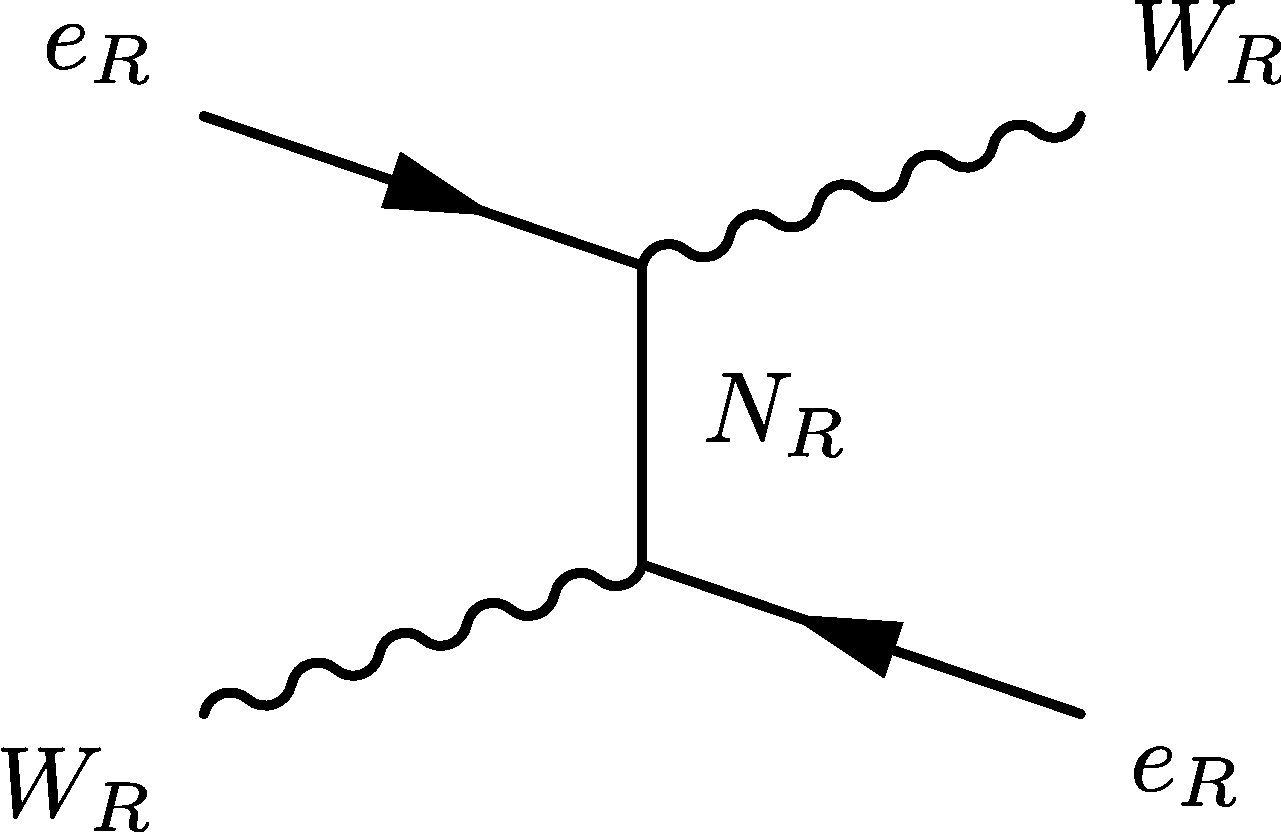}&
\\
\raisebox{1cm}{(c)}&
\includegraphics[width=3.2cm]{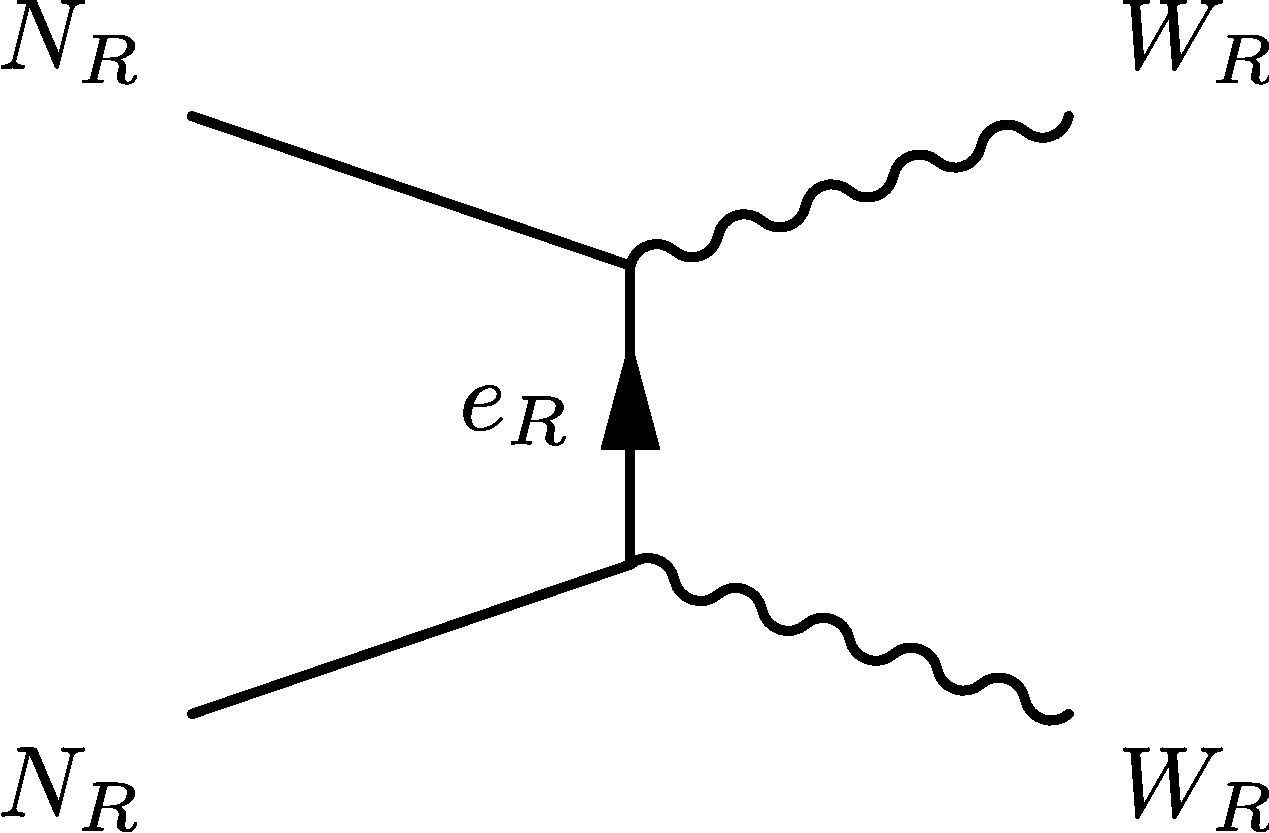}&
\includegraphics[width=3.2cm]{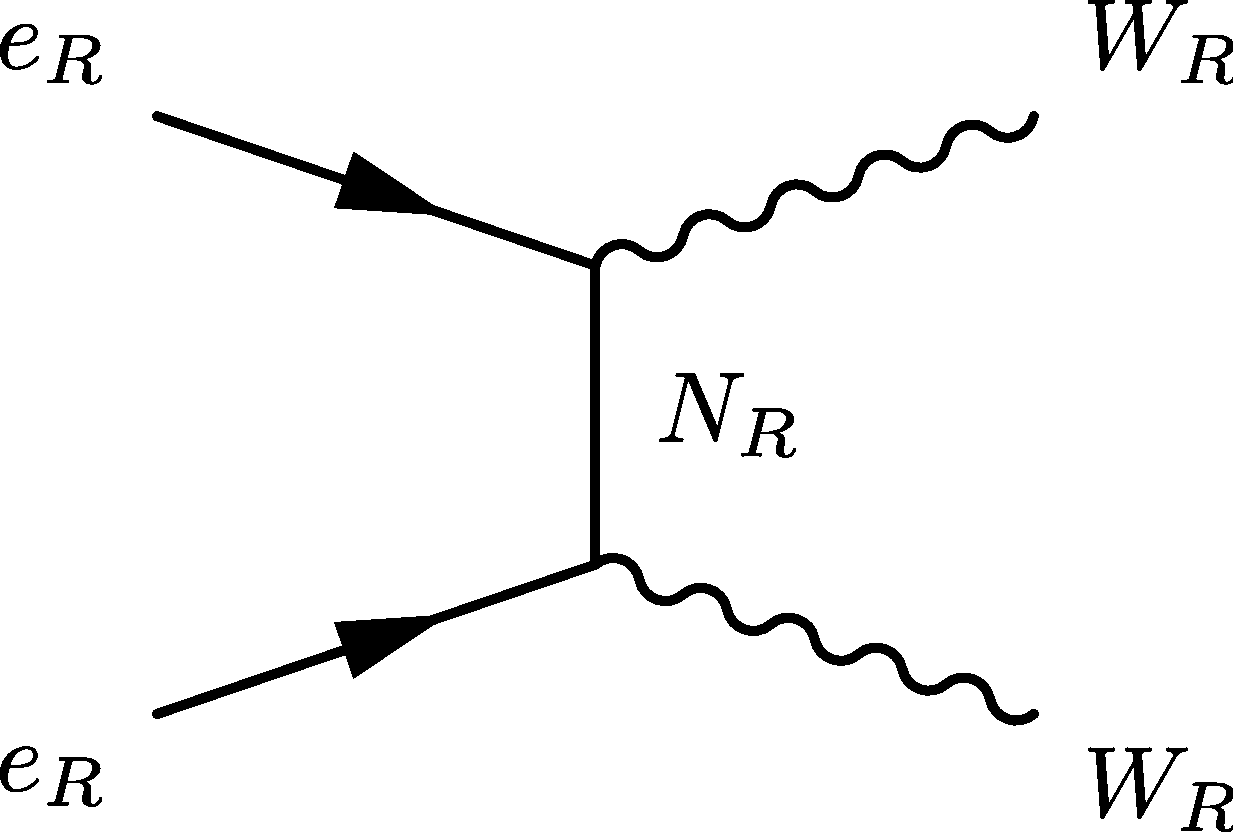}&
&&
\\
\raisebox{1cm}{(d)}&
\includegraphics[width=3.2cm]{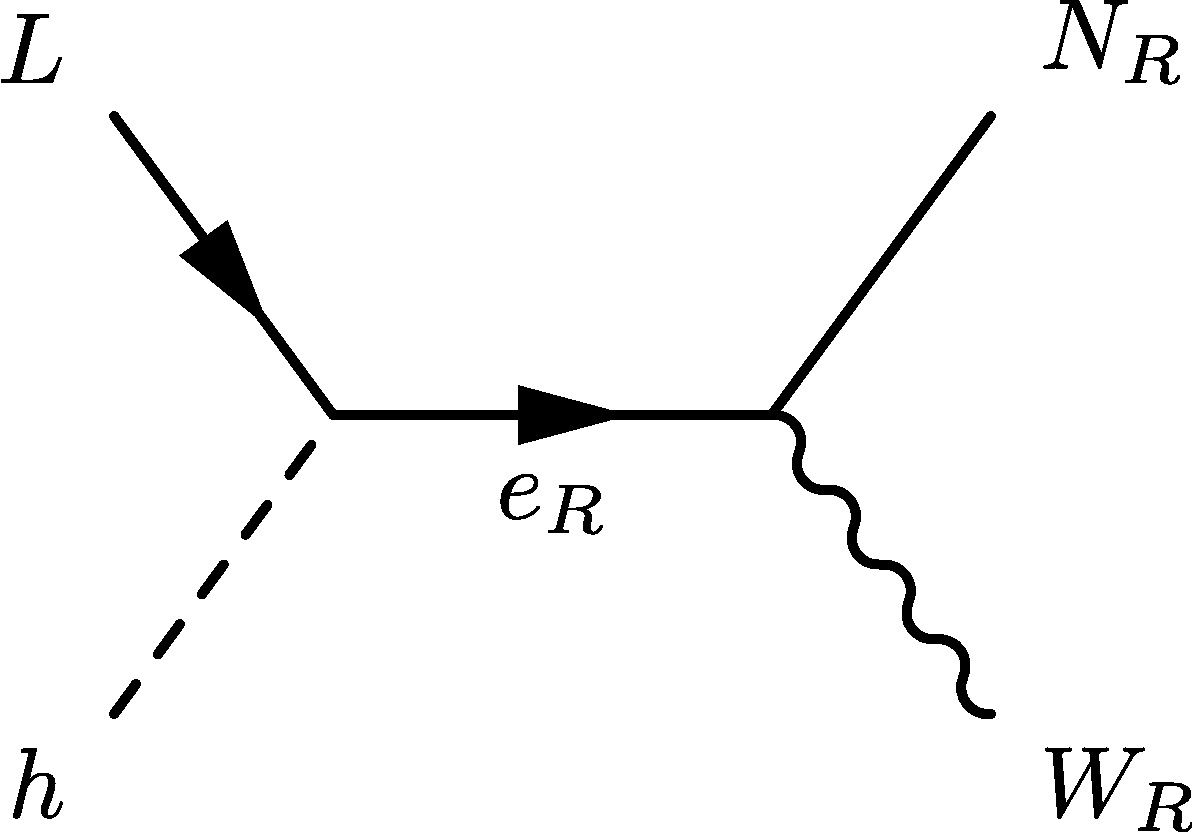}&
\includegraphics[width=3.2cm]{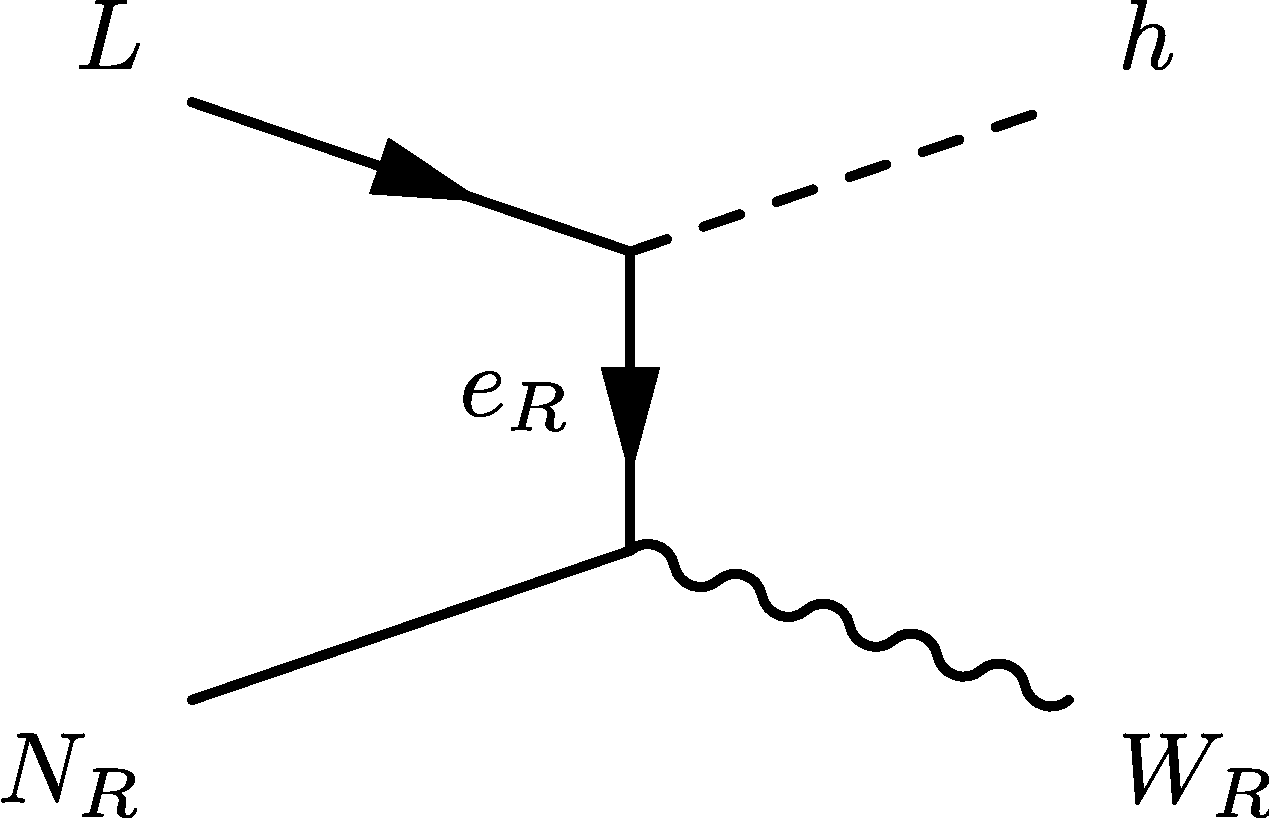}&
\includegraphics[width=3.2cm]{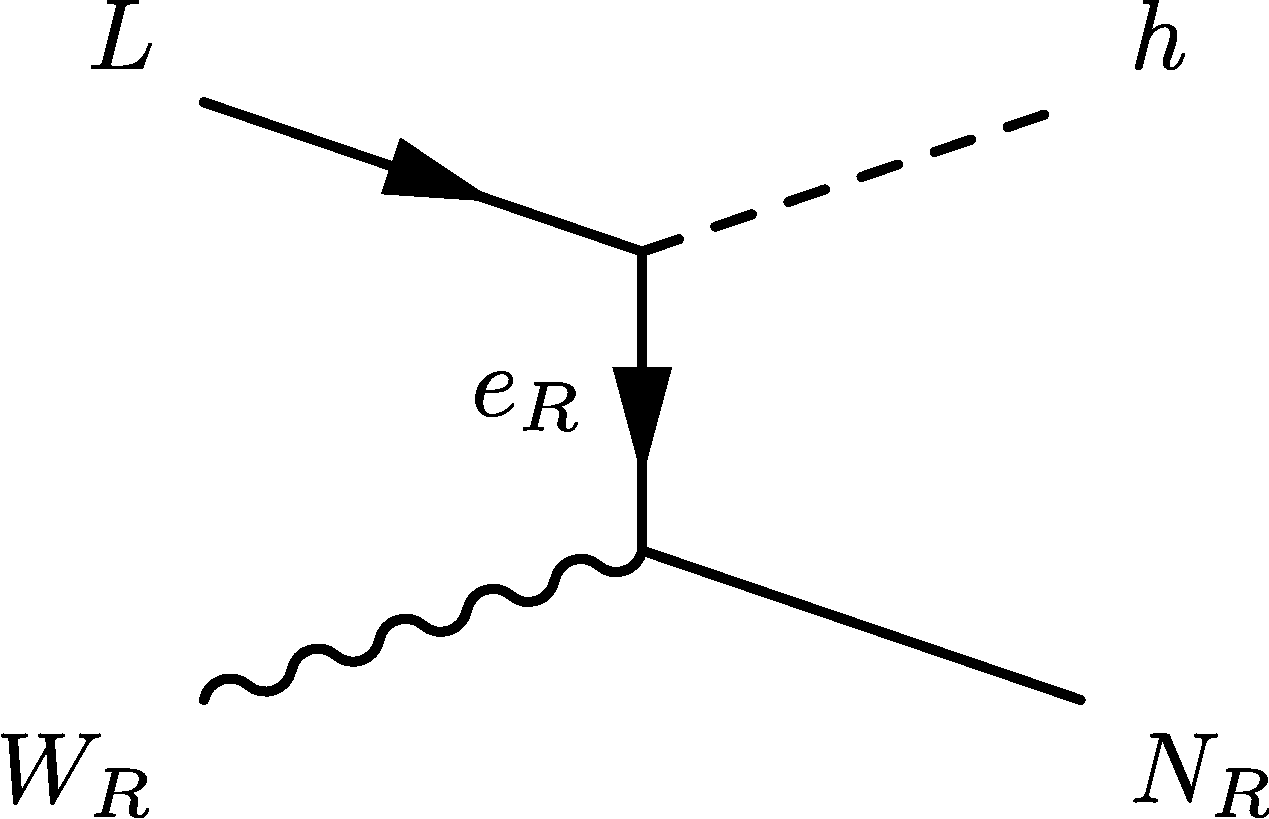}&&&
\\
&
\includegraphics[width=3.2cm]{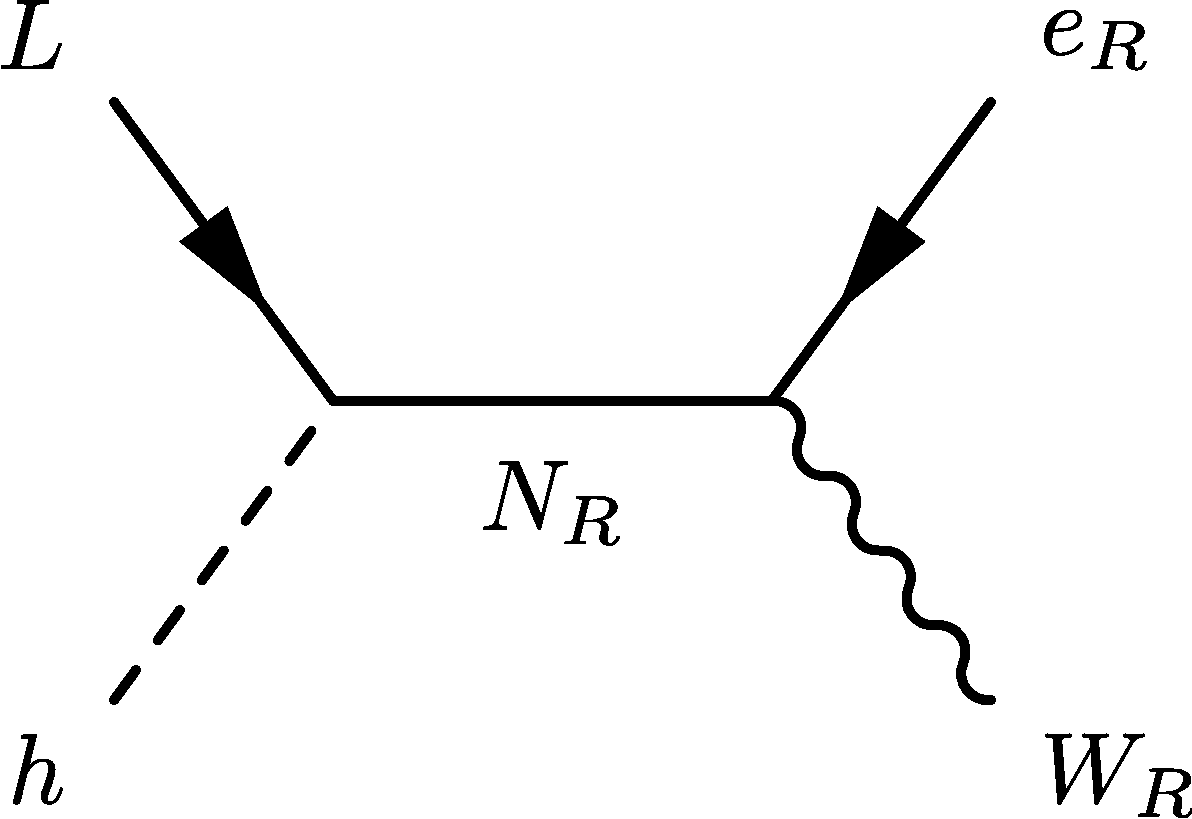}&
\includegraphics[width=3.2cm]{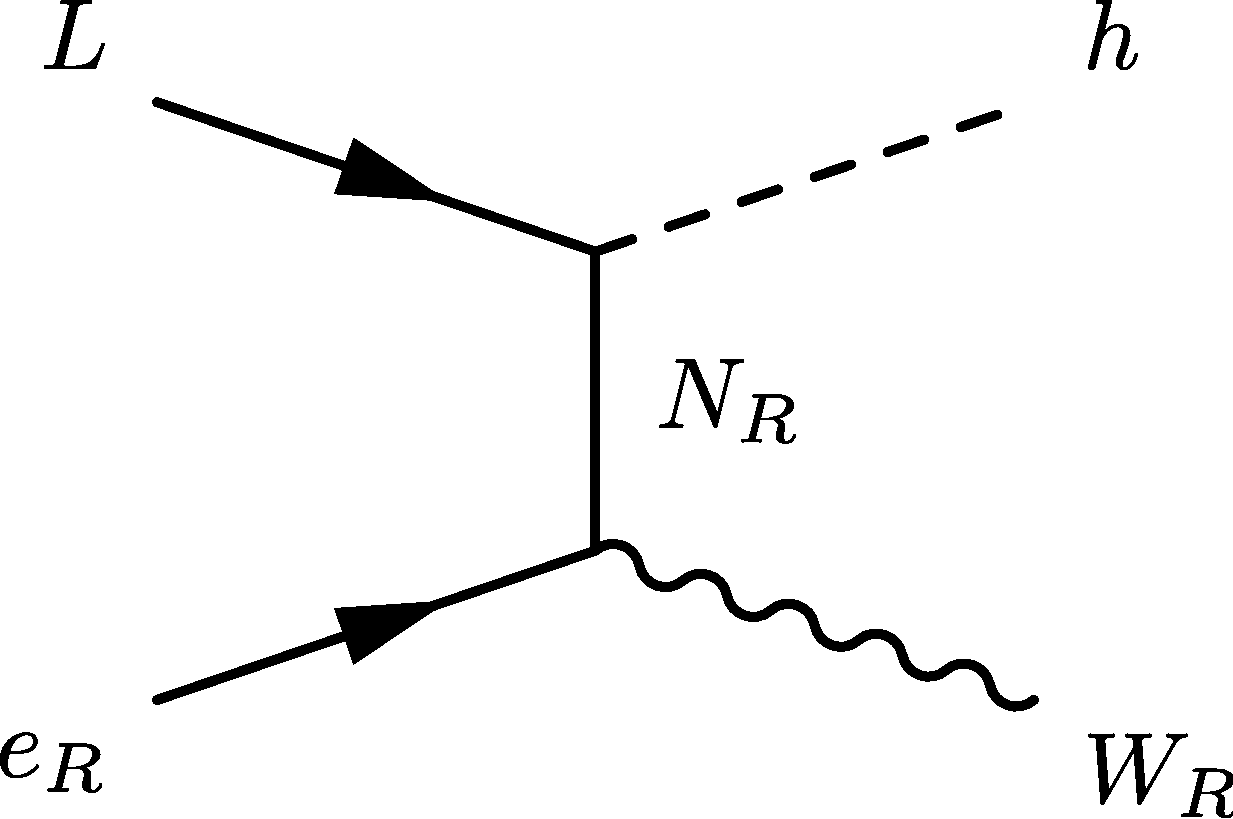}&
\includegraphics[width=3.2cm]{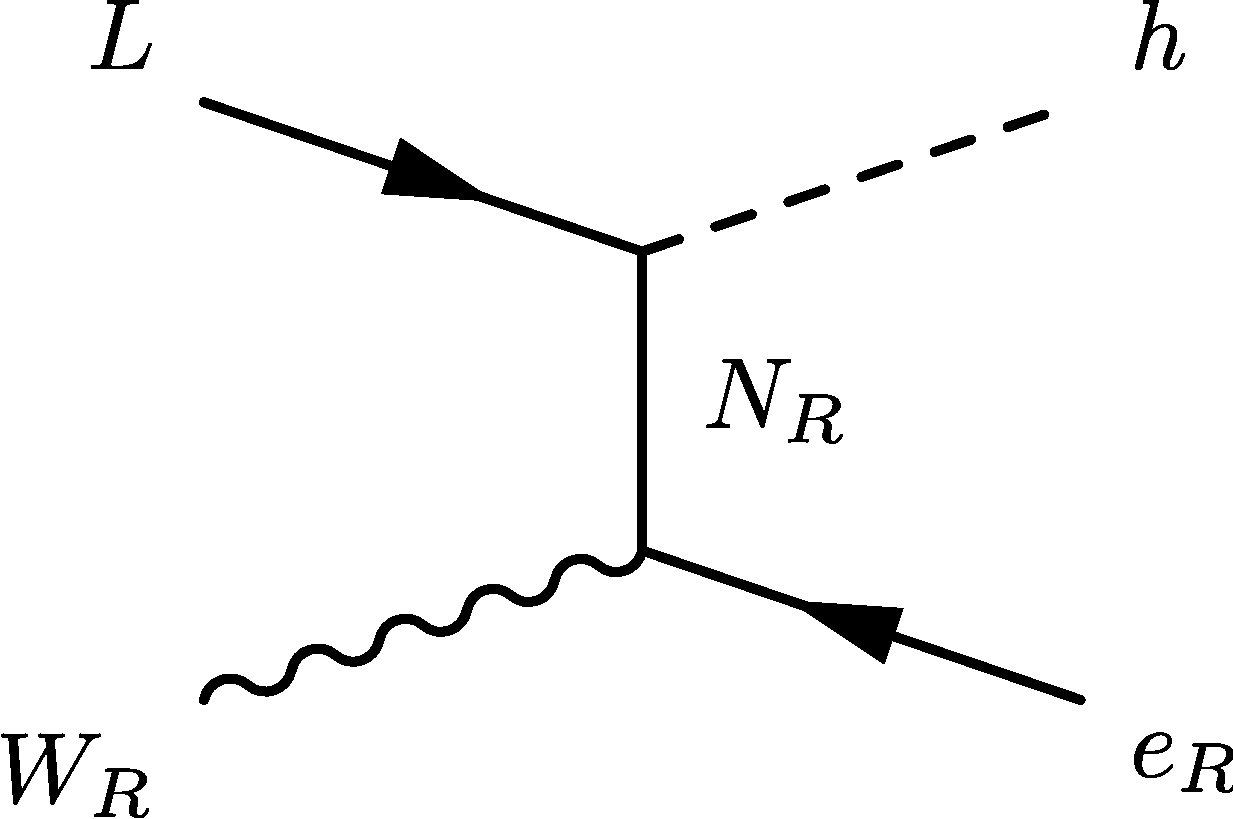}&
\end{tabular}
\caption{Scatterings involving the $W_R$.}
\label{gauge_scatterings}
\end{figure}

Right-handed gauge interactions induce a long series of scatterings, given in Fig.~1. To explain their effects let us first consider scatterings which do not involve any external $W_R$, Fig.~1.a.
The density reaction rates $\gamma_{Nu}$, $\gamma_{Nd}$, $\gamma_{Ne}$, $\gamma_{NN}$ can be computed from the following reduced cross sections:
\begin{eqnarray}
\hat{\sigma}(Ne_R\rightarrow \bar{u}_R d_R)&=& \frac{9 g_R^4}{8\pi s[(s-m_{W_R}^2 )^2+m_{W_R}^2 \Gamma_{W_R}^2]} \left( \frac{m_N^6}{6} - \frac{m_N^2 s^2}{2} + \frac{s^3}{3} \right)
\label{gammaNu}\\
\hat{\sigma}(N\bar{u}_R\rightarrow e_R \bar{d}_R)&=& \frac{9 g_R^4}{8\pi s} \int_{m_N^2-s}^0 dt ~\frac{(s+t)(s+t-m_N^2)}{(t-m_{W_R}^2)^2}\\
\hat{\sigma}(N d_R\rightarrow e_R u_R)&=& \frac{9 g_R^4}{8\pi} \frac{\left(m_N^2 - s \right)^2}{m_{W_R}^2 \left(s+m_{W_R}^2-m_N^2 \right)}
\label{gammaNe}\\
\hat{\sigma}(N N\rightarrow e_R \bar{e}_R)&=& \frac{g_R^4}{8\pi s} \int_{t_0}^{t_1} dt ~\left( \frac{(s+t+-m_N^2)^2}{(t-m_{W_R}^2)^2} + \frac{(m_N^2-t)^2}{(2 m_N^2 -s-t-m_{W_R}^2)^2} \right. \nonumber \\ && \hspace{2.5cm} \left.- \frac{m_N^2 s}{(t-m_{W_R}^2)(2 m_N^2 -s-t-m_{W_R}^2)}\right)
\label{gammaNN}
\end{eqnarray}
Among these scatterings the three first ones involving only one external $N$
have a peculiar property. Unlike in ordinary pair annihilation or in
coannihilation with a heavier particle, their decoupling in the
$Y_N$ Boltzmann equation does not proceed with a Boltzmann
suppression of their rate. The decoupling condition is:
\begin{equation}
\frac{\gamma_A}{n_N^{eq} H} \lesssim 1
\label{decoupl}
\end{equation}
with $H$ the Hubble constant and $\gamma_A=
\gamma_{Nu}+\gamma_{Nd}+\gamma_{Ne}$. For $T$ well below
$m_N$ the reaction density, Eq.~(\ref{ScatRates}), is Boltzmann
suppressed (i.e.~in $e^{-m_N/T}$) but so is also $n_N^{eq}$ in the
denominator. Therefore, decoupling comes at low temperature only
from the approximately linear in $T$ behaviour of
$\frac{\gamma_A}{n_N^{eq} H}$ for small $T$. This can be understood
from the fact that what sets the thermal equilibrium of $Y_N$ is the
number of interactions per $N$, not the number of interactions
irrespective of the number of $N$. In other words these processes
are important because the abundance of the other particles involved
is large with respect to the $N$ density.

It is useful to compare this behaviour with the one of ordinary left-handed gauge
scatterings which have been considered for leptogenesis
from the decay of a scalar triplet \cite{typeIIleptoeffic} or
of a fermion triplet \cite{typeIIIlepto}.
In these models these scatterings necessarily involve two external
heavy-states (i.e.~annihilation or creation of a pair of scalar triplets or a
pair of fermion triplet respectively) and therefore are
doubly Boltzmann suppressed (which leads to an exponential Botzmann type decoupling:
 $\frac{\gamma}{n_T^{eq} H}\sim e^{-m_T/T}$) .

The right-handed gauge interaction induced scatterings remain
therefore in thermal equilibrium down to temperatures much lower
than the left-handed gauge triplet interactions for equal decaying
state and gauge boson masses. Their decoupling also doesn't occur so
sharply (compare for example $\gamma_A$ with $\gamma_{NN}$  in
Fig.~\ref{rates} below or with the left-handed gauge scattering rates
of Fig.~3 of Ref.~\cite{typeIIleptoeffic} or of Fig.~6 of
Ref.~\cite{typeIIIlepto}).

For $m_{W_R}$ and $m_N$ of order TeV, one observes from a numerical analysis that the decoupling temperature which follows from Eq.~(\ref{decoupl}) is $\sim  15$ orders of magnitude below these masses. At this temperature the number of $N$ remaining is hugely Boltzmann suppressed, so that no sizeable asymmetry can be created. However, due to the fact that their decoupling is not sharp, these scatterings still allow the creation of a highly suppressed but non-vanishing lepton asymmetry at temperature well above this value (see numerical results below). In all cases the later the $N$ decays with respect to $m_{W_R}$, the less the gauge scatterings will be in thermal equilibrium at the time of the decays, and the smaller will be the suppression effect from them.

Note also that unlike the left-handed gauge interactions, the
suppressions from the scatterings of
Eqs.~(\ref{gammaNu})-(\ref{gammaNe}) also operate in the $Y_{\cal L}$
Boltzmann equation, Eq.~(\ref{LBoGauge}). This can lead to several
orders of magnitude further suppression (see below). The decoupling of these scatterings in the $Y_{\cal L}$ Boltzmann equation results from a Boltzmann suppression when $\gamma_A/(n_l^{eq} H) \lesssim 1$. In Ref.~\cite{Cosme} these effects of gauge scatterings
(as well as of three body inverse decays)
in the $Y_{\cal L}$ Boltzmann equation have been omittted. In the region of parameters considered in this reference, these effects are nevertheless moderate, see below.

Beside the gauge scattering of Fig.~1.a there are also scatterings
with one external $W_R$ changing the number of $N$ and/or violating
lepton number,  Fig.~1.b. Since a substantial asymmetry can be
created only at temperature as low as possible, well below
$m_{W_R}$ for $m_{W_R} \gtrsim m_N$, all these scatterings are suppressed
with respect to the ones with no external $W_R$,
Eqs.~(\ref{gammaNu})-(\ref{gammaNe}). The relative suppression effect
is $e^{-m_{W_R}/m_N}$. Similarly the scatterings with two external
$W_R$, Fig.~1.c are further suppressed. Finally the scatterings
of Fig.~1.d are suppressed by powers of the Yukawa couplings. As a
result we will neglect all the scatterings of Fig.~1.b-1.d and keep
only the ones of Fig.~1.a.\footnote{These scatterings can only
further suppress leptogenesis, which as we will see is anyway
already far too suppressed to be successful.}

\subsection{Efficiency results}

\begin{figure}
\begin{tabular}{c}
\includegraphics[width=5cm]{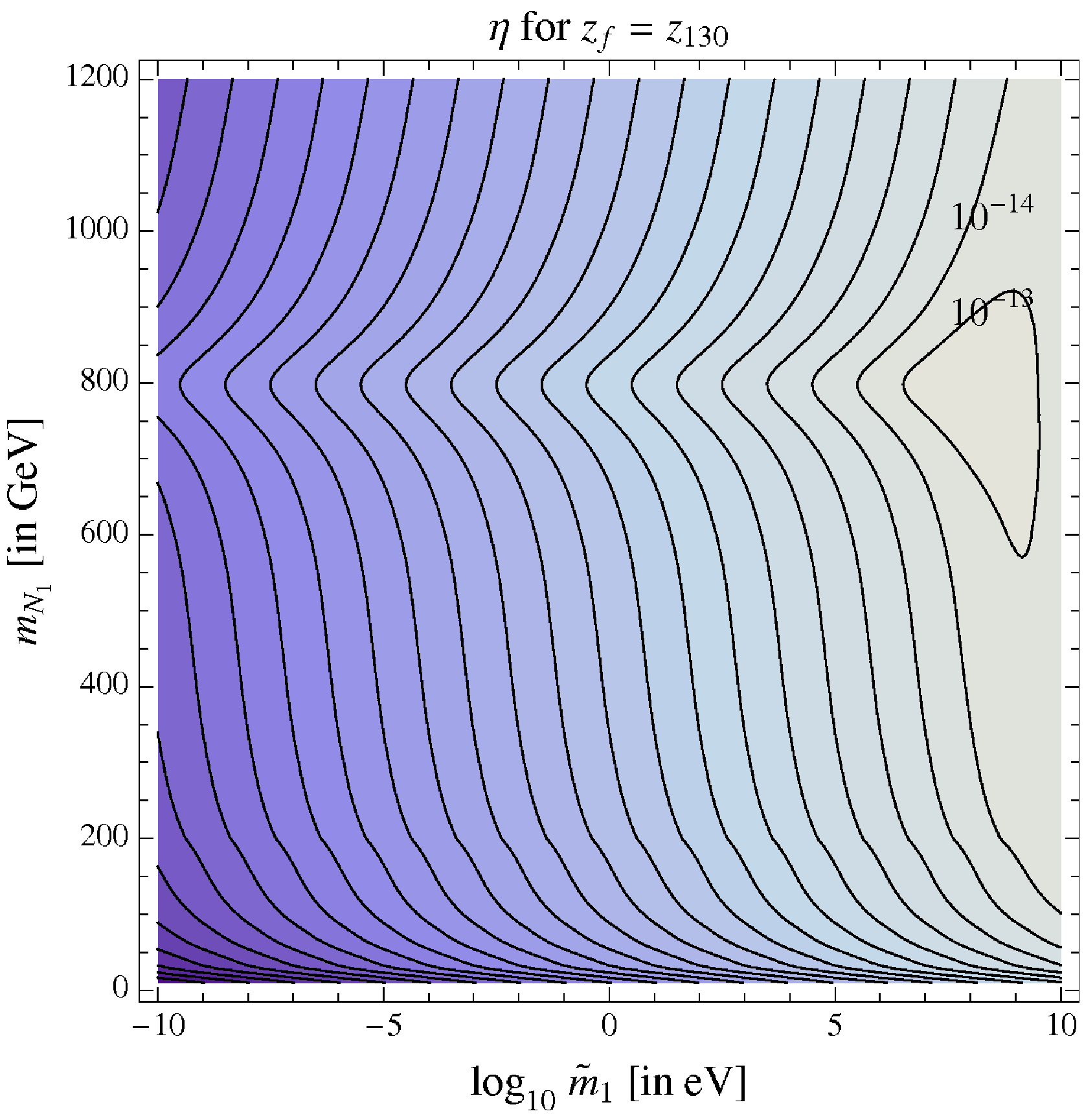}
\includegraphics[width=5cm]{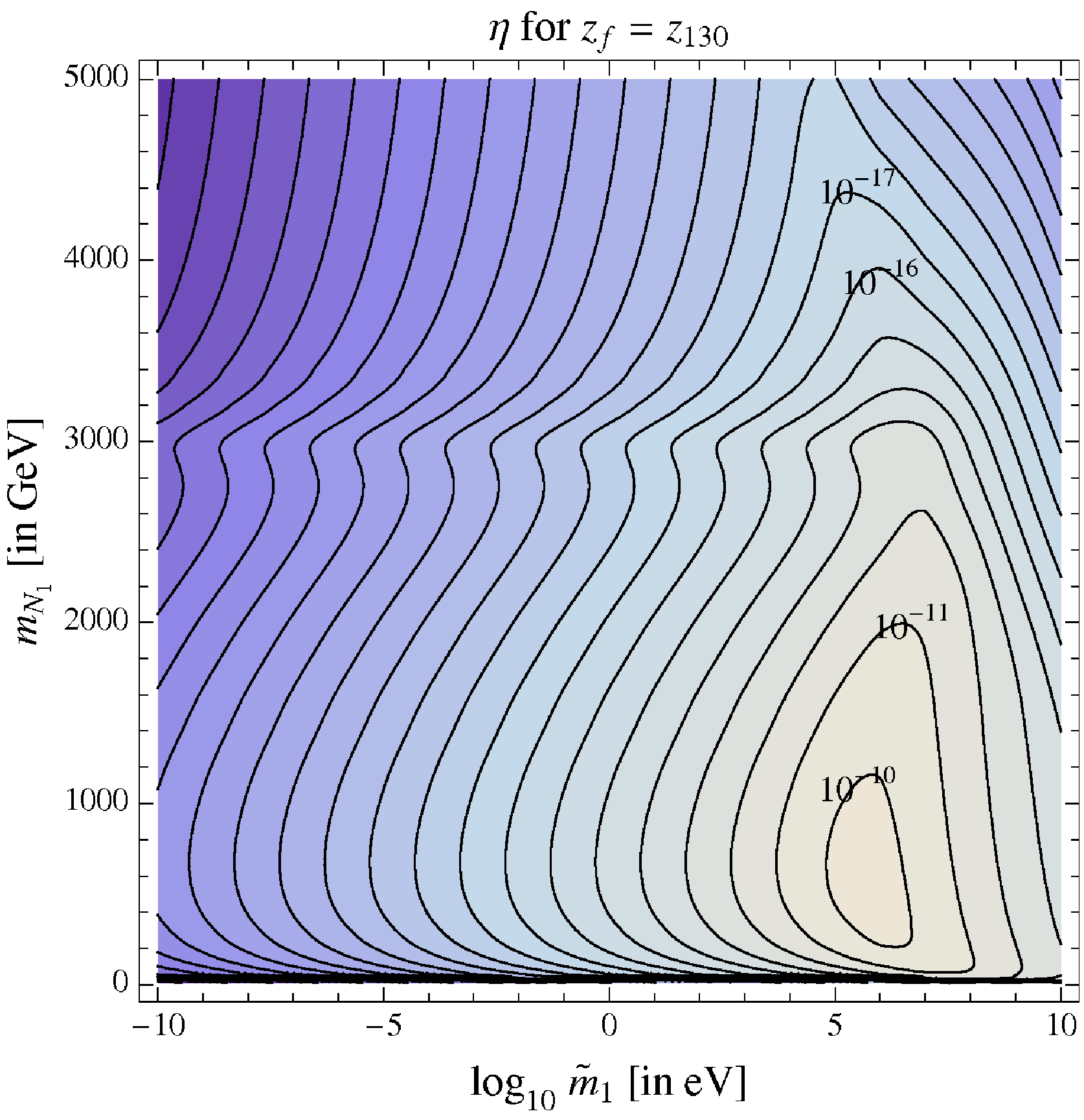}
\includegraphics[width=5cm]{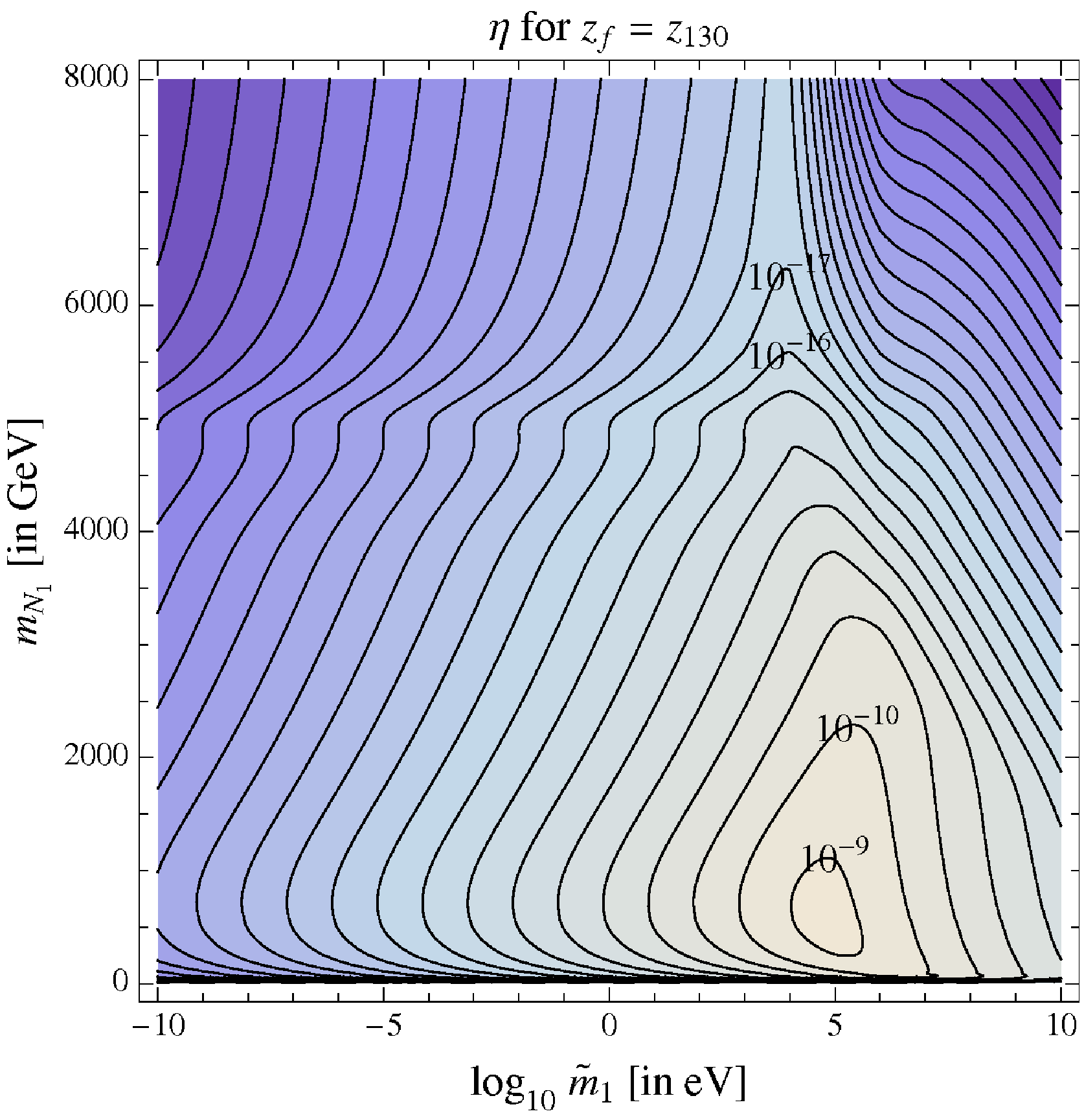}
\end{tabular}
\caption{For values of the right-handed gauge boson mass which could be probed at LHC, $m_{W_R}=0.8, \,3,\, 5$ TeV,  iso-efficiency curves as a
function of $\tilde{m}$ and $m_N$. As expected the efficiency decreases from right to left panel, and is always too suppressed to obtain successful baryogenesis.} \label{efficiencies}
\end{figure}

All in all the efficiency we obtain numerically is given in Fig.~\ref{efficiencies}, as a function of $m_N$ and $\tilde{m}=v^2{Y_\nu^\dagger Y_\nu}/{m_N}=\Gamma_N^{(l)}{8 \pi v^2}/{m_N^2}$ for
various values of $m_{W_R}= 800 \,\hbox{GeV},\,3\,\hbox{TeV},\,5\,\hbox{TeV}$ with $v=174$\,GeV.
$m_{W_R}=800$\,GeV corresponds essentially to the lower experimental limit \cite{PDG}, while $m_{W_R}= 3$~TeV corresponds essentially to the value LHC could reasonably reach \cite{LHCstudies}.
Motivated by the analysis of Ref.~\cite{Burnier}, these figures are based on the approximation that all $L$ asymmetry produced above $T\sim 130$\;GeV (for $m_h\sim 120$\,GeV) has been converted to a $B$ asymmetry (with conversion factor as given in Eq.~(\ref{LtoBfactor})), but none of it afterwards.
In all cases we get an efficiency factor far below $\sim 7\cdot 10^{-8}$ which is the minimum value necessary to get the observed baryon asymmetry $Y_{\cal B}= (6-9)\cdot 10^{-11}$ (with maximal CP-asymmetry).

\begin{figure}
\begin{tabular}{cc}
\includegraphics[width=7.5cm]{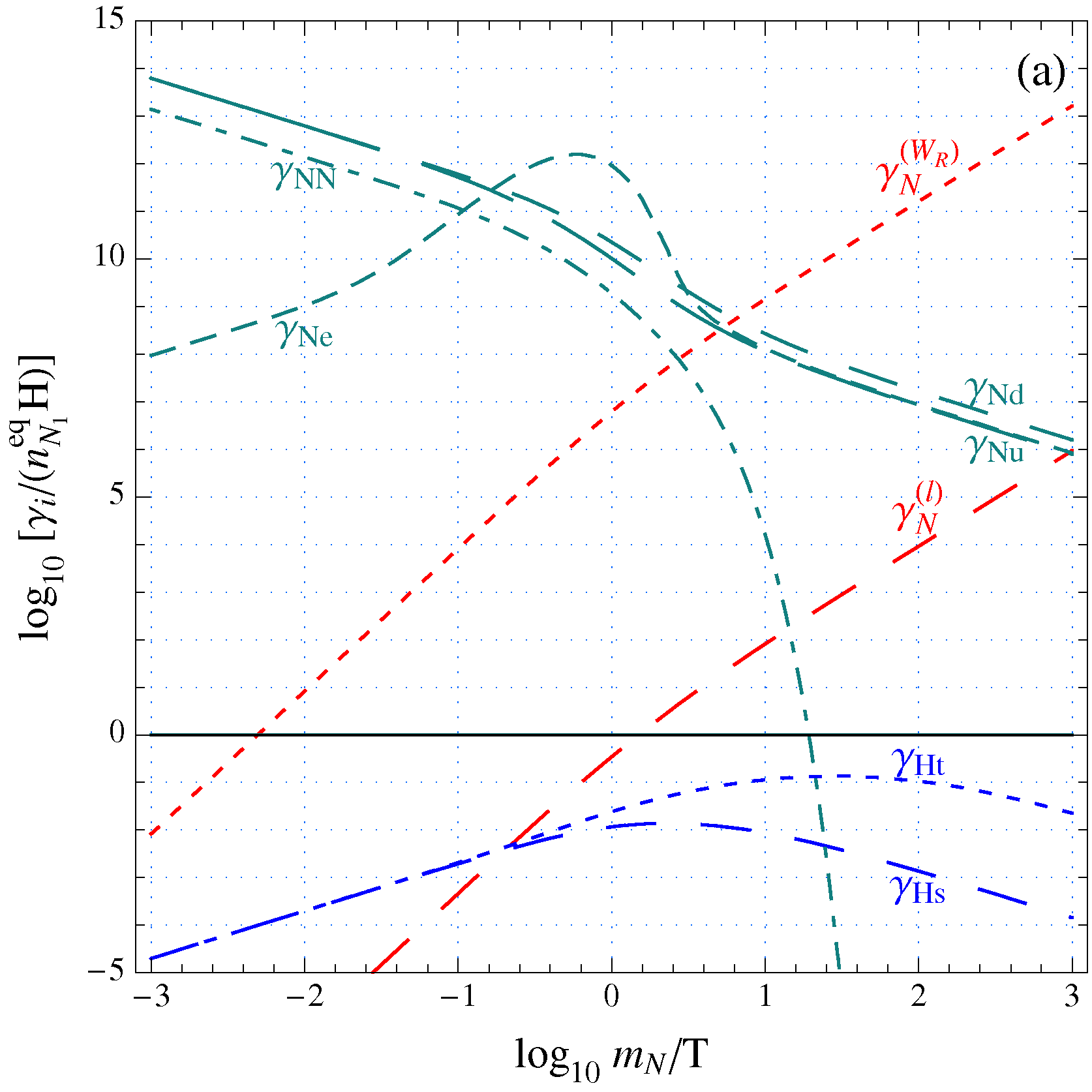}
\includegraphics[width=7.5cm]{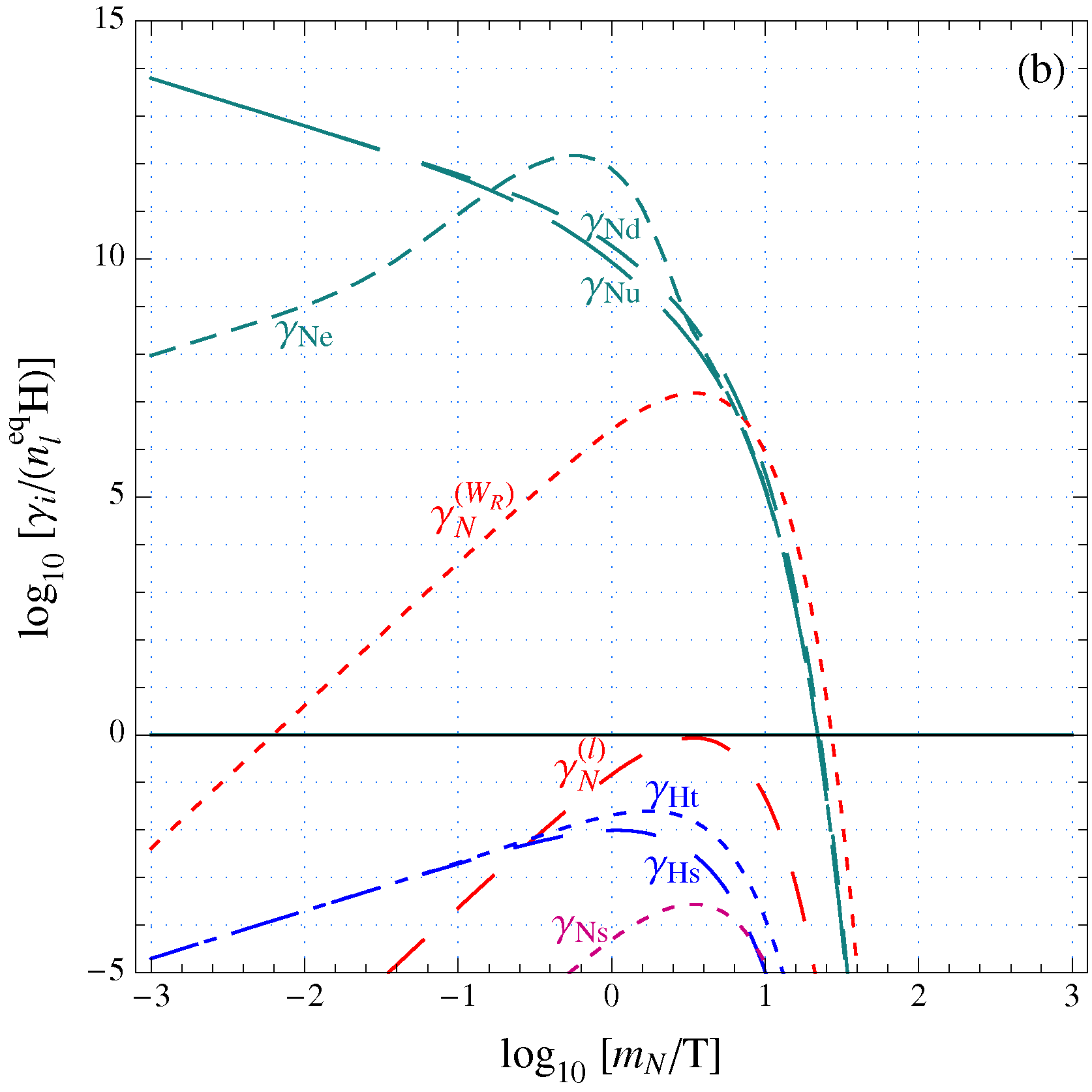}
\end{tabular}
\caption{Evolution of the reaction densities (a) $\gamma/(n_N^{eq} H)$ and (b) $\gamma/(n_l^{eq} H)$ with $z$ for $m_N=500$ GeV, $m_{W_R}=3$ TeV and $\tilde{m}=10^{-3}$ eV.}
\label{rates}
\end{figure}

\begin{figure}
\begin{tabular}{cc}
\includegraphics[width=5cm]{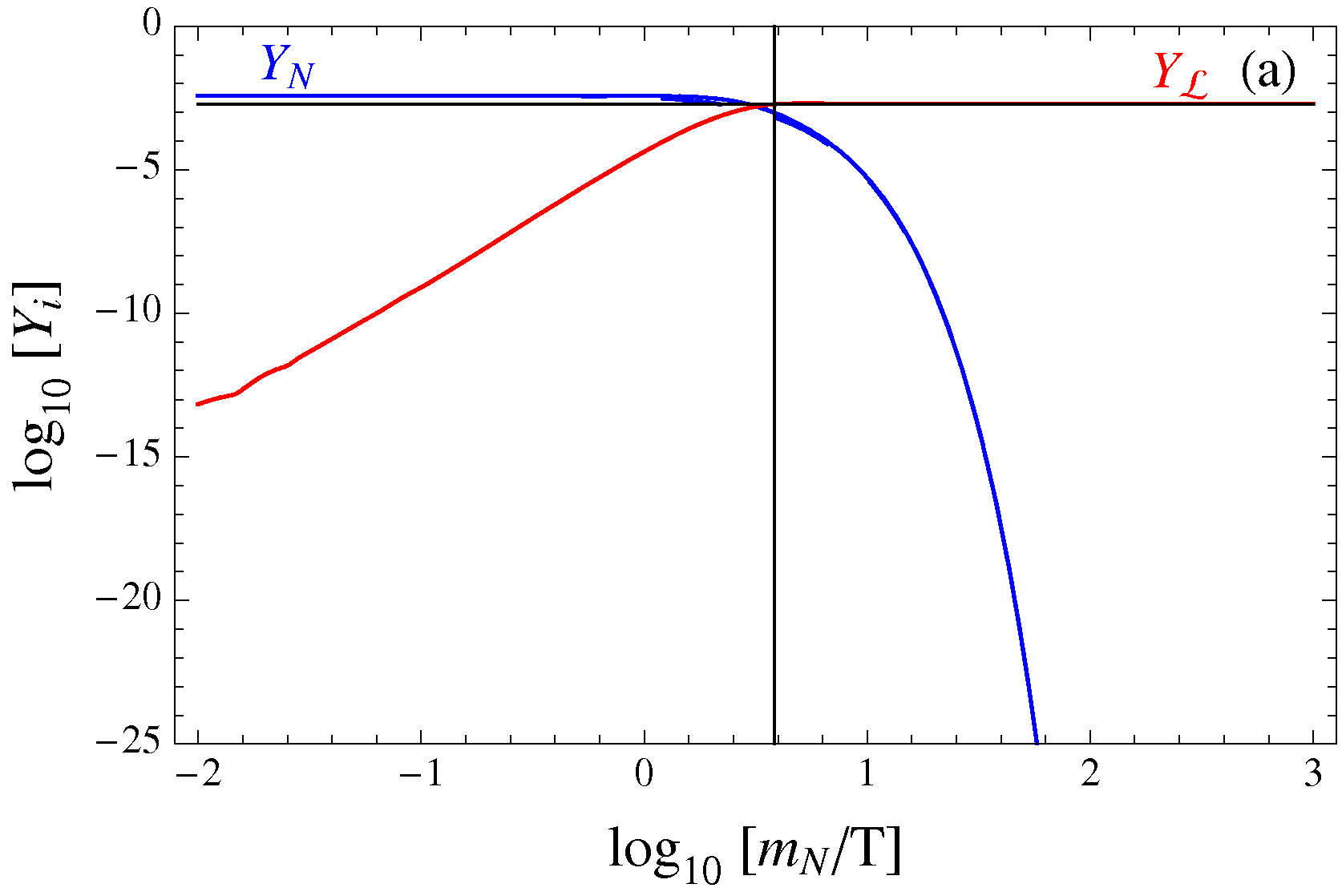}
\includegraphics[width=5cm]{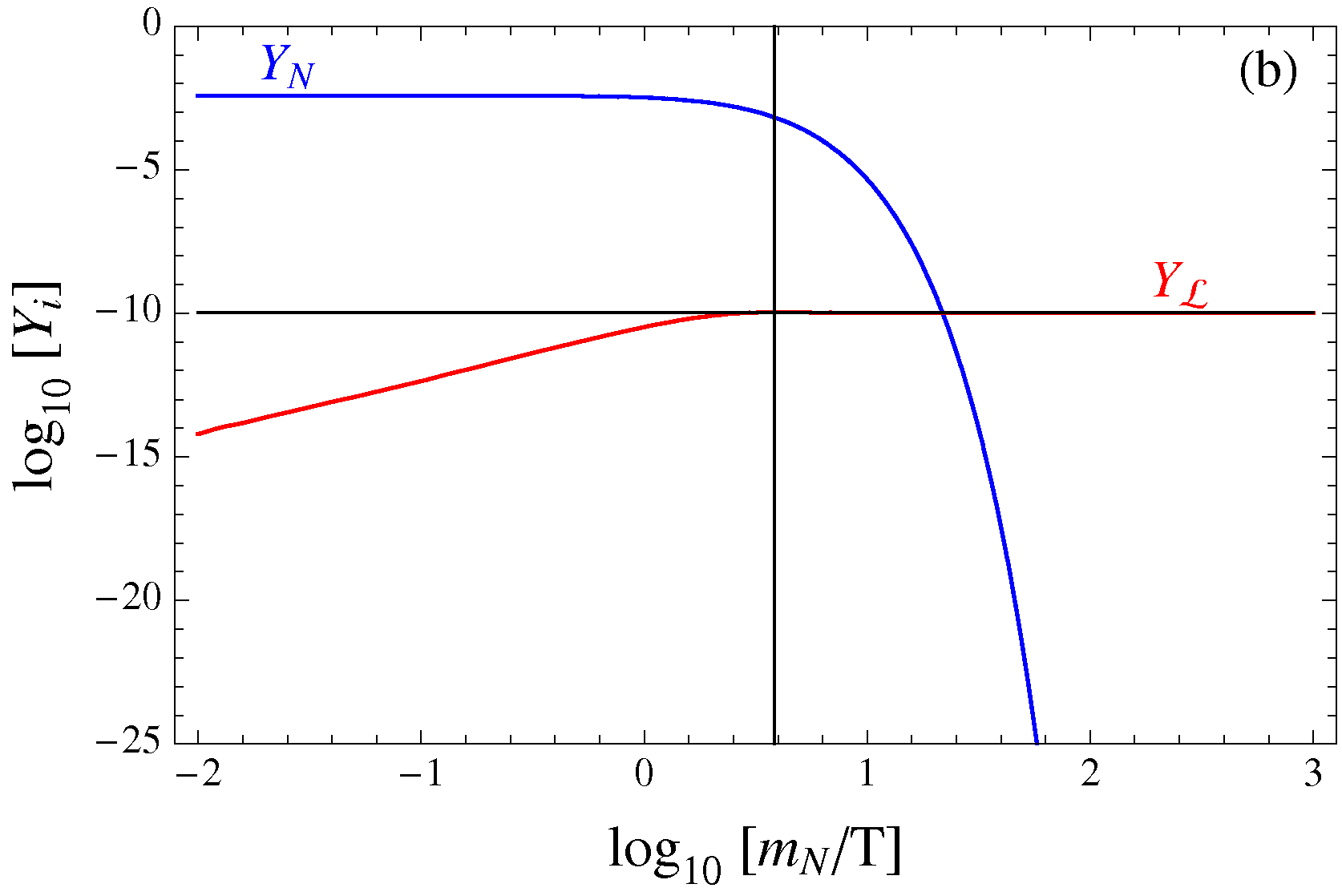}
\includegraphics[width=5cm]{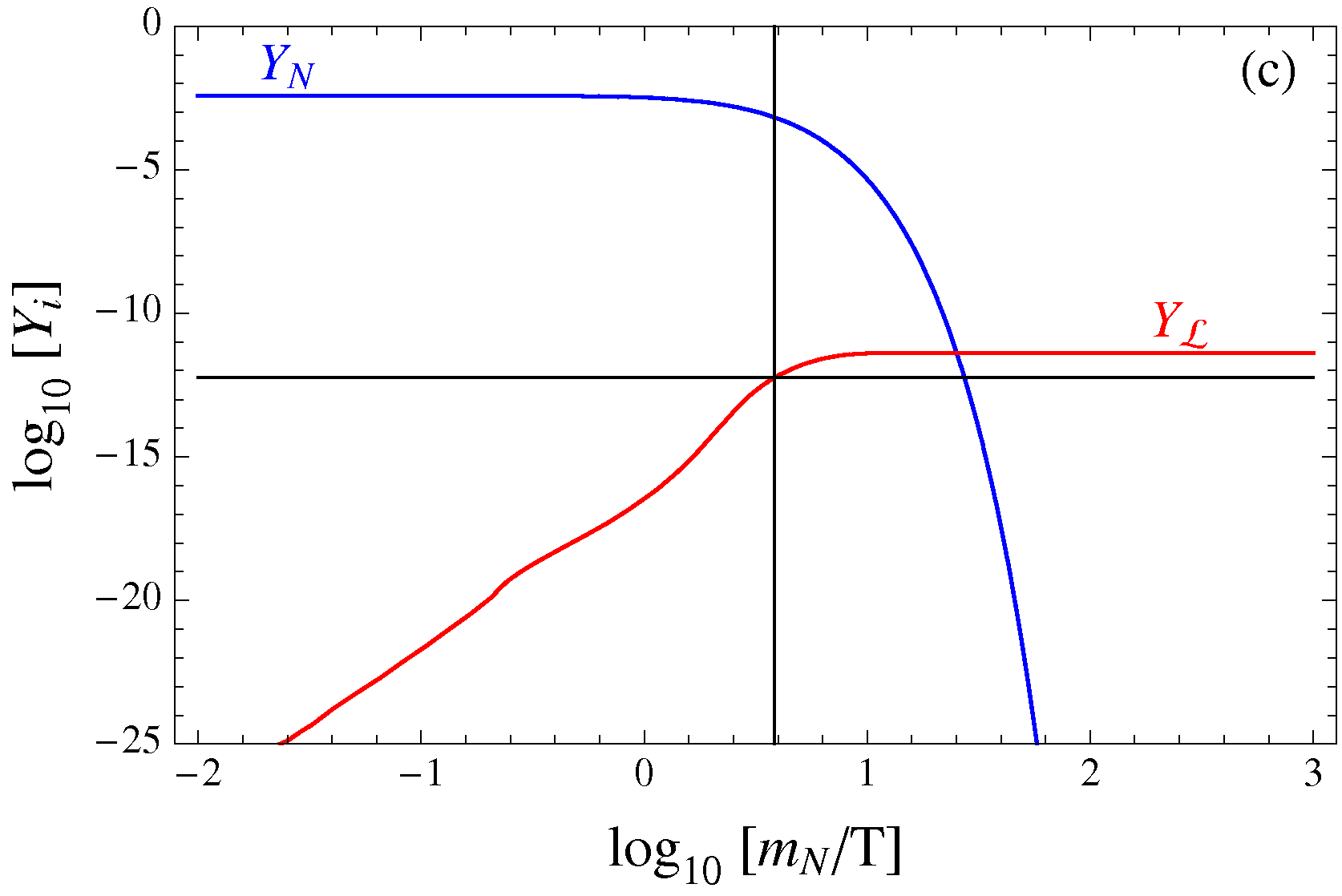}\\
\includegraphics[width=5cm]{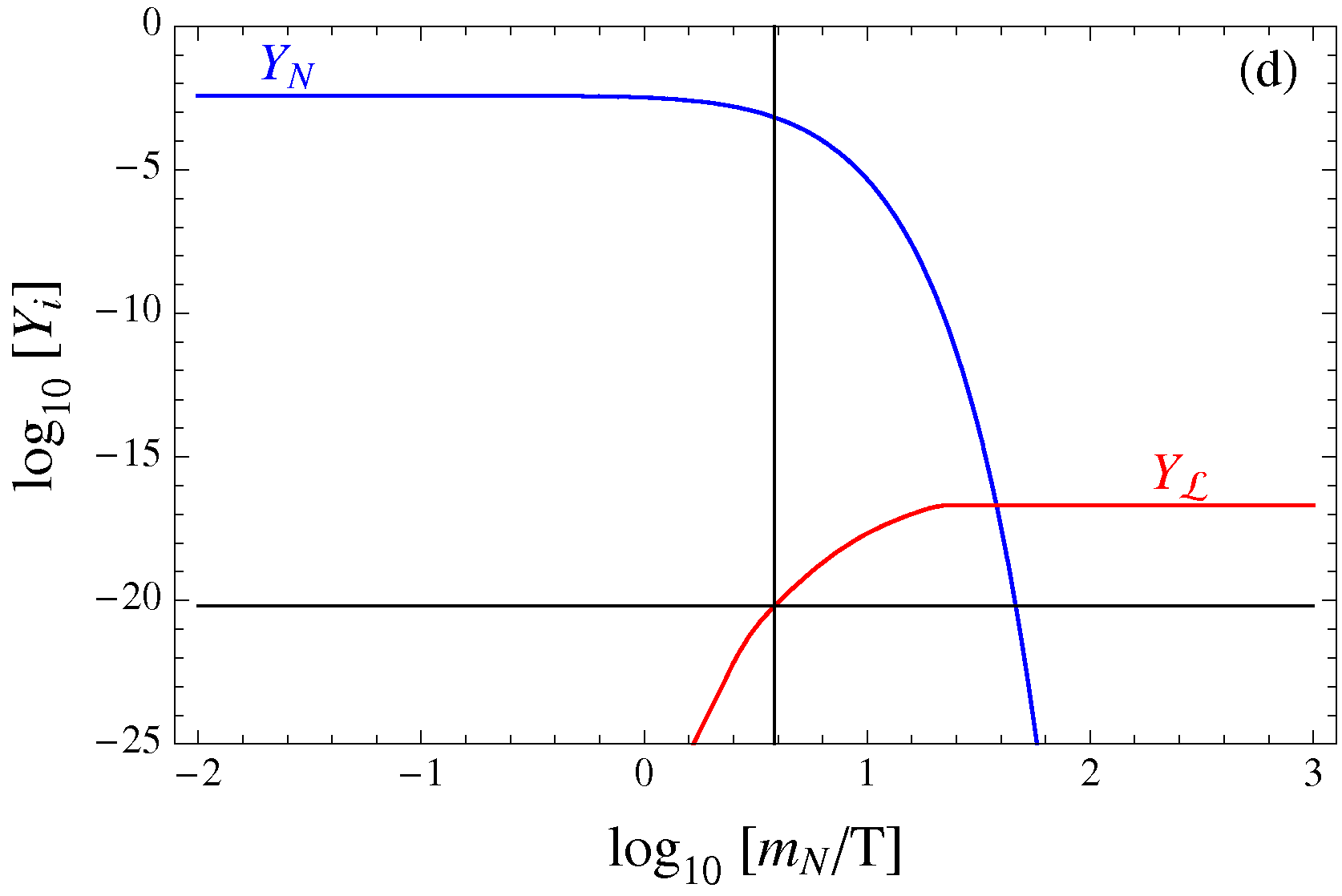}
\includegraphics[width=5cm]{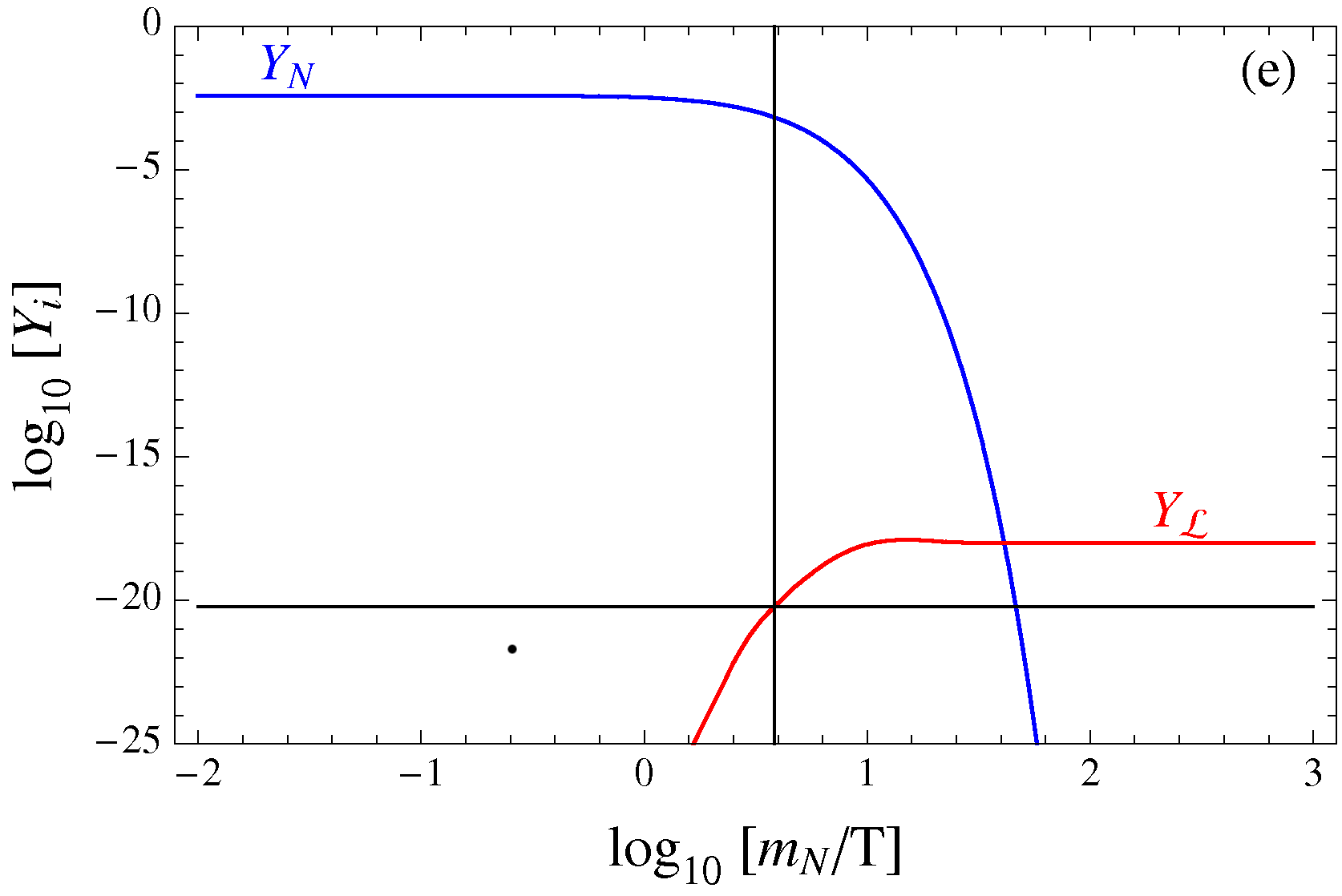}
\end{tabular}
\caption{Evolution of $Y_N$ and $Y_{\cal L}$ abundances as a
function of $z=m_N/T$ for $m_N=500$ GeV, $m_{W_R}=3$ TeV and
$\tilde{m}=10^{-3}$ eV, including various effects in the Boltzmann
equations as explained in the text. The straight lines indicate the
value of $z$ and $Y_{\cal L}$ at sphaleron decoupling.}
\label{abundances}
\end{figure}

To understand these results it is useful to discuss the effect of
the various terms step by step. For this, we take as example the set
of parameters: $m_N=500$~GeV, $m_{W_R}=3$~TeV,
$\tilde{m}=10^{-3}$~eV. Fig.~\ref{rates} provides the various
reaction densities divided by $n_N^{eq} H$ and $n_l^{eq} H$, as
relevant for discussing thermal equilibrium in the $Y_N$ and
$Y_{\cal L}$ Boltzmann equation respectively. Fig.~\ref{abundances}
gives the $Y_N$ and $Y_{\cal L}$ abundances as a function of $z$. As
well known, omitting all $W_R$ interactions,
Fig.~\ref{abundances}.a, there is no large efficiency suppression
for $\tilde{m}=10^{-3}$ eV, we get $\eta\simeq 0.5$, i.e.~$Y_{\cal
B}= 6.2\cdot10^{-4}$ (with $\varepsilon_N =1$). Adding to this case
only the effect of the 3 body decay in the $Y_N$ Boltzmann equation,
Fig~\ref{abundances}.b, leads to the dilution effect explained
above: $\eta \simeq \gamma_N^{(l)}/\gamma_N^{(W_R)}\simeq 2.8\cdot10^{-8}$,
i.e. $Y_{\cal B}\simeq 3.6\cdot10^{-11}$. Adding the gauge scattering
terms in the $Y_N$ Boltzmann equation leads to a even more suppressed
result for any $z < 6.5$ because in this range $\gamma_A >
\gamma_N^{(W_R)}$. Given the fact that the sphaleron decoupling
temperature corresponds to $z\simeq 4$ we do get an extra
suppression: $\eta\simeq 1.5\cdot10^{-10}$, i.e.~$Y_{\cal B}
\simeq1.8\cdot10^{-13}$, Fig.~\ref{abundances}.c. The efficiency is
roughly given by the value of $\gamma_A/\gamma_N^{(l)}$ a bit before
sphaleron decoupling.
Note that the result is sensitive to the sphaleron
decoupling temperature. For smaller decoupling temperatures where
$\gamma_A$ is smaller the efficiency would have been larger and
would have lead to about the same result as in
Fig.~\ref{abundances}.b. Adding furthermore the $\Delta L=1$ gauge
scattering effects in the $Y_{\cal L}$ Boltzmann equation,
Fig.~\ref{abundances}.d,  leads to further suppression because for
$T> 130$ ~GeV, these scatterings turn out to be fast enough to put
leptons close to chemical equilibrium, i.e. $\gamma_A/n_l^{eq} H >
1$, see Fig.~\ref{rates}.b. We get: $\eta\simeq1.6\cdot10^{-18}$,
i.e. $Y_{\cal B}\simeq2.1\cdot10^{-21}$. Finally adding the 3 body
decay effect to the $Y_{\cal L}$ Boltzmann equation doesn't lead to
further sizable suppression at $T=130$ GeV because above this
temperature $\gamma_A > \gamma_N^{(W_R)}$. Only between $z\simeq6.5$
(when $\gamma_N^{(W_R)}$ becomes larger than $\gamma_A$) and $z=30$
(when $\gamma_N^{(W_R)}/n_l^{eq} H$ becomes smaller than 1) it could have had
an effect, compare Fig.~\ref{abundances}.d and
Fig.~\ref{abundances}.e. Alltogether at $T=130$ GeV we get
$\eta\simeq1.6\cdot10^{-18}$ as given in Fig.~\ref{efficiencies}.

\begin{figure}[!t]
\center
\includegraphics[width=5cm]{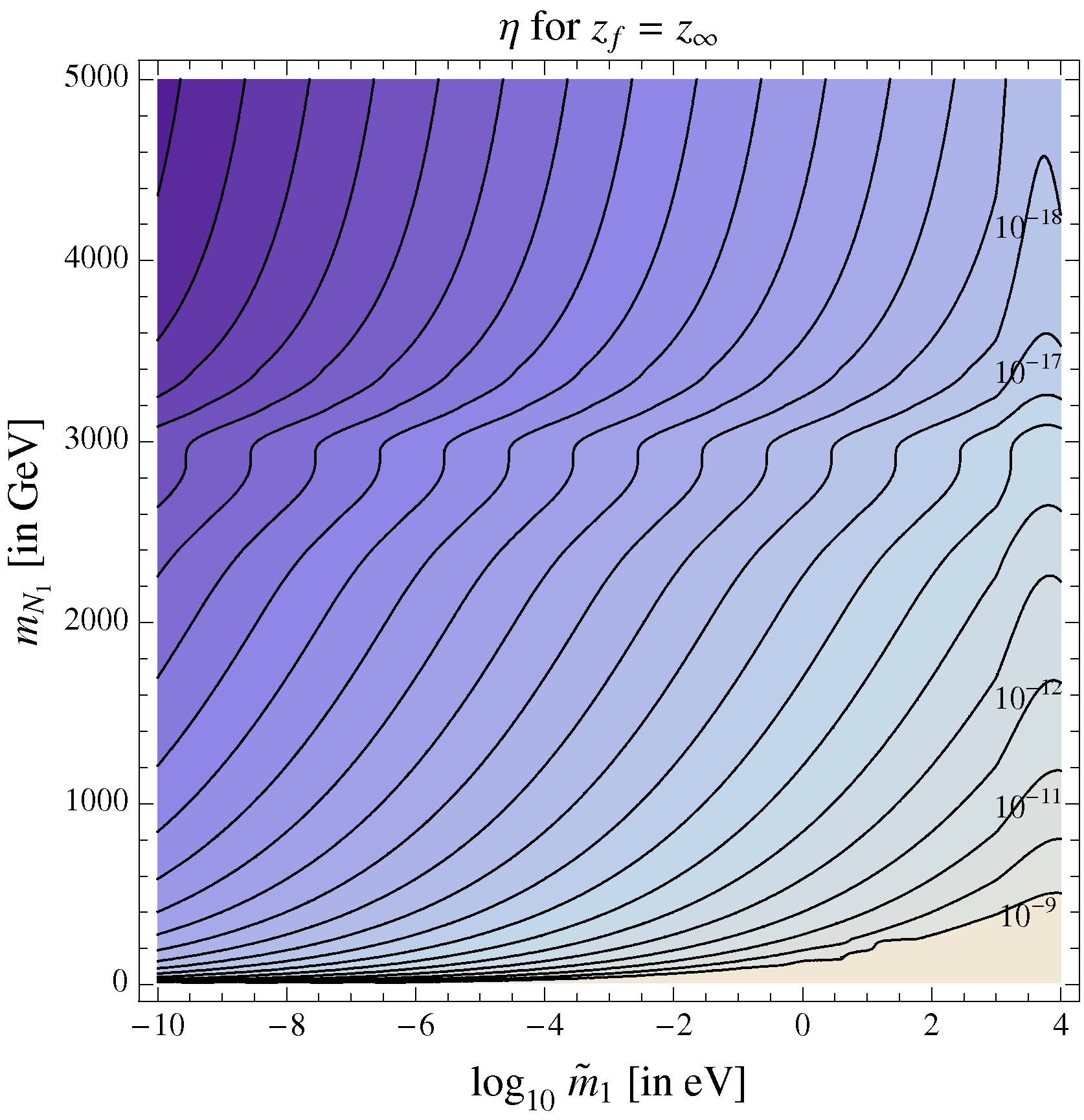}\\
\caption{Efficiencies without sphaleron decoupling for $m_{W_R}= 3$ TeV. (For values of $\tilde{m}$ beyond $10^{5}$~eV, Yukawa driven $\Delta L=2$ scatterings are so fast that the efficiency collapses.)}
\label{nosphaleron}
\end{figure}

Note that for $m_{W_R}=3$ TeV, the values $m_N\simeq 500$~GeV and
$\tilde{m} \simeq  10^5$~eV appear to be the ones which maximize the
efficiency. Larger values of $m_N$ lead to more suppression from the
$W_R$. Smaller values lead to a creation of the asymmetry occurring
too late to be converted by the sphalerons. The important effect of
sphaleron decoupling  for low $N$ mass can be seen by comparing
Fig.~2.b with Fig.~\ref{nosphaleron} where no sphaleron decoupling temperature cut
has been applied. Similarly smaller values of $\tilde{m}$ leads to
more suppressed efficiency from larger $\gamma_A/\gamma_N^{(l)}$ and
$\gamma_N^{(W_R)}/\gamma_N^{(l)}$ ratios in the $Y_N$ Boltzmann
equation. Large values of $\tilde{m}$ lead though to very large
suppression from Yukawa driven inverse decays and $\Delta L=2$
scatterings. Those effects start to dominate over the $W_R$ effects
for $\tilde{m} \simeq 10^5$~eV, which explains why in Fig.~2.a
maximum is got around this value of $\tilde{m}$: $\eta \simeq
10^{-10}$.

Note also that, for $m_N \sim m_{W_R}$, in Fig. \ref{efficiencies}, there is a local enhancement of the efficiency because,
as $m_N$ approaches $m_{W_R}$ from below, the $\gamma_A$ rate becomes more and more insensitive
to the $W_R$ resonance.
However as $m_N$ gets larger than $m_{W_R}$ the $N \rightarrow W_R l_R$ decay opens
up and the efficiency gets again suppressed.

\begin{figure}[!h]
\center
\begin{tabular}{c}
\includegraphics[width=5cm]{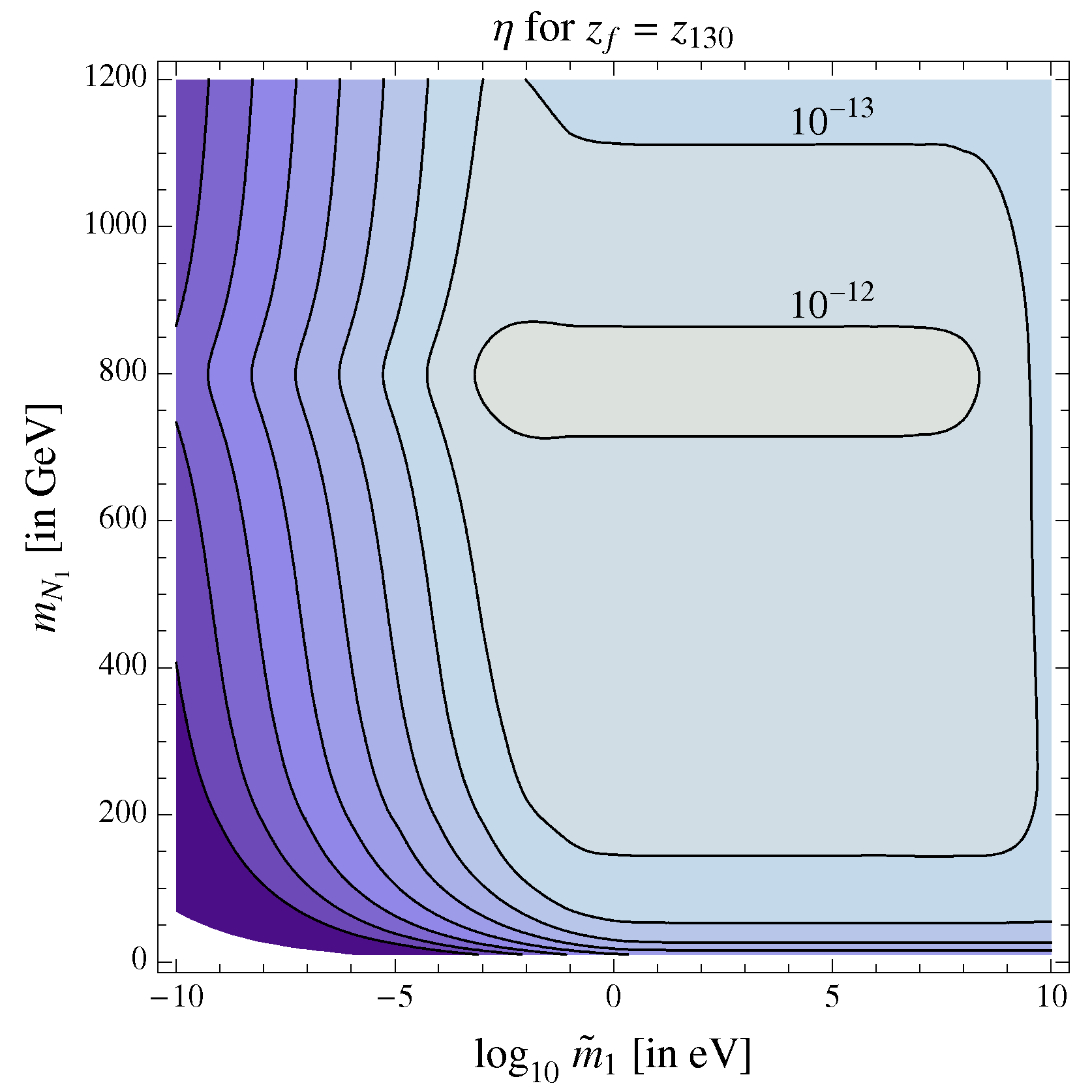}
\includegraphics[width=5cm]{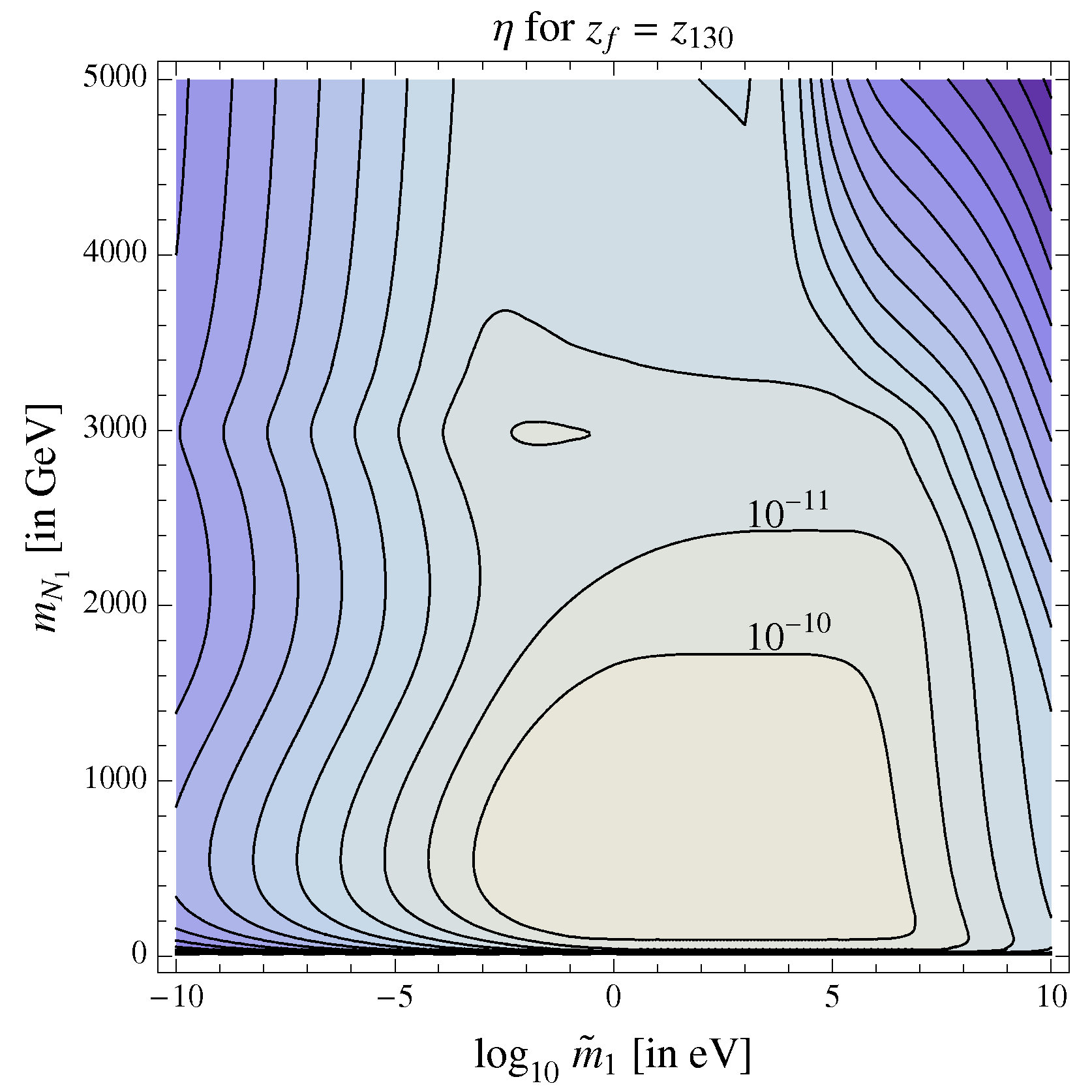}
\includegraphics[width=5cm]{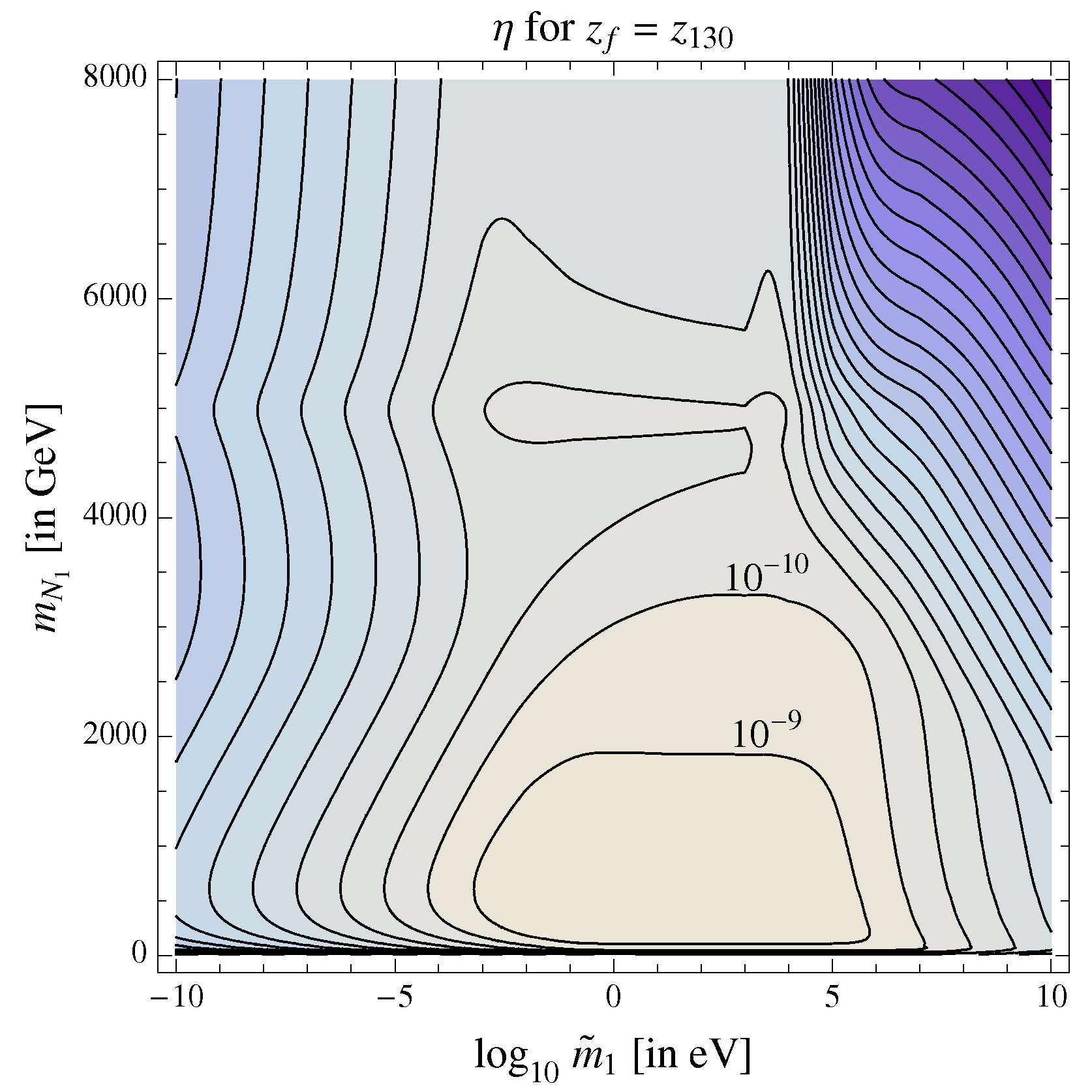}
\end{tabular}
\caption{Iso-efficiency curves  for $m_{W_R}=0.8, \,3,\, 5$ TeV as a
function of $\tilde{m}$ and $m_N$ when gauge interactions are only present in the $Y_N$ Boltzmann equation.} \label{efficienciesWonlyN}
\end{figure}

One additional question one must ask is whether our results depend on the fact that  we considered only the evolution of the total lepton number asymmetry. The results can indeed largely depend on the flavour structures of the Yukawa couplings as well as on the flavour of the $SU(2)_R$ light partner of the $N$, but not enough to allow successful leptogenesis. For example even if $N$ could create an asymmetry
only in flavours orthogonal to the flavour of its $SU(2)_R$ partner, leptogenesis still wouldn't work.
In this case the asymmetry produced wouldn't be washed-out by any $W_R$ interaction appearing in the $Y_{\cal L}$ Boltzmann equation, but still the $W_R$ thermalization effects in the $Y_N$ Boltzmann equation would be fully effective since they do not depend on flavour.\footnote{We neglect effects of charged leptons Yukawa couplings which are much less important.} We have checked over the full $\tilde{m}$ and $m_N$ parameter space that even in this extreme case we would get a far too suppressed efficiency to have successful leptogenesis. Our results for this case are given in Figure.~\ref{efficienciesWonlyN}, see also the example of Fig.~\ref{abundances}.

One more question to ask is whether the results obtained above could
sizeably depend on the initial distribution of $N$ before they
decay. The answer is simply no, due to the fact that, starting from
any number of $N$ at temperature above $m_N$ (from no $N$ to only
$N$ in the universe) the $W_R$ interactions very quickly put the
$N$'s in deep thermal equilibrium.

Note finally that since we neglected the scatterings of Fig.~\ref{gauge_scatterings}.b and
Fig.~\ref{gauge_scatterings}.c, strictly speaking our result is valid only for $m_N < m_{W_R}$. But this is where the
maximum efficiency is obtained and elsewhere these scatterings can
only suppress even more leptogenesis.

\section{Bounds on $m_{W_R}$ and $m_N$}

\begin{figure}
\center
\begin{tabular}{cc}
\includegraphics[width=7.5cm]{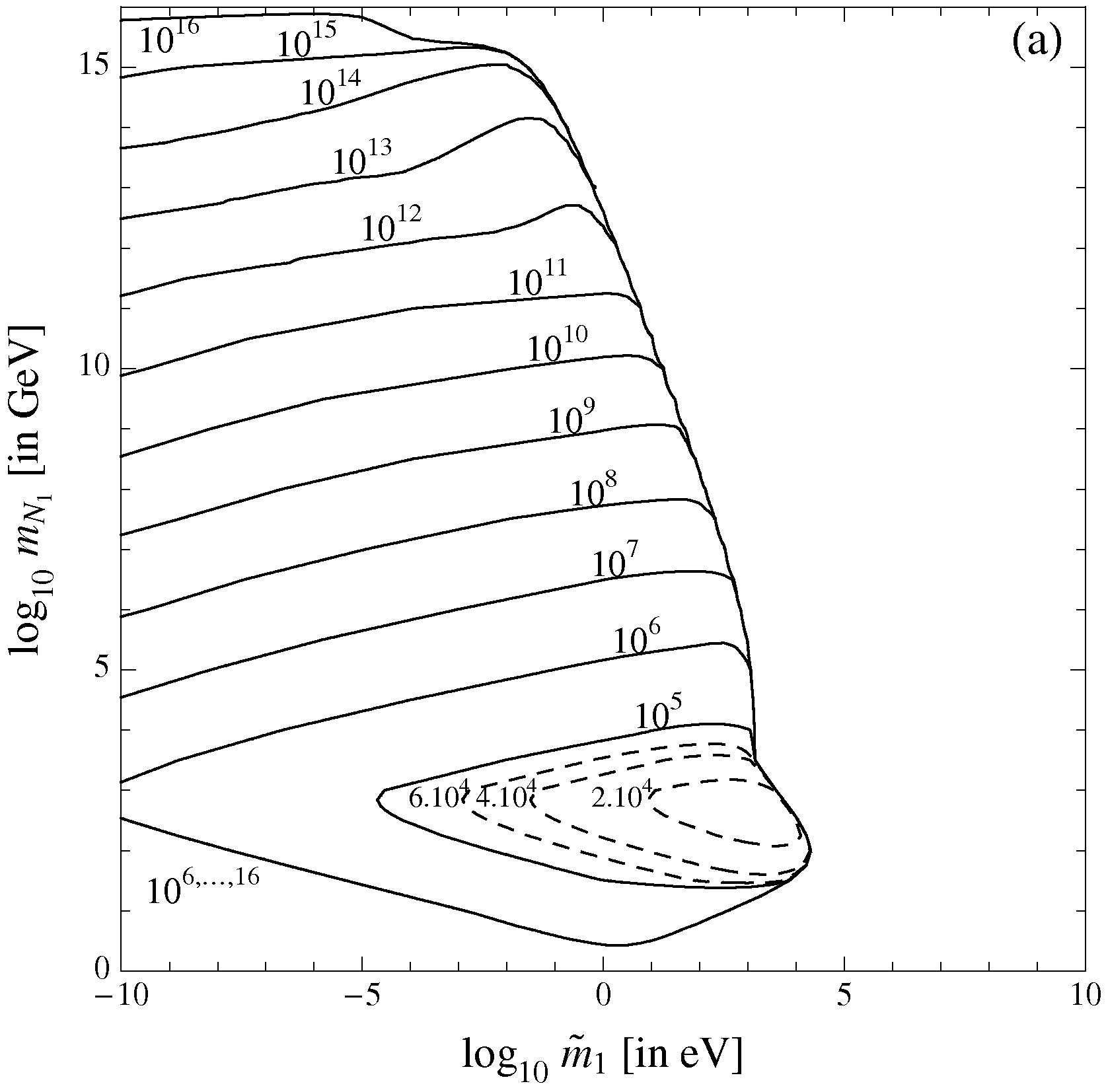}
\raisebox{.17cm}{\includegraphics[width=7.25cm]{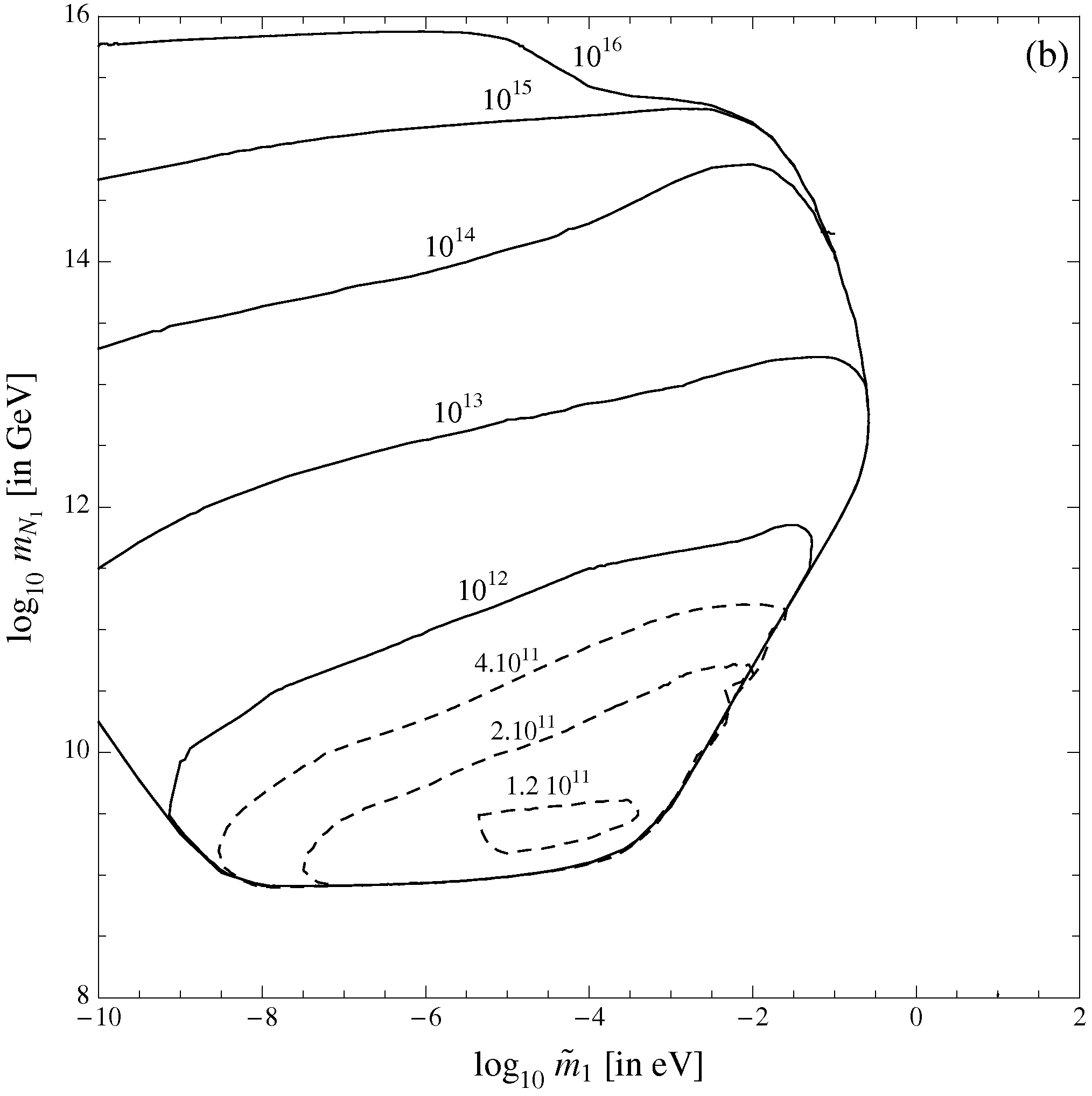}}
\end{tabular}
\caption{For various values of $m_{W_R}$ (in GeV), the inner part of each curve gives the values of $\tilde{m}$ and $m_N$ which can lead to successful leptogenesis (i.e. $Y_{\cal B}=9\cdot10^{-11}$). Left (right) pannel is obtained for $\varepsilon_N = 1$ ((3/16$\pi$)$ \,m_N\sqrt{\Delta m^2_{atm}}/v^2$). The dependance in $m_{W_R}$ of the lower bound on $m_N$ is totally negligible, except for $m_{W_R}< 10^6$ (left panel) and $m_{W_R}< 2\cdot10^{11}$ (right panel). }
\label{Wbounds}
\end{figure}

In the previous section we have seen that for $m_{W_R}$ reachable at
LHC, successful leptogenesis from $N$ decays is not possible. Larger
values of $m_{W_R}$ lead however to better efficiencies. It is
useful to determine what are the bounds on $m_{W_R}$ for a given
value of $m_N$ and vice versa. These can obtained from Fig.~\ref{Wbounds}.a
which for fixed values of $m_{W_R}$ gives the allowed range of $m_N$
and $\tilde{m}$ taking the maximum value $\varepsilon_N=1$. One
observes that the absolute lower bound on $m_{W_R}$ is  $18$~TeV. It is obtained for
$m_N=500$ GeV and $\tilde{m}=3\cdot10^{2}$ eV. This value of $\tilde{m}$ requires
large cancellations between large Yukawa couplings in the neutrino
masses. More usual values lead to a more severe bounds, we get
\begin{equation}
m_{W_R}>110, \,60, \,35 ~ \hbox{TeV} \quad \hbox{for} \quad \tilde{m}=10^{-5, -3, -1} \,\hbox{eV}
\label{boundsreso}
\end{equation}
Note also that as can be seen in Fig.~\ref{Wbounds}.a for successful leptogenesis 
we get the bound
\begin{equation}
m_N >  2.6 \, \hbox{GeV}
\end{equation}
which holds even for the case where $W_R$ effects are negligible. This gives an absolute lower bound on $m_N$ which is another tantalizing target for excluding leptogenesis.

For completeness we also give in Fig.~\ref{Wbounds}.b the results we obtain taking the lower bound $\varepsilon_{N}< (3/16
\pi) \,m_N \sqrt{\Delta m^2_{atm}}/v^2$ \cite{di} which holds for a
hierarchical spectrum of right-handed neutrinos.
We obtain the absolute bound $m_{W_R}> 10^{11}$\,GeV which requires $m_N=2.6\cdot10^9$\,GeV and
$\tilde{m}=5\cdot 10^{-5}$\,eV. We also get
\begin{equation}
m_{W_R}>1.1\cdot 10^{11}, \,1.3\cdot 10^{11},\,1.1\cdot 10^{12} \, \hbox{GeV} \quad \hbox{for} \quad \tilde{m}=10^{-5,-3,-1} \, \hbox{eV}\,.
\label{boundshier}
\end{equation}

The flavour dependance of the results of this section is relatively moderate. For the extreme case above where all $W_R$ have been omitted in the $Y_{\cal L}$ Boltzmann equation, instead of equation Eq.~(\ref{boundsreso}), we get $m_{W_R}>39, \, 13, \, 8.8 ~ \hbox{TeV}$, while the absolute lower bound on $m_{W_R}$ becomes $8.7$\,TeV which we obtained for $\tilde{m}=10^1$\,eV. The bounds of Eq.~(\ref{boundshier}) in this case are relaxed by less than 10 percent, while the lower bounds on $m_N$, as well as the upper bounds on $\tilde{m}$, are negligibly affected in Figs.~\ref{Wbounds}.a and \ref{Wbounds}.b. As for the upper bounds on $m_N$ in these figures, they are relaxed by up to one order of magnitude. The results of Fig.~\ref{Wbounds}.b agree with the one of \cite{Cosme} for what can be compared, modulo these flavour effects, since the $W_R$ effects are neglected in the $Y_{\cal L}$ Boltzmann equation in this reference. 

Note that we do not expect that the results of Fig.~\ref{Wbounds} could be largely affected by the (neglected) scatterings of Fig.~\ref{gauge_scatterings}.b-c, because
all bounds in these figures are obtained with $m_N \lesssim m_{W_R}$ (except in corners of parameters space for large $m_{W_R}$ and large $\tilde{m}$ where it is not excluded that these scatterings could reduce the bounds on $m_N$ by up to a few times).

\section{Generalization to several right-handed neutrinos}

The results obtained above are strictly valid only if the lepton asymmetry is produced by a single 
right-handed neutrino, the effects of the other heavy states being present only in the CP asymmetry $\varepsilon_N$ and in the $\Delta L =2$ washout.\footnote{In $\gamma_{Ns}^{sub}$ and $\gamma_{Nt}$ above we took into account the contributions from $N_{2,3}$ proportional to the neutrino masses, as given in Eqs.~(92, 93) of Ref.~\cite{gammas2} with $\xi=\sqrt{\Delta m^2_{atm}}/\tilde{m}$, because these contributions are relevant anyway (even for hierarchical $N$'s) for very large $m_N$ and/or very large $\tilde{m}$.} 
Consequently these results assume that the heavier states do not create their own asymmetry and do not induce any washout besides this $\Delta L =2$ one. However, we are not aware of any model where $\varepsilon_N$ can be obtained as large as unity, the upper bound we considered above, 
and where the above assumption can be fully justified.
For instance, as said above, one possibility to have large CP asymmetries at low scale is through quasi-degeneracy of at least 2 right-handed neutrinos leading to a resonant enhancement of the self-energy diagram. In this case to a very good approximation both right-handed neutrinos have equal CP-asymmetries and equal masses, which means that both $N_{1,2}$ must be considered in the Boltzmann equations.
In the Appendix \ref{appendix} we show that this does not change though our conclusions.
The point is that the asymmetry produced by two neutrinos is bounded by the sum of both asymmetries we get in the single $N$ case with $\tilde{m}=\tilde{m}_1$ and with $\tilde{m}=\tilde{m}_2$ (with $\tilde{m}_i$ refering to the value of $\tilde{m}$ of $N_i$), Eq.~(\ref{ineqYL}). From the results of Figs.~\ref{efficiencies} and \ref{efficienciesWonlyN} this shows that the lepton asymmetry produced will be always too small to produce enough asymmetry if $m_{W_R}$ is as low as in these figures, as relevant for the LHC. Furthermore from this inequality, if both $\tilde{m}_i$ lie ouside the range of values 
allowed by Fig.~7.a, a large enough baryon asymmetry cannot be produced. Moreover it can be checked numerically that this figure remains also valid to a good approximation for the case $\tilde{m}=\tilde{m}_1=\tilde{m}_2$.
It is in this sense that this figure has to be interpreted for the several $N$ case.

\section{Other possible suppression effects}

\subsection{Effects of a $Z'$ associated to a $U(1)$ symmetry}

\begin{figure}
\begin{tabular}{c}
\includegraphics[width=5cm]{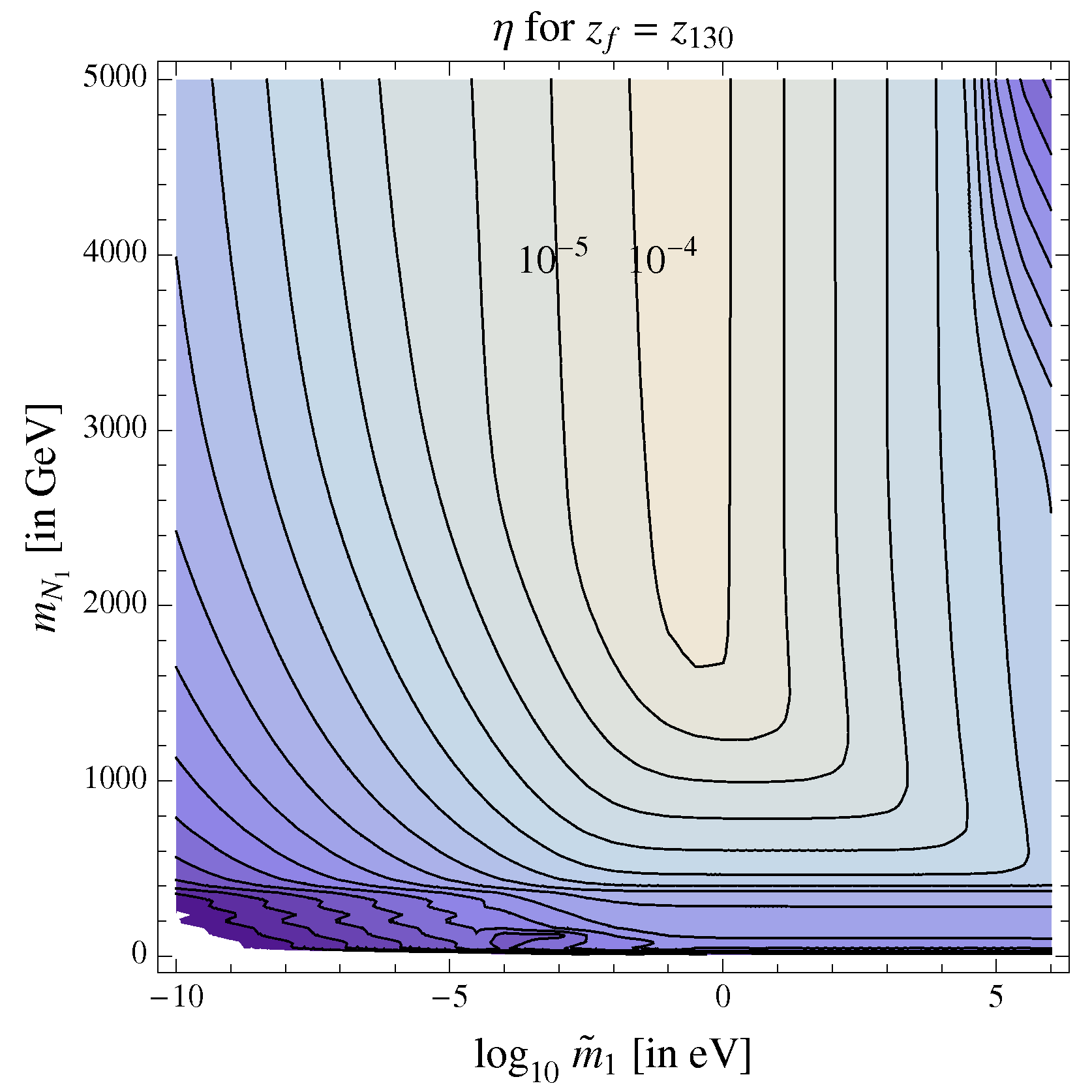}
\includegraphics[width=5cm]{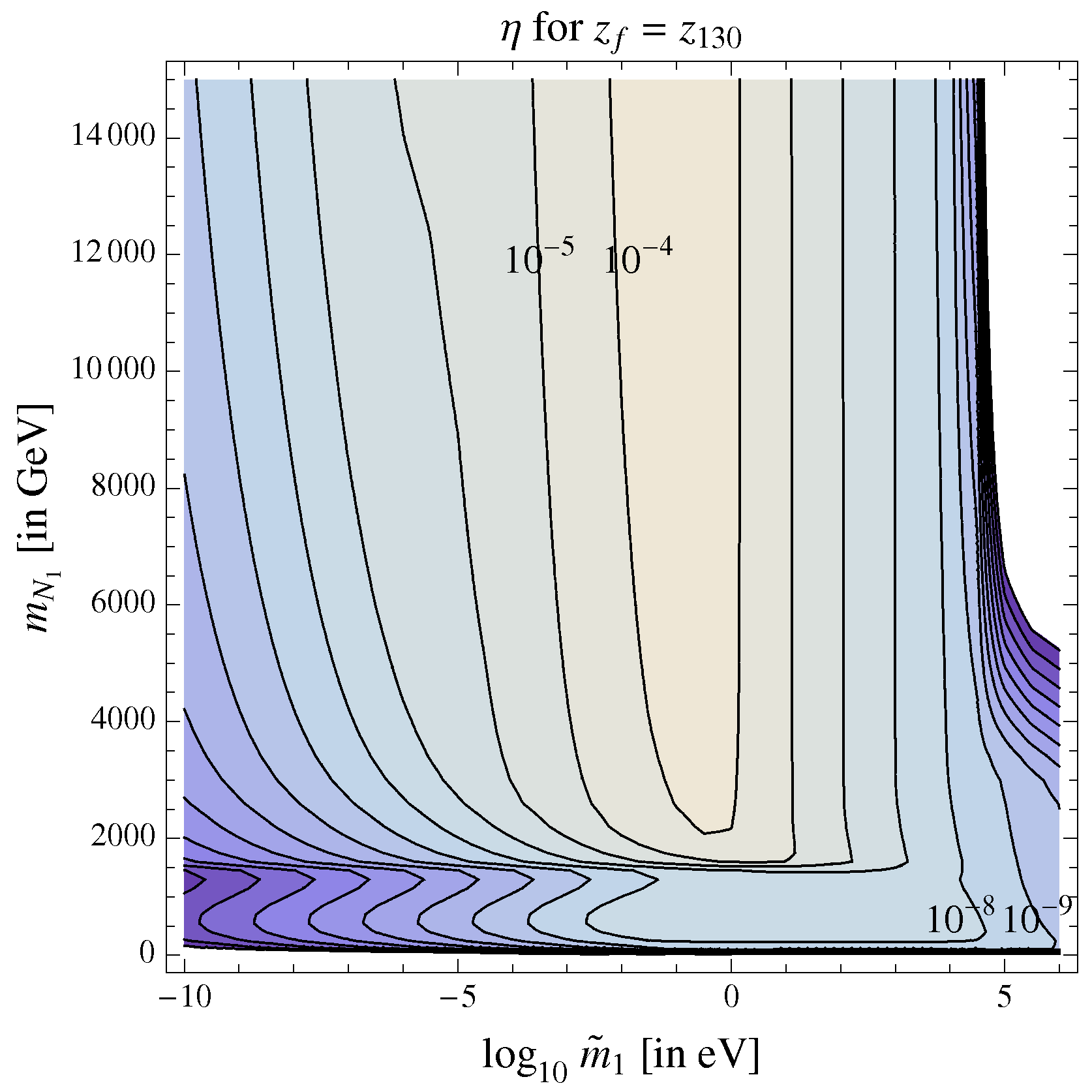}
\includegraphics[width=5cm]{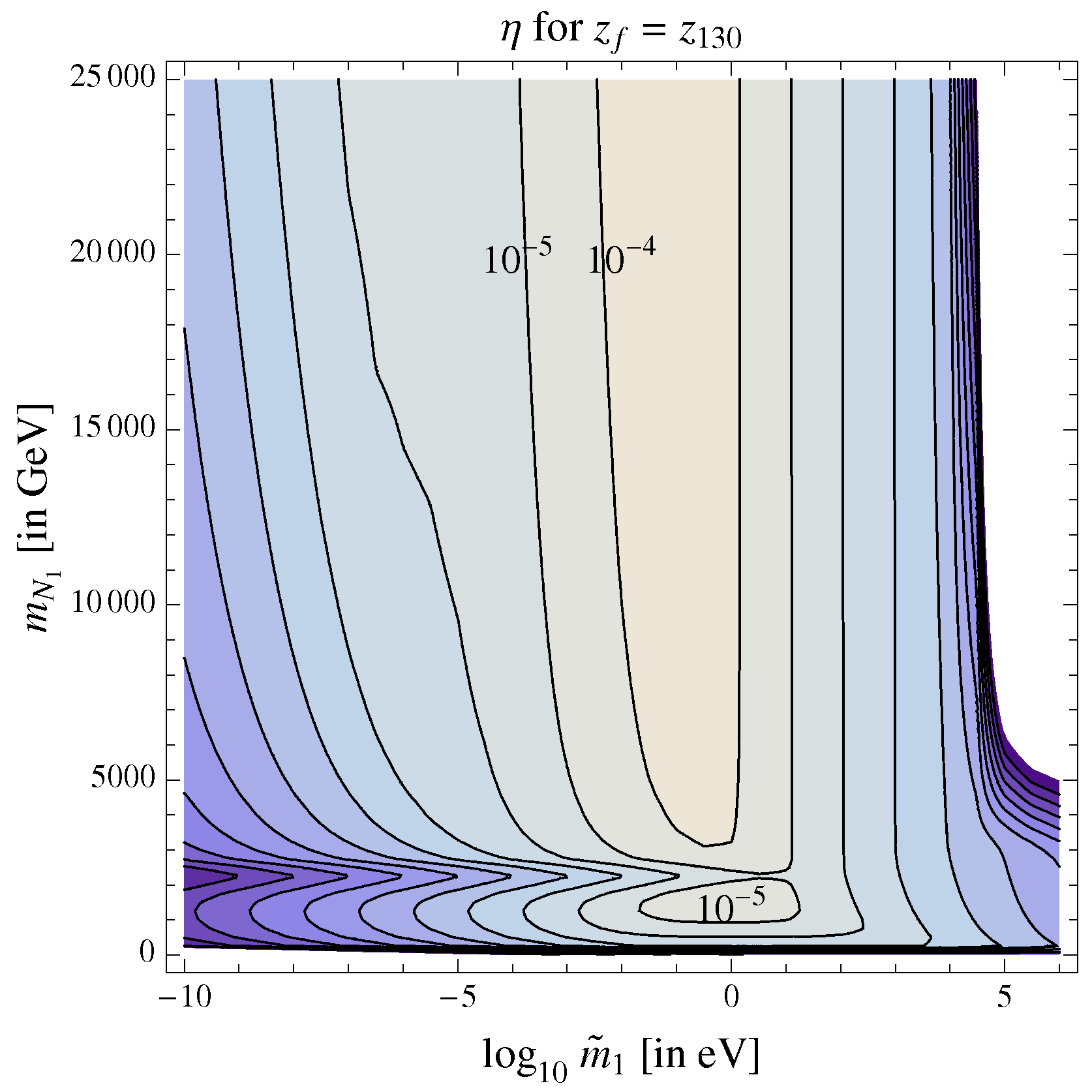}
\end{tabular}
\caption{Iso-efficiency curves  for $m_{Z'}=0.8, \,3,\, 5$ TeV as a
function of $\tilde{m}$ and $m_N$.} \label{Zefficiencies}
\end{figure}

A $Z'$ associated to an extra low energy $U(1)$
could be discovered at LHC up to $\sim 3$-5 TeV \cite{cmsreport,Wlhckras}. If it couples to $N$ through the
$Z'^\mu (\bar{N} \gamma_\mu N)$ interaction it has effect on the efficiency through the $Y_N$ Boltzmann equation. Since this interaction involves 2 $N$ it doesn't induce  any relevant 2 or 3 body decays which could cause dilution, and the associated scatterings decouple through a Boltzmann suppression. As a result the suppression effect is not as large as with a $W_R$.
For example considering a $U(1)_{Y'}$ as it has been considered in \cite{pluma}, see also \cite{kanu},
including all associated scatterings (i.e. the effect of $NN \leftrightarrow f \bar{f},\,HH$ scatterings), the efficiency we obtain for $M_{Z'}=0.8,\,3,\,5$ TeV is given in Fig.~\ref{Zefficiencies}. It shows that the discovery of a $Z'$ would not necessarily rule out leptogenesis depending on the values of $\tilde{m}$, but would require very large values of $\varepsilon_N$.

\subsection{Effects of a $Z'$ associated to a $SU(2)_R$ symmetry}

The neutral gauge boson associated to $SU(2)_R$ symmetry could
also be discovered at LHC up to $\sim 3$-5 TeV \cite{cmsreport,Wlhckras}. Since it is in the same multiplet as the $W_R$, its effect should be included
in the analysis above together with the effects of the $W_R$.
As it also couples only to 2 $N$, the suppression effects due to this neutral gauge boson will nevertheless
be negligible with respect to the ones of the $W_R$ when the asymmetry is created: the $N$ will have an interaction involving a $W_R$ before having one involving the $Z'$ (as long as $m_{Z'} \simeq m_{W_R}$ as expected in the left-right symmetric models).

\subsection{Effects of a right-handed triplet}

The consequences of the discovery of one or several components
of a right-handed scalar triplet $\Delta_R=(\delta_R^{++},\delta_R^+,\delta_R^0)$
could be dramatic for leptogenesis in some cases.

The easiest state to discover at LHC is the doubly charged one,
$\delta_R^{++}$, due to suppressed background in the same sign
dilepton channel \cite{tripletLHC}. As this state couples only to 2
right-handed charged leptons \cite{SU2R}, and doesn't couple directly to the
$N$, it has no sizable effect on the $Y_N$ Boltzmann equation but can
have an effect on the second one through L-violating $l_R l_R  H H$
interactions mediated by the $\delta_R^{++}$. This effect
can be large if the couplings involved are
of order $\sim10^{-4}$ or larger depending on the masses.
The presence of the $\delta_R^{++}$ would be however indicative of the
existence of other triplet members.

A $\delta_R^+$ (e.g.~more difficult to see at LHC because it doesn't
produce same sign dilepton channels in as direct a way as the
$\delta_R^{++}$), can couple to a $N$ and a $l_R$ as the $W_R$. It
can therefore induce dilution effect from the $N\rightarrow
\delta_R^+ l_R$ decay if kinematically allowed, or from $N
\rightarrow  l_R H^+H^0$ decays otherwise (i.e. through a
$\delta_R^+ H^- H^0$ coupling with $H$ any lighter scalar particle,
e.g.~from the bidoublet in LR models \cite{SU2R}). Similarly it
induces dangerous scatterings similar to the one of Fig.~1.a,
replacing the $W_R$ by a $\delta_R$ and the quark pair by a $H^+
H^0$ pair. For couplings in these processes as large as the $W_R$
gauge couplings, the suppression of the efficiency is expected to be
similar to the one caused by the $W_R$ in section 2, which would
rule out leptogenesis. For smaller couplings however the suppression
decreases quickly. In the later case leptogenesis can be
successfully produced from $N \rightarrow \delta_R^+ l_R$ decays if
kinematically allowed \cite{FHM}.\footnote{The observation of a
$W_R$ would rule out this leptogenesis mechanism in the same way as
in section 2.}

Finally the $\delta_R^0$ couples to 2 $N$ and therefore is expected to have
effects roughly similar to the ones of a $Z'$, if the Yukawa couplings are as large as the gauge couplings, less otherwise.

\subsection{Effects of a neutral or charged $SU(2)_L$ scalar singlet}

In large varieties of models, e.g.~non left-right, a $SU(2)_L$ scalar singlet can couple to 2 $N$
if it is neutral or to a $N$ and a $e_R$ if its electromagnetic charge is unity.
These states, if they also couple to right-handed quarks, can be dangerous for leptogenesis in a similar way as
the above $\delta^0_R$ and $\delta_R^+$ states respectively.

\section{Suppression effects in other frameworks : scalar and fermion triplet leptogenesis, electroweak baryogenesis }

In the above we have shown that a $W_R$ discovered at current or
future colliders would exclude any possibility to create a large enough baryon asymmetry from the
decay of a $N$. However there exist other ways to induce successfully the baryon asymmetry through leptogenesis. In seesaw models this can be achieved from the decay of a scalar triplet to 2 leptons or from the decay of a fermion triplet to a lepton and Higgses, through diagrams involving another heavy state \cite{typeIIlepto,typeIIleptoeffic,typeIIIlepto}.
In these models there are
washout effects from $SU(2)_L$ interactions.
These effects have been calculated in Refs.~\cite{typeIIleptoeffic,typeIIIlepto} and show that they are not large enough to rule-out leptogenesis even for masses as low as few TeV. For such low masses
leptogenesis appears to be possible though only for asymmetries of order unity (i.e.~assuming almost perfect resonance which requires e.g.~large fine-tuning).

Since a $W_R$ (or more generally any right-handed gauge boson)
does not couple to left-handed triplets, its discovery at low scale
would have no direct consequences for
the triplet number density Boltzmann equation.

The discovery of a $W_R$ at low scale would nevertheless provide a
strong hint for the existence of $N$'s at low energy, see section 7.
This would lead to 2 additional washout effects on the asymmetry
produced by the triplet decays. First, $\Delta L\neq 0$ scatterings
involving both the $W_R$ and $N_R$, Fig.~1, will be important (in
the flavour channels coupling to the $N$'s) if both these particles
have masses smaller or of order the triplet mass. Second, these $N$,
through their Yukawa interactions, and together with sphalerons,
could easily wash-out any previously produced lepton and baryon
asymmetry, unless some of their Yukawa couplings are so suppressed
that they preserve to a very good approximation at least one flavor
number combination (which has not to be preserved in the triplet
decay).

Putting all these effects together it can be checked that, the discovery of a $W_R$ and a $N$ would rule out the possibility to have any successful thermal leptogenesis from triplet decays at any scale as well, except for such kind of extreme flavour pattern.

Note that in the case of very low triplet mass a direct
discovery of the triplets is possible through Drell-Yan pair production \cite{tripletLHC,typeIIILHC}.

Finally leptogenesis is also possible in
more exotic models from the decay of $SU(2)_{L,R}$ singlets,
in case all the gauge interaction induced suppression effects considered
 in the above would be irrelevant for the decaying particle Boltzmann equation but still
 would be relevant for the $Y_{\cal L}$ one.
Similarly, electroweak baryogenesis with first order phase
transition from the presence of particles beyond the standard model
around the electroweak scale, can be affected by the L violating interactions
driven by a light $W_R$ and/or light $N$,
but could survive because these
cannot erase the $B$ asymmetry produced in this case.
For electroweak baryogenesis at the right-handed scale \cite{frere2} the effects could be large, and this would require
a specific analysis. 

\section{$N$ and $W_R$ at colliders}
We have shown this far to which (huge) extent the discovery of
gauge interactions affecting the right-handed sector would cripple
leptogenesis, offering - at least in the case of canonical neutrino decay leptogenesis
- a rare opportunity of falsifying an otherwise particularly sturdy
mechanism. This should provide additional motivation for this quest.

The discovery potential of LHC has been investigated for both
massive right-handed neutrinos and gauge bosons associated to
$SU(2)_R$; in particular sensitivity plots corresponding to various
stages of LHC operation can be found in \cite{Wlhcferrari,Wlhckras,cmsreport}, and
scales of the order of $4$-$5$\, TeV in the best case are reached for $W_R$.
Some attention should however be paid to the generality of the
search. The "benchmark" just mentioned is reached under the
assumption that at least one right-handed neutrino $N$ is lighter
than the $W_R$, and therefore that the process: $ p + p \rightarrow
X + W_R \rightarrow  X +N + l^-$ leads to an
on-shell $N$, which can be reconstructed. Being a Majorana state,
the $N$ can decay indifferently into the channels $l^-  + u +
\overline{d}$ or $l^+  + \overline{u} + d$, which, in connection
with the production reaction leads to (non-resonant) dilepton
signals of like or opposite charge in equal quantities. Same sign dilepton channels are particularly clean for background and its observation would establish the Majorana character of neutrino and N masses \cite{dileptonGK}.

Given the importance for excluding leptogenesis, it may thus be
worthwhile to go beyond this benchmark, and to examine the cases
where either the $W_R$, the $N$ or both are virtual.

The case of virtual $N$ still gives a striking signature: namely, in
equal amounts, 2 charged leptons of same or opposite sign + 2 jets,
no missing energy, with the invariant mass resonating at $m_{W_R}$.
The case of $W_R$ heavier than the $N$ is however of particular interest to
us, even if the $W_R$ only intervenes in a virtual way. In this
case, the above process keeps the same overall signature, in
particular equal amounts of like and opposite-sign dileptons, but
resonance is only observed in the (lepton + 2 jets)- branch.

Only in the case where both $N$ and $W_R$ are both above threshold
is the signature reduced to 2 jets + equal amounts of like or
opposite charge dileptons.

It may also be worth pursuing other channels for detection of the
$W_R$, in particular if the $N$'s are heavy. For this purpose, it is
useful to note that, even if heavy $N$'s make the $W_R$ leptonic
decay impossible, it still couples to right-handed quarks whose mass
is known. These quarks, being massive, also link to the left-handed
sector. Hence the process   $p+p \rightarrow X + W_R ^*$ followed by
$W_R ^* \rightarrow t + \overline{b} \rightarrow \overline{b} + b +
l^+ + \nu_L$, the last decay occurring through an ordinary
$W_L$ ($W_R ^*$ stands here for either a real or a virtual
$W_R$)\cite{Frere:1990qm}. This possibility has been used at the
Tevatron detectors \cite{tevatron} but not yet studied for LHC
detectors. The interest in focusing on the top quark in the process
is that it decays without having time to hadronize, and therefore
keeps the helicity correlations. In particular, the final lepton
energy distribution is markedly softer \cite{Frere:1990qm} than in
the similar process where both production and decay occur via $W_L$. A discovery through the top channel would not prove nevertheless that the $W_R$ actually couples to the $N$ but would be a strong hint for it.\footnote{Models where the $W_R$ (or the $Z'$) does not couple to the $N$, and therefore where it has little effect on leptogenesis, are with the $SU(2)'_R$ ($U(1)_N$) subgroup of $E_6$, instead of the ususal $SU(2)_R$ \cite{HMRS}.}
We should finally mention the  case where the right-handed neutrinos
are (nearly) massless, in which case they cannot induce
leptogenesis, but also  cannot interfere with baryogenesis from
another source. This case is difficult to characterize, as the
right-handed closely resembles a heavier left-handed in most
processes. Here again, the above-mentioned top quark intermediary
channel, with its polarization effects would come to help.

\section{Conclusion}

We have shown that the discovery at LHC or future accelerators, of a $W_R$ coupling to a right-handed neutrino and a right-handed charged lepton, would
rule out the possibility to create any relevant lepton asymmetry from the decay of right-handed neutrinos,
see Fig.~2.
A $W_R$ induces extra $N$ decay channels inducing large dilution and washout effects, as well as very fast gauge scatterings (whose decoupling doesn't occur through Boltzmann suppression).
We determined bounds on $m_{W_R}$ and $m_N$ for successful leptogenesis, given in Fig.~\ref{Wbounds} and Eqs.~(\ref{boundsreso}) and (\ref{boundshier}).
Similarly we discussed how the discovery of other particles generally expected in presence of right-handed gauge interactions, or of a $Z'$, could also affect leptogenesis, ruling it out too in some cases.
Leptogenesis from the decay of scalar or fermion triplet would be also basically ruled out in presence
of a $N$ or both a $N$ and a $W_R$ around the TeV scale, unless there is a flavour symmetry to protect one flavour combination from the washout due to these states.

\section*{Acknowledgments}
The authors received partial support from the Belgian Science Policy (IAP VI-11), IISN, as well as from the NSF/PHY05-51164 grant.
  T.H. thanks the FNRS-FRS for
support.

\newpage
\appendix
\section{Several right-handed neutrino case}
\label{appendix}

\begin{figure}[h!]
\centering
\begin{tabular}{ccccccccccccccc}
\includegraphics[width=3.2cm]{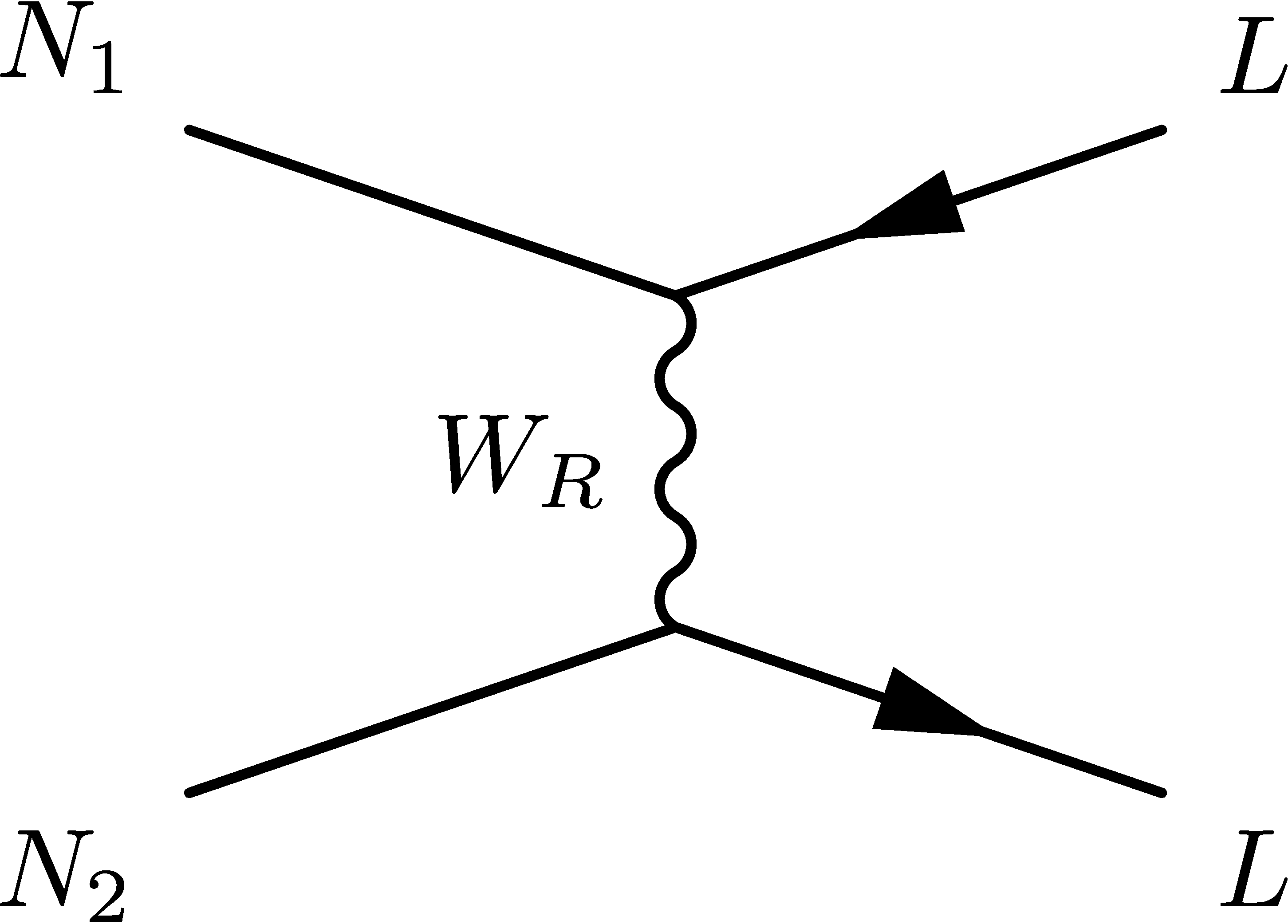}&
\includegraphics[width=3.2cm]{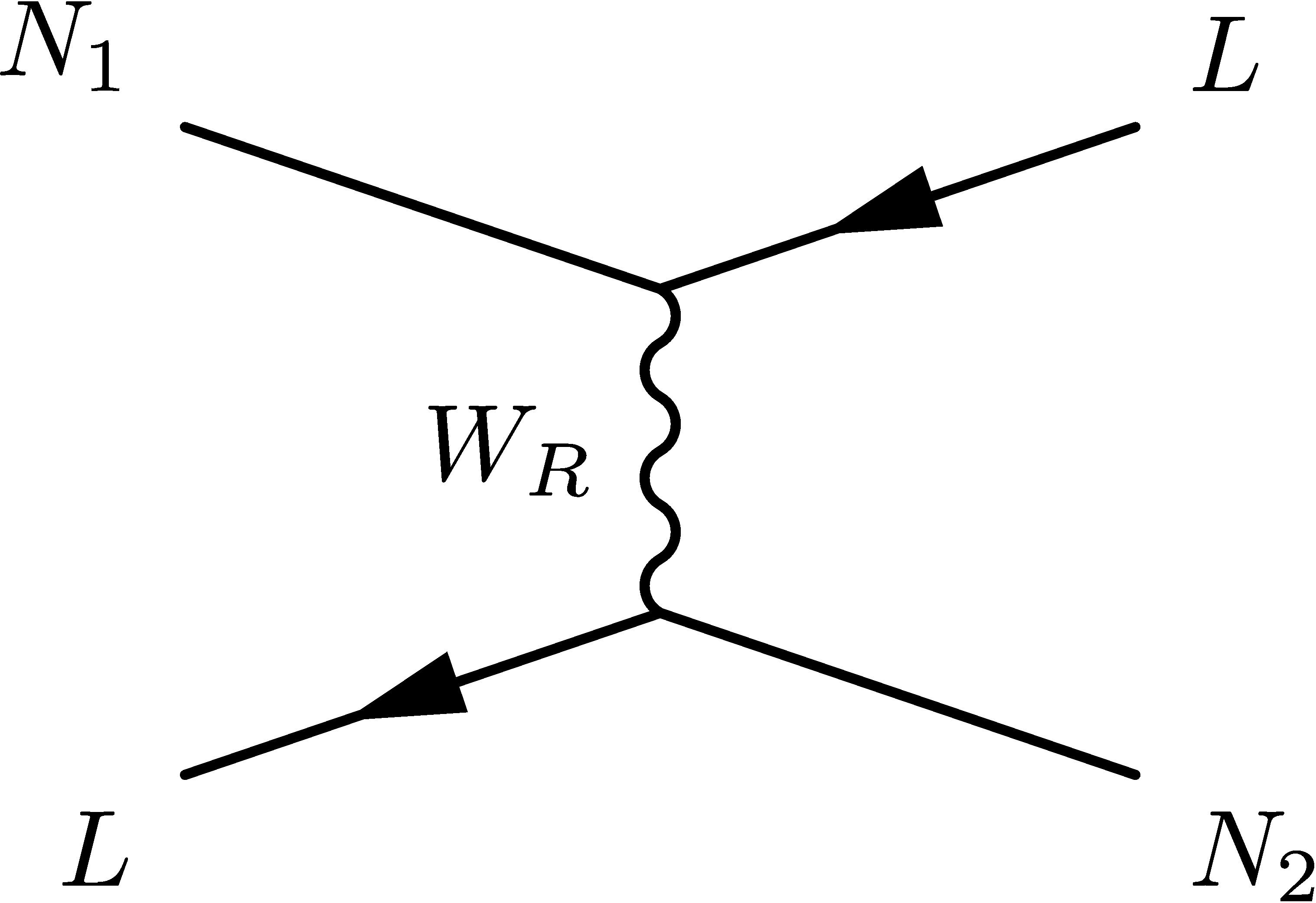}&
\includegraphics[width=3.2cm]{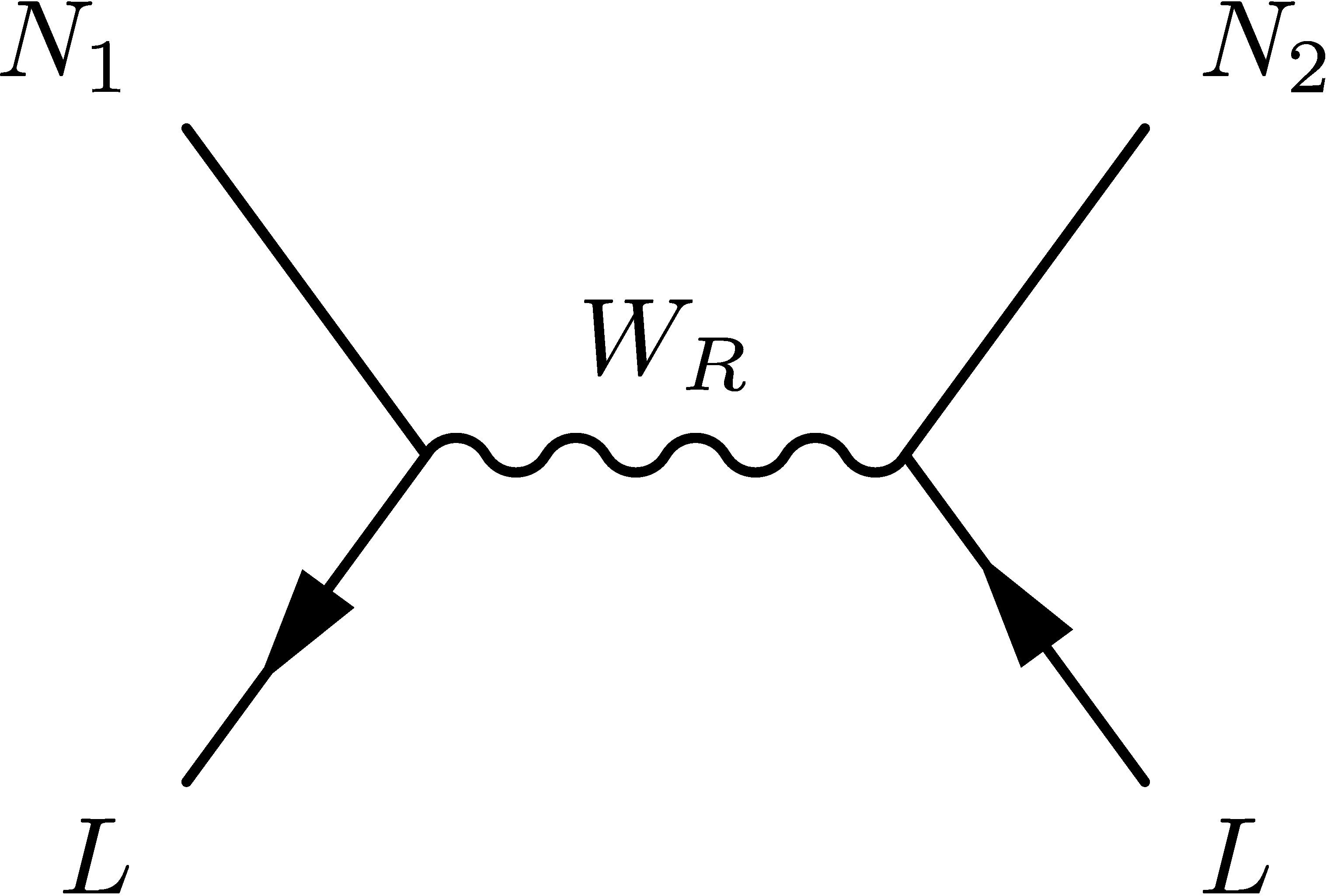}&
\end{tabular}
\caption{Scatterings involving 2 N.}
\label{2N_scatterings}
\end{figure}

\hspace{0cm}With 2 right-handed neutrinos, and at the same level of approximation as for Eqs.~(\ref{NBoGauge}, \ref{LBoGauge}) \footnote{See footnote 4.},  we get the following Boltzmann equations:
\begin{eqnarray}
zH(z)s\, Y'_{N_1} &=& -\left(\frac{Y_{N_1}}{Y_{N_1}^{\rm eq}}-1 \right) \left(\gamma_{N_1}^{(l)} + \gamma_{N_1}^{(W_R)} + 2 \gamma^{N_1}_{Hs} + 4\gamma^{N_1}_{Ht}+ 2 \gamma_{N_1u} + 2 \gamma_{N_1d} + 2 \gamma_{N_1e} \right)  \nonumber\\
& &- \left(\frac{Y_{N_1}^2}{\left.Y_{N_1}^{eq}\right.^2} - 1 \right) \gamma_{N_1N_1}^{(W_R~t)} - \left(\frac{Y_{N_1}Y_{N_2}}{Y_{N_1}^{eq}Y_{N_2}^{eq}} - 1 \right) \gamma_{N_1N_2}^{(W_R~t)} \nonumber \\
&&- \left(\frac{Y_{N_1}}{Y_{N_1}^{eq}} - \frac{Y_{N_2}}{Y_{N_2}^{eq}} \right)\left( \gamma_{N_1N_2}^{(W_R~s)}+\gamma_{N_1N_2}^{(H,L)}\right) \label{N1BoGauge_new}\\
zH(z)s\, Y'_{N_2} &=& -\left(\frac{Y_{N_2}}{Y_{N_2}^{\rm eq}}-1 \right) \left(\gamma_{N_2}^{(l)} + \gamma_{N_2}^{(W_R)} + 2 \gamma^{N_2}_{Hs} + 4\gamma^{N_2}_{Ht}+ 2 \gamma_{N_2u} + 2 \gamma_{N_2d} + 2 \gamma_{N_2e} \right)\nonumber\\
& &- \left(\frac{Y_{N_2}^2}{\left.Y_{N_2}^{eq}\right.^2} - 1 \right) \gamma_{N_2N_2}^{(W_R~t)}  - \left(\frac{Y_{N_2}Y_{N_1}}{Y_{N_2}^{eq}Y_{N_1}^{eq}} - 1 \right) \gamma_{N_2N_1}^{(W_R~t)} \nonumber \\
&&- \left(\frac{Y_{N_2}}{Y_{N_2}^{eq}} - \frac{Y_{N_1}}{Y_{N_1}^{eq}} \right) \left(\gamma_{N_2N_1}^{(W_R~s)} + \gamma_{N_2N_1}^{(H,L)} \right)
\label{N2BoGauge_new}\\
zH(z)s\, Y'_{{\cal L}} &=&\gamma_{N_1}^{(l)} \varepsilon_{N_1} \left(\frac{Y_{N_1}}{Y_{N_1}^{\rm eq}}-1\right) + \gamma_{N_2}^{(l)} \varepsilon_{N_2} \left(\frac{Y_{N_2}}{Y_{N_2}^{\rm eq}}-1\right)\nonumber \\
&&- \left(\gamma_{N_1}^{(l)}+ \gamma_{N_1}^{(W_R)} +\gamma_{N_2}^{(l)}+ \gamma_{N_2}^{(W_R)}   \right)\frac{Y_{{\cal L}}}{2\,Y_{L}^{\rm eq}}\nonumber\\
&&
-\frac{Y_{{\cal L}}}{Y_{L}^{\rm eq}}\left(2\,\gamma_{Ns}^{\rm sub}+2\,\gamma_{Nt}+2\,\gamma^{N_1}_{Ht} + 2\,\gamma^{N_1}_{Hs}\,\frac{Y_{N_1}}{Y_{N_1}^{\rm eq}}
+ \,\gamma_{N_1u}
+ \,\gamma_{N_1d}
+ \,\gamma_{N_1e}\,\frac{Y_{N_1}}{Y_{N_1}^{\rm eq}}\right.
\nonumber\\
&&
\qquad \quad
\left.+2\,\gamma^{N_2}_{Ht} + 2\,\gamma^{N_2}_{Hs}\,\frac{Y_{N_2}}{Y_{N_2}^{\rm eq}}
+ \,\gamma_{N_2u}
+ \,\gamma_{N_2d}
+ \,\gamma_{N_2e}\,\frac{Y_{N_2}}{Y_{N_2}^{\rm eq}}\right)
\,\,\label{LBoGauge_new}
\end{eqnarray}
$\gamma_{Ns}^{\rm sub}$
and $\gamma_{Nt}$ take into account the effects of the $\Delta L=2$ channels  $LH \leftrightarrow \bar{L}H$ and $LL (\bar{L}\bar{L})\leftrightarrow HH$ from both $N_1$ and $N_2$. 
$\gamma_{N_iN_j}^{(W_R ~t)}$ and $\gamma_{N_i N_j}^{(W_R ~s)}$ parametrize the effects of the $W_R$ mediated processes with 2 external $N$, $N_i N_j \leftrightarrow L\bar{L}$ and $N_i L \leftrightarrow N_j L$ respectively, as illustrated in Fig.~\ref{2N_scatterings}. Similarly $\gamma_{N_iN_j}^{(H,L)}$ parametrizes the effects of the Yukawa induced $N_i L \leftrightarrow N_j L$ and $N_i H \leftrightarrow N_j H$ scatterings mediated by a $H$ and a $L$ respectively.  
In these equations it is a very good approximation for the resonant case  to take $m_{N_1}=m_{N_2}$, $\varepsilon_{N_1}=\varepsilon_{N_2}$, $Y_{N_1}^{\rm eq}=Y_{N_2}^{\rm eq}$, $\gamma_{N_1N_2}^{(H,L)}=\gamma_{N_2N_1}^{(H,L)}$, as well as all gauge induced processes equal:
$\gamma_{N_{1}u,d,e}=\gamma_{N_{2}u,d,e}$,  $\gamma^{(W_R ~t,s)}_{N_1N_2}=\gamma^{(W_R ~t,s)}_{N_2 N_1}=\gamma^{(W_R ~t,s)}_{N_1 N_1} =\gamma^{(W_R ~t,s)}_{N_2 N_2}$. $N_1$ and $N_2$ can have significantly different effects only through their Yukawa coupling contributions. 

To compare Eqs.~(\ref{NBoGauge}, \ref{LBoGauge}) and Eqs.~(\ref{N1BoGauge_new}, \ref{N2BoGauge_new}, \ref{LBoGauge_new})
let us first note that the $Y_{N_{1,2}}$ equations differ from the $Y_N$ equation only through the $\gamma_{N_i N_j}^{(W_R ~t,s)}$ and $\gamma_{N_i N_j}^{(H,L)}$ terms. As in the one $N$ case it can be checked 
that the $\gamma_{N_i N_j}^{(W_R ~t)}$ terms have very little effects because their reaction rates are smaller than the $\gamma_{Nu,d,e}$ ones (compare for example in Fig.~3.a $\gamma_{NN}$ with $\gamma_{Ne}+\gamma_{Nu}+\gamma_{Nd}$). The $\gamma_{N_i N_j}^{(W_R ~s)}$ terms on the other hand have a size similar to the one of $\gamma_{Nu,d,s}$ but they are multiplied by $Y_{N_2}-Y_{N_1}$.
This means that their effect is suppressed because those terms could be important only 
as long as the $W_R$ effects ($\gamma_{Nu,d,s}$  and $\gamma_N^{(W_R)}$) dominate the thermalization of the $N's$ (with respect to the Yukawa induced processes), but these $W_R$ effects equally affect $Y_{N_1}$ and $Y_{N_2}$.
Similarly it can be checked that the $\gamma_{N_iN_j}^{(HL~s)}$ are of little importance. They are relevant only for very large values of both $\tilde{m}_1$ and $\tilde{m}_2$, beyond the values of interest for our purpose.
As a result all these terms can be neglected in Eqs.~(\ref{N1BoGauge_new}, \ref{N2BoGauge_new}) and the evolution of $Y_{N_1}$ and $Y_{N_2}$ are essentially the same as the one of $Y_N$ in
Eq.~(\ref{NBoGauge}) replacing $\tilde{m}$ by $\tilde{m}_{1}$ and $\tilde{m}_{2}$ respectively. There are no important differences at this level. 
Differences however can come from Eq.~(\ref{LBoGauge_new}) because this equation involves source and washout terms from both $N_1$ and $N_2$. To discuss this equation it is useful to split it in two parts as follows
\begin{eqnarray}
zH(z)s\, Y'_{{\cal L}a} &=&\gamma_{N_1}^{(l)} \varepsilon_{N_1} \left(\frac{Y_{N_1}}{Y_{N_1}^{\rm eq}}-1\right) - \left(\gamma_{N_1}^{(l)}+ \gamma_{N_1}^{(W_R)} +\gamma_{N_2}^{(l)}+ \gamma_{N_2}^{(W_R)}   \right)\frac{Y_{{\cal L}a}}{2\,Y_{L}^{\rm eq}}\nonumber\\
&&
-\frac{Y_{{\cal L}a}}{Y_{L}^{\rm eq}}\left(2\,\gamma_{Ns}^{\rm sub}+2\,\gamma_{Nt}+2\,\gamma^{N_1}_{Ht} + 2\,\gamma^{N_1}_{Hs}\,\frac{Y_{N_1}}{Y_{N_1}^{\rm eq}}
+ \,\gamma_{N_1u}
+ \,\gamma_{N_1d}
+ \,\gamma_{N_1e}\,\frac{Y_{N_1}}{Y_{N_1}^{\rm eq}}\right.
\nonumber\\
&&
\qquad \quad
\left.+2\,\gamma^{N_2}_{Ht} + 2\,\gamma^{N_2}_{Hs}\,\frac{Y_{N_2}}{Y_{N_2}^{\rm eq}}
+ \,\gamma_{N_2u}
+ \,\gamma_{N_2d}
+ \,\gamma_{N_2e}\,\frac{Y_{N_2}}{Y_{N_2}^{\rm eq}}\right)
\,\,\label{LaBoGauge}\\
zH(z)s\, Y'_{{\cal L}b} &=&\gamma_{N_2}^{(l)} \varepsilon_{N_2} \left(\frac{Y_{N_2}}{Y_{N_2}^{\rm eq}}-1\right) - \left(\gamma_{N_1}^{(l)} + \gamma_{N_1}^{(W_R)} + \gamma_{N_2}^{(l)} + \gamma_{N_2}^{(W_R)}\right)\frac{Y_{{\cal L}b}}{2\,Y_{L}^{\rm eq}}\nonumber\\
&&
-\frac{Y_{{\cal L}b}}{Y_{L}^{\rm eq}}\left(2\,\gamma_{Ns}^{\rm sub}+2\,\gamma_{Nt}+2\,\gamma^{N_1}_{Ht} + 2\,\gamma^{N_1}_{Hs}\,\frac{Y_{N_1}}{Y_{N_1}^{\rm eq}}
+ \,\gamma_{N_1u}
+ \,\gamma_{N_1d}
+ \,\gamma_{N_1e}\,\frac{Y_{N_1}}{Y_{N_1}^{\rm eq}}\right.\nonumber\\
&&
\qquad \quad
\left.+2\,\gamma^{N_2}_{Ht} + 2\,\gamma^{N_2}_{Hs}\,\frac{Y_{N_2}}{Y_{N_2}^{\rm eq}}
+ \,\gamma_{N_2u}
+ \,\gamma_{N_2d}
+ \,\gamma_{N_2e}\,\frac{Y_{N_2}}{Y_{N_2}^{\rm eq}}\right)
\,\,\label{LbBoGauge}
\end{eqnarray}
with $Y_{{\cal L}}=Y_{{\cal L}a}+Y_{{\cal L}b}$. Clearly comparing the $Y_{{\cal L}a}$ ($Y_{{\cal L}b}$) Boltzmann equations with the one $N$ corresponding equation, Eq.~(\ref{LBoGauge}), one observes that these equations are the same except that Eqs.~(\ref{LaBoGauge}, \ref{LbBoGauge})  involve additional washout terms from $N_2$ ($N_1$).
Since these terms can only decrease\footnote{Except if in $\gamma_{Ns}^{\rm sub}$ and $\gamma_{N t}$ there is a destructive interference between the contribution of $N_1$ and $N_2$ but even so, from the effects of all other terms, the following inequalities hold (except for very large $m_N$ close to $10^{15}$~GeV which is not of interest for our purpose).}
the absolute value of the lepton asymmetry obtained\footnote{Note that, due to the $W_R$ effects it is a good approximation to start from thermal distributions of $N_{1,2}$, as explained above. Therefore  there is no change of sign of $Y_{\cal L}$ and the argument applies.} one consequently gets
\begin{eqnarray}\label{Lbound}
Y_{{\cal L}a} (m_N,\varepsilon_N, \tilde{m}_1,\tilde{m}_2 )&<& Y_{\cal L}^{(1)}(m_N,\varepsilon_N,\tilde{m}_1)\\
Y_{{\cal L}b} (m_N,\varepsilon_N, \tilde{m}_1,\tilde{m}_2 )&<& Y_{\cal L}^{(1)}(m_N,\varepsilon_N,\tilde{m}_2)
\end{eqnarray}
which gives
\begin{equation} \label{ineqYL}
Y_{\cal L}(m_N,\varepsilon_N, \tilde{m}_1,\tilde{m}_2 )< Y_{\cal L}^{(1)}(m_N,\varepsilon_N,\tilde{m}_1) + Y_{\cal L}^{(1)}(m_N,\varepsilon_N,\tilde{m}_2) 
\end{equation}
with $Y_{\cal L}^{(1)}$ which refers to the lepton number asymmetry obtained from Eqs.~(\ref{NBoGauge}, \ref{LBoGauge}).
This inequality has several consequences. (i) It means that if leptogenesis is ruled out in the one $N$ case taking $\varepsilon_N<1$ (as above) it will be also ruled out in the 2 $N$ case if we take $\varepsilon_{N_{1,2}}<1/2$ (which is the bound to be considered in this case, see Ref.~\cite{typeIIIlepto}). One just need to apply the results of Figs.~2 and 5 to both terms of Eq.~(\ref{ineqYL}). 
(ii) As Eq.~(\ref{ineqYL}) obviously also holds for the case where 
we neglect the $W_R$ effects in the lepton number Boltzman equation, this conclusion remains true even if we play with flavour (applying to Eq.~(\ref{ineqYL}) the results of Fig.~6). (iii) 
If, for a given value of $m_N=m_{N_1}\simeq m_{N_2}$ and $m_{W_R}$, both $\tilde{m}_1$ and $\tilde{m}_2$ are outside the allowed range of $\tilde{m}$ given in Fig.~7.a, the lepton asymmetry produced will be too small. Numerically it can be checked also that this Figure remains valid to a good approximation for the $\tilde{m}=\tilde{m}_1=\tilde{m}_2$ case. For
$m_{W_R}$ above $\sim50$~TeV the allowed region is shrinked by a hardly visible amount. As for the absolute lower bound on $m_{W_R}$ it is larger in the 2 $N$ case than in the one $N$ case (i.e.~than the value $18$~TeV above) but not by more than a few TeV.
With more than 2 right-handed neutrinos these conclusions remain valid.


\end{document}